\documentclass[preprintnumbers,apsrev4-2,prd,showpacs,floatfix,superscriptaddress,nofootinbib,letterpaper]{revtex4-2}
\usepackage[utf8]{inputenc}
\usepackage{graphicx} % Required for inserting images
\usepackage{xcolor}
\usepackage{bm}
\usepackage{amsmath, amsthm, amssymb}

\usepackage{hyperref}
\usepackage[top = 2cm, bottom = 2cm, right = 2cm, left =2cm]{geometry}
\usepackage{subcaption}
\usepackage{float}
\usepackage{orcidlink}

\newcommand{\beq}{\begin{equation}}
\newcommand{\eeq}{\end{equation}}
\newcommand{\bea}{\begin{eqnarray}}
\newcommand{\eea}{\end{eqnarray}}

\newcommand{\wtm}{\widetilde{M}}
\begin{document}

\title{Damour–Solodukhin Wormholes: Matter Sources, Observational Constraints, and Accretion Dynamics}

\author{G. Alencar \orcidlink{0000-0002-3020-4501}}
\email{geova@fisica.ufc.br }
\affiliation{Departamento de Física, Universidade Federal do Ceará, Caixa Postal 6030, Campus do Pici, 60455-760 Fortaleza, Ceará, Brazil}
\author{V. H. U. Borralho \orcidlink{0009-0009-0085-6152}}
\email{victorborralho@fisica.ufc.br} 
\affiliation{Departamento de Física, Universidade Federal do Ceará, Caixa Postal 6030, Campus do Pici, 60455-760 Fortaleza, Ceará, Brazil}
\author{T. M. Crispim \orcidlink{0009-0004-6604-4695}}
\email{tiago.crispim@fisica.ufc.br} 
\affiliation{Departamento de Física, Universidade Federal do Ceará, Caixa Postal 6030, Campus do Pici, 60455-760 Fortaleza, Ceará, Brazil}

\author{Marcos V. de S. Silva \orcidlink{0000-0002-8080-9277}%\footnote{Author to whom any correspondence should be addressed.}
}
\email{marcos.sousa@uva.es}
\affiliation{Department of Theoretical Physics, Atomic and Optics, and Laboratory for Disruptive Interdisciplinary Science (LaDIS), Campus Miguel Delibes, \\ University of Valladolid UVA, Paseo Bel\'en, 7,
47011 - Valladolid, Spain}

\begin{abstract}
We investigate the physical realization and observational properties of Damour--Solodukhin wormholes within general relativity. We first construct an explicit matter source supporting this geometry, consisting of a phantom scalar field coupled to nonlinear electrodynamics, and examine both electric and magnetic configurations. We then study the propagation of photons and derive constraints on the wormhole deformation parameter $\lambda$ using Event Horizon Telescope observations of Sgr A*. We show that, for $\lambda^2 \leq 1/2$, the shadow radius coincides exactly with the Schwarzschild value, although the radial geometry continues to affect gravitational lensing and the structure of photon rings. Using a thin accretion disk with a Johnson SU emissivity profile, we investigate the resulting optical images and examine the effects of the wormhole throat on their structure. Finally, we analyze stationary relativistic accretion and demonstrate that, for identical boundary conditions, the energy density and particle number density are independent of $\lambda$ throughout the common exterior region, while the radial velocity retains an explicit dependence on the deformation parameter. We also derive the condition $0 < w < (1+4\lambda^2)^{-1}$ for the existence of an exterior sonic point. These results highlight the observational degeneracy between Damour--Solodukhin wormholes and Schwarzschild black holes in the small-deformation regime, while identifying the geometric and dynamical effects through which this degeneracy may be investigated.
\end{abstract}

%
%\pacs{72.80.Le, 72.15.Nj, 11.30.Rd}
%
\maketitle
\section{Introduction}

Wormholes are solutions of General Relativity (GR) that can describe regular, asymptotically flat spacetimes connecting two asymptotic regions through a throat \cite{Morris:1988cz}. A pioneering construction was introduced by Einstein and Rosen, giving rise to the concept of the Einstein–Rosen bridge \cite{Einstein:1935tc}. However, this bridge is not traversable in the conventional interpretation of the maximally extended Schwarzschild spacetime. Later, in 1973, Ellis \cite{Ellis:1973yv} obtained a regular, traversable wormhole solution, known as the Ellis drainhole, supported by a scalar field with a reversed kinetic-energy sign. In the same year, Bronnikov \cite{Bronnikov:1973fh} independently investigated exact solutions in scalar–tensor theories of gravity, contributing to the development of this class of geometries. A systematic treatment of traversable wormholes was subsequently presented by Morris and Thorne \cite{Morris:1988cz}, who established the geometric and physical conditions required for their existence. Among their results, they showed that, within the framework of GR, maintaining a traversable throat generally requires a violation of the null energy condition (NEC), motivating the introduction of what is commonly referred to as exotic matter.

Since then, numerous studies have investigated the matter sources capable of supporting wormhole geometries \cite{Crispim:2024dgd,deSSilva:2025ncg,Crispim:2024lzf,Su:2024gxp,Matyjasek:2020cmi,Simpson:2018tsi,Bronnikov:2012ch,Bronnikov:2015pha,Bolokhov:2012kn,Bronnikov:2016xvj,Lobo:2005us,Lobo:2009ip,Mehdizadeh:2015jra,Garcia:2010xb,Lobo:2017cay,deSSilva:2024gdc}. Among the approaches explored in the literature are nonlinear electrodynamics (NED) and scalar fields with a reversed kinetic-energy sign, commonly known as ghost scalar fields. Depending on the specific model and field configuration, NED can modify the effective energy–momentum tensor and contribute to the regularity of the spacetime, whereas ghost scalar fields can provide the energy-condition violation required to sustain an open throat \cite{Bronnikov:2017sgg,Bronnikov:2021uta,Alencar:2026kqx,Alencar:2026hjf,deSSilva:2025ncg}.

%A scalar field can be described by the action\begin{equation}S_{\phi} = \int d^4x \sqrt{-g}\left[-\epsilon \partial_{\mu}\phi\partial^{\mu}\phi+V(\phi)\right],\end{equation}where $\epsilon$ determines the nature of the scalar field. With this convention, $\epsilon>0$ corresponds to a canonical scalar field, whereas $\epsilon<0$ describes a phantom scalar field. Another widely studied approach is the so-called trapped-ghost model, in which the phantom behavior of the scalar field is confined to a limited region of spacetime, typically in the vicinity of the wormhole throat, while the field can recover its canonical behavior farther away \cite{Bronnikov:2018vbs,Bronnikov:2010hu,Bronnikov:2022bud}. In this model, the constant $\epsilon$ is replaced by a field-dependent function $\epsilon(\phi)$, allowing the sign of the kinetic term to vary throughout spacetime. However, this construction does not automatically eliminate the stability problems associated with ghost degrees of freedom. Moreover, at points where $\epsilon(\phi)=0$, the scalar-field kinetic term vanishes, potentially leading to a degeneracy in the field dynamics. Therefore, the regularity of the spacetime geometry and the stability of perturbations must be investigated separately \cite{Bronnikov:2015kea}.

On the other hand, the Damour–Solodukhin Wormhole (DSWH) \cite{Damour:2007ap} is a model constructed by deforming the Schwarzschild spacetime through a parameter $\lambda$, which replaces the event horizon with a wormhole throat. The primary motivation behind this model is to investigate how closely a wormhole can mimic the properties of a black hole, and to determine the extent to which black hole geometries and their associated physical features can be reproduced by varying $\lambda$. For a Schwarzschild black hole, a particle that crosses the event horizon cannot escape back to the exterior. In the DSWH spacetime, however, the parameter $\lambda$ can be exponentially small, making the classical ``wormhole bounce'' timescale comparable to the quantum Hawking evaporation time of a black hole of the same mass. For instance, for the supermassive object at the center of our galaxy to be modeled as a DSWH, the deformation parameter must be of the order of $\lambda \ll \exp(-10^{15})$ \cite{Damour:2007ap}. Consequently, the wormhole becomes practically indistinguishable from a black hole over standard astrophysical timescales. More recently, extensions of this model to include electric charge and rotation have been developed \cite{Solodukhin:2026mux}. This spacetime has also been extensively studied regarding gravitational lensing and shadow images \cite{Tsukamoto:2023rzd,Tsukamoto:2020uay,Tsukamoto:2026beg}, timelike orbits \cite{Ho:2026gmy}, scalar perturbations \cite{Bueno:2017hyj}, accretion dynamics \cite{Yusupova:2025niy}, and late-time tails and echoes arising from perturbations \cite{Qian:2025lau,Qian:2024zvq}.

In this work, we investigate the matter sources supporting this spacetime by combining NED with a scalar field. While wormholes share spherical symmetry with black holes, their fundamental distinction lies in the complete absence of an event horizon. Literature has established that the shadows generated by horizonless configurations, including naked singularities \cite{Bambi:2008jg,Ortiz_2015,Shaikh_2018} and various wormhole geometries \cite{Bambi:2013nla,Azreg-Ainou:2014dwa,Ishkaeva:2023xny,Alloqulov:2024olb,Guerrero:2021ues}, can closely resemble those of black holes in specific regimes \cite{Shaikh:2018kfv}. This observational degeneracy, coupled with the landmark imaging of black hole shadows by the Event Horizon Telescope (EHT) \cite{EventHorizonTelescope:2019dse,EventHorizonTelescope:2022wkp}, has reignited considerable interest in probing these compact entities and searching for signatures capable of discerning them in GR and modified gravity frameworks \cite{DeFalco:2020afv,DeFalco:2023twb,Jana:2024pdh}.

Building upon the theoretical indistinguishability proposed by Damour and Solodukhin, we carry out classical observational tests to evaluate how effectively the DSWH geometry mimics a black hole. Specifically, we study photon motion in this background and utilize EHT observational data to set $1\sigma$ and $2\sigma$ constraints on the parameter $\lambda$. Additionally, we model the visual appearance of the wormhole encircled by a thin accretion disk via a Johnson SU profile across various deformation parameter values, alongside a detailed analysis of radial accretion dynamics. Through these classical diagnostics, we assess whether the optical and thermodynamic signatures of the DSWH asymptotically converge to those of a Schwarzschild black hole in the small-$\lambda$ regime.

This paper is organized as follows: In Sec.~\ref{sec2}, we introduce the theoretical background of the DSWH spacetime, while the underlying field equations supported by NED and a scalar field are presented in Sec.~\ref{sec3}. In Sec.~\ref{sec4}, we examine photon geodesics. In Sec.~\ref{sec5} we determine the shadow radius, and constrain the parameter $\lambda$ using EHT observational data. In Sec.~\ref{sec6}, we analyze the optical appearance of the wormhole surrounded by a thin accretion disk using the Johnson SU distribution, followed by the study of radial accretion dynamics in Sec.~\ref{sec7}. Finally, in Sec.~\ref{sec8}, we present our final remarks and main conclusions. Throughout this work, we adopt units where $c=8\pi G = 1$ and use the metric signature $(-,+,+,+)$.

%The resulting phantom scalar field is expressed in terms of elliptic integrals and remains regular throughout the spacetime when written in terms of the proper radial coordinate $\ell$. An interesting feature of this wormhole is that its total energy density vanishes identically, while the pressure remains nonzero, so that the null energy condition (NEC) is still violated. Thus, although the NED and scalar-field sectors each contribute a nonzero energy density, their contributions cancel exactly, yielding a vanishing total energy density. 

\section{DSWH spacetime}\label{sec2}
The most general way to describe the geometry of a traversable wormhole is through the Morris-Thorne metric \cite{Morris:1988cz}
\begin{equation}\label{morristhorne}
    ds^2=-e^{2\Phi(r)}dt^2+\left(1-\frac{b(r)}{r}\right)^{-1}dr^2+r^2d\theta^2 +r^2 \sin^2 \theta d\varphi^2,
\end{equation}
where $b(r)$ is the shape function and $\Phi(r)$ is the redshift function. For the geometry to represent a traversable wormhole, the following conditions must be satisfied:
\begin{equation}
    b(r_{th})=r_{th},
\end{equation}
where $r_{th}$ is the throat radius. In addition, the flare-out condition must be satisfied:
\begin{equation}
    b'(r_{th})<1.
\end{equation}
Finally, the redshift function must remain finite throughout the spacetime, i.e., $|\Phi(r)|<\infty$.

For this metric, the components of the Einstein tensor are:
\begin{align}
    G_0^0&=-\frac{b'(r)}{r^2}, \\
     G_1^1&=-\frac{b(r)}{r^3}-\frac{2 b(r) \Phi '(r)}{r^2}+\frac{2 \Phi '(r)}{r} ,\\
     G_2^2&=G_3^3=-\frac{b'(r)}{2 r^2}-\frac{b'(r) \Phi '(r)}{2 r}+\frac{b(r)}{2 r^3}-\frac{b(r) \Phi '(r)}{2 r^2}-\frac{b(r) \Phi ''(r)}{r}-\frac{b(r) \Phi '(r)^2}{r}+\Phi ''(r)+\Phi '(r)^2+\frac{\Phi '(r)}{r}.
\end{align}
For the DSWH, the functions $\Phi(r)$ and $b(r)$ take the particular form:
\begin{equation}\label{functionsmoristhorne}
    \Phi(r)=\frac{1}{2}\ln\left(1-\frac{b(r)}{r}+\lambda^2\right) 
    \quad\text{and} \quad b(r) = 2  \wtm,
\end{equation}
leading to the line element \cite{Damour:2007ap}:
\begin{equation}
    ds^2=-\left(1-\frac{2 \wtm}{r}+\lambda^2\right)dt^2
    +\left(1-\frac{2 \wtm}{r}\right)^{-1}dr^2+r^2d\Omega^2.
\end{equation}
Indeed, it satisfies the Morris-Thorne conditions:
\begin{equation}
    b(2 \wtm)=2 \wtm \quad\text{and}\quad b'(2 \wtm)=0<1.
\end{equation}
Furthermore,
\begin{equation}
    \Phi(2 \wtm)=\ln|\lambda| 
    \quad \text{and} \quad 
    \Phi(r\to\infty)=\frac{1}{2}\ln(1+\lambda^2).
\end{equation}
For $\lambda\neq0$, the redshift function remains finite at $r=2 \wtm$. This is the main distinction from the Schwarzschild spacetime: the surface $r=2 \wtm$ is no longer an event horizon, but instead corresponds to a wormhole throat connecting two asymptotically flat regions. Although the spatial part of the metric is identical to that of Schwarzschild, its temporal component is modified by the parameter $\lambda$, resulting in a different causal structure.

To ensure the standard normalization of the temporal coordinate at spatial infinity, we introduce the rescaling \cite{Ho:2026gmy,Amir:2018pcu}
\begin{equation}
    \tilde{t}=\sqrt{1+\lambda^2}\,t.
\end{equation}
The metric then takes the form
\begin{equation}
    ds^2=
    -\left(1-\frac{2 \wtm}{(1+\lambda^2)r}\right)d\tilde{t}^{\,2}
    +\left(1-\frac{2 \wtm}{r}\right)^{-1}dr^2
 +r^2d\theta^2 +r^2 \sin^2 \theta d\varphi^2,
\end{equation}
For convenience, we further introduce the reparametrization \cite{Ho:2026gmy,Amir:2018pcu}
\begin{equation}
    M=\frac{\wtm}{1+\lambda^2},
\end{equation}
so that the metric can be written as
\begin{equation}\label{metric}
    ds^2=
    -\left(1-\frac{2M}{r}\right)d\tilde{t}^{\,2}
    +\left(1-\frac{2(1+\lambda^2)M}{r}\right)^{-1}dr^2
    +r^2d\theta^2 +r^2 \sin^2 \theta d\varphi^2,
\end{equation}
where $\lambda$ determines the throat radius $r_{th}=2(1+\lambda^2)M$.

It is convenient to perform a coordinate transformation to write the metric in terms of the proper radial distance $\ell$, since the radial coordinate $r$ is defined only for $r\geq r_{th}$. This transformation allows us to extend the radial coordinate across the throat and describe the complete wormhole geometry:
\begin{equation}\label{dl}
        d\ell^2=\left(1-\frac{2(1+\lambda^2)M}{r}\right)^{-1}{dr^2}.
\end{equation}
Thus, we obtain the proper radial distance $\ell(r)$:
\begin{equation} \label{proper}
\ell(r)=\pm \left(r \sqrt{1-\frac{2 M \left(\lambda^2+1\right)}{r}}+2 M \left(\lambda^2+1\right) \text{arctanh}\left(\sqrt{1-\frac{2 M \left(\lambda^2+1\right)}{r}}\right)\right).
\end{equation}
The coordinate transformation is defined for $r\geq 2M(\lambda^2+1)$, with the throat located at $r=2M(\lambda^2+1)$ and corresponding to $\ell=0$. The two branches in Eq.~\eqref{proper} correspond to the two asymptotically flat regions of the wormhole, with $\ell=0$ at the throat. Hence, the proper radial coordinate spans the entire range $-\infty<\ell<+\infty$.

Since $r_{th}=2M(\lambda^2+1)$, the throat corresponds to the minimum value of the radial coordinate. In particular, $\ell=0$ at the throat, while $r\rightarrow\infty$ corresponds to $\ell\rightarrow\pm\infty$ on the two respective branches. The coordinate $\ell$ therefore provides a continuous description of both sides of the wormhole.

Due to the functional form of $\ell(r)$, it is not possible to invert the relation analytically to obtain $r(\ell)$. We therefore write $r(\ell)$ implicitly and express the metric as
\begin{equation}\label{metricproper}
    ds^2=-\left(1-\frac{2M}{r(\ell)}\right)d\tilde{t}^{\,2}+d\ell^2+r^2(\ell)d\Omega^2.
\end{equation}

Here, $r(\ell)$ denotes the areal radius as a function of the proper radial coordinate, satisfying $r(\ell)\geq r_{th}$ and $r(0)=r_{th}$. The two asymptotically flat regions are represented by $\ell>0$ and $\ell<0$, which are smoothly connected at the throat $\ell=0$. 

\subsection{Embedding Diagram}
To visualize more clearly the structure of this spacetime, we consider the embedding diagram. This consists of embedding the spatial geometry of the spacetime given by Eq.~\eqref{metric}, by fixing $\tilde{t}=\mathrm{const.}$ and $\theta=\pi/2$, into a two-dimensional cylindrical surface embedded in a three-dimensional Euclidean space. Thus, the induced two-dimensional metric is
\begin{equation}\label{metric2D}
    ds^2=
    \left(
    1-\frac{2(1+\lambda^2)M}{r}
    \right)^{-1}dr^2
    +r^2d\varphi^2,
\end{equation}
while the Euclidean metric with cylindrical symmetry is given by
\begin{equation}\label{embeddingmetric}
    ds^2_{3D}
    =d\tilde r^2+\tilde r^2d\varphi^2+dz^2
    =
    \left[
    \left(\frac{d\tilde r}{dr}\right)^2
    +
    \left(\frac{dz}{dr}\right)^2
    \right]dr^2
    +\tilde r^2d\varphi^2.
\end{equation}
Comparing Eqs.~\eqref{metric2D} and \eqref{embeddingmetric}, we obtain
\begin{equation}
    \tilde r=r
    \quad\text{and}\quad
    \left(\frac{dz}{dr}\right)^2
    =
    \left(
    1-\frac{2(1+\lambda^2)M}{r}
    \right)^{-1}-1,
\end{equation}
which leads to
\begin{equation}
    z(r)
    =
    \pm2
    \sqrt{
    2(1+\lambda^2)M
    \left[
    r-2(1+\lambda^2)M
    \right]
    }.
\end{equation}
The two signs correspond to the two sides of the wormhole, which meet at the throat. The function $z(r)$ is well defined throughout the physical domain and satisfies
\begin{equation}
    z(r_{th})=0.
\end{equation}

\begin{figure}[H]
    \centering
    \includegraphics[width=0.45\linewidth]{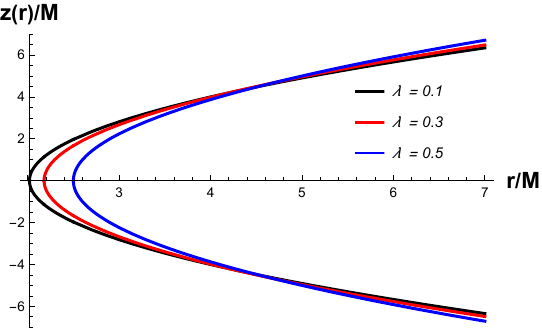}
    \includegraphics[width=0.45\linewidth]{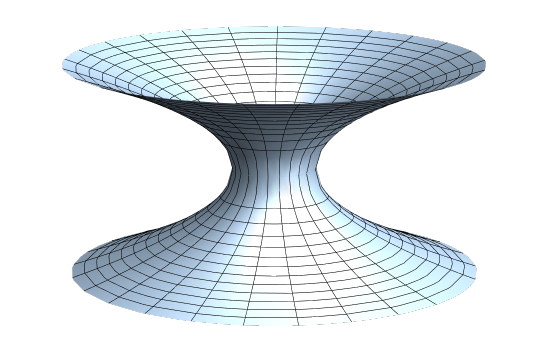}
    \caption{Plot of the function $z(r)$ (left) for different values of $\lambda$ and the corresponding embedding diagram for the DSWH  with $\lambda=0.1$ (right).}
    \label{embedding}
\end{figure}

In Fig.~\ref{embedding}, we show the function $z(r)$ as well as the corresponding embedding diagram obtained by constructing a surface of revolution around the $z$-axis. The embedding diagram is identical to that of Schwarzschild up to the corresponding rescaling of the mass parameter, because the DS deformation affects only the temporal component of the metric. Nevertheless, the causal structure is fundamentally different: in Schwarzschild, the corresponding Schwarzschild radius is an event horizon, where the temporal and radial coordinates exchange their causal character, whereas for $\lambda\neq0$ this horizon is removed in the DS spacetime, allowing causal curves to cross the throat and connect the two asymptotically flat regions.

\section{Field Sources}\label{sec3}
The matter content required to support the DSWH has previously been investigated from a quantum-field perspective. In particular, Matyjasek \cite{Matyjasek:2020cmi} studied the renormalized stress--energy tensor of massive quantized fields in this background, examining whether their contributions could satisfy the conditions necessary to sustain the wormhole geometry. Here, we pursue a different approach by constructing an explicit classical matter source within GR. Specifically, we show how the DS geometry can be supported by a phantom scalar field coupled to NED, considering both electric and magnetic configurations.

More generally, the traversability of the wormhole is directly related to the energy conditions at the throat. In particular, the flare-out condition requires a violation of the NEC. Within GR, this violation manifests itself as a sufficiently negative radial pressure at the throat, providing the necessary tension to keep the wormhole throat open. We investigate whether this behavior can be realized through a combination of a scalar field and NED, a framework widely used in the literature to study this type of geometry \cite{Bronnikov:2021uta,Bronnikov:2023aya,Crispim:2024lzf,Crispim:2024dgd}. To this end, we start from the Einstein--Hilbert action in the presence of a scalar field and a NED Lagrangian, given by
\begin{equation}\label{action}
S[g_{\mu\nu},\phi,A_{\mu}] = \int\sqrt{-g}\left[\frac{R}{2}-\epsilon\partial_{\mu}\phi\partial^{\mu}\phi +V(\phi)+\mathcal{L}(\mathcal{F})\right]d^4x,
\end{equation}
where $\phi$ is the scalar field, $V(\phi)$ is the potential associated with the scalar field, $\mathcal{L}(\mathcal{F})$ is the NED Lagrangian that depends on the electromagnetic invariant $\mathcal{F}=F_{\mu\nu}F^{\mu\nu}$, and the parameter $\epsilon=1,-1$ defines whether the scalar field is canonical or phantom, respectively. By minimizing the action above, we obtain the following field equations:
\begin{equation}
    2\epsilon\Box\phi+\frac{d V(\phi)}{d \phi}=0,
\end{equation}
\begin{equation}\label{MaxwellNED}
    \nabla_\mu\left(\mathcal{L}_{\mathcal{F}} F^{\mu\nu}\right)=0,
\end{equation}
\begin{equation}
    G_{\mu\nu}= T_{\mu\nu}^{\phi}+ T_{\mu\nu}^{NED},
\end{equation}
where $\mathcal{L}_{\mathcal{F}}=d\mathcal{L}/d\mathcal{F}$, $\Box=\nabla_\mu\nabla^\mu$ is the d'Alembertian operator and  $T_{\mu\nu}^\phi$ and $T_{\mu\nu}^{NED}$ are, respectively, the energy-momentum tensors of the scalar field and the electromagnetic field, given by:
\begin{align}
&T_{\mu\nu}^\phi=2\epsilon\partial_\mu\phi\partial_\nu\phi-g_{\mu\nu}\left[\epsilon\partial_\alpha\phi\partial^\alpha\phi-V(\phi)\right],\\
&T_{\mu\nu}^{NED}=g_{\mu\nu}\mathcal{L}(\mathcal{F})-4\mathcal{L}_{\mathcal{F}}(\mathcal{F})F_{\mu\alpha}F_{\nu}^{\alpha}.
\end{align}

For the metric \eqref{metric}, the components of the Einstein tensor are:

\begin{align}
    G_0^0&=0,  \label{g00}\\
     G_1^1&=-\frac{2 M \lambda^2}{r^2 (r-2 M )}, \label{g11}\\
     G_2^2&=G_3^3=\frac{M  \lambda^2}{r (r-2 M )^2}-\frac{M ^2 \lambda^2}{r^2 (r-2 M )^2}.
\end{align}

The vanishing of the temporal component of the Einstein tensor, $G_0^0=0$, implies that the total energy density associated with the field sources must vanish. Therefore, the contributions from the scalar and NED fields must exactly compensate each other, such that
\begin{equation}
    T_0^0{}^{\phi}+T_0^0{}^{NED}=0.
\end{equation}
Thus, although the individual field sources may contribute nontrivially to the energy density, their combined contribution is identically zero. 

The violation of the NEC can be directly verified from Eqs.~\eqref{g00} and \eqref{g11}. Since $G_0^0=0$ and $G_1^1=-2M\lambda^2/[r^2(r-2M)]$, their sum is negative for $M>0$, $\lambda\neq 0$, and $r>2M$. Therefore, the NEC is violated throughout the considered radial domain.

\subsection{Magnetic Source}
In order to obtain a magnetically charged source, we take the following ansatz for the components of the Faraday tensor:
\begin{equation}
F_{23}=P\sin\theta \implies \mathcal{F}=\frac{2P^2}{r^4}.
\end{equation}
 In this way, given the spherical symmetry of the solution, we consider a purely radial scalar field, $\phi=\phi(r)$. Together with the magnetic ansatz introduced above, this specifies the nonvanishing components of the energy-momentum tensors, which can be written as

\begin{align}
    T_{\mu}^{\nu \;NED}&= \text{diag}\left(\mathcal{L}(r),\mathcal{L}(r),\mathcal{L}(r)-\frac{4P^2\mathcal{L}_\mathcal{F}(r)}{r^4},\mathcal{L}(r)-\frac{4P^2\mathcal{L}_{\mathcal{F}}(r)}{r^4} \right) \\
    T_{\mu}^{\nu \;\phi}&=\delta_\mu^{\nu}V(r)+\frac{\epsilon  [2M (1+\lambda^2)-r] \phi '(r)^2}{r}\text{diag}\left(1,-1,1,1 \right),
\end{align}

Substituting these expressions into the Einstein equations and using the metric in Eq.~\eqref{metric}, we obtain the following set of equations:
\begin{equation}\label{mFE11}
    \mathcal{L}(r)+V(r)+\frac{\epsilon  [2 M (1+\lambda^2)-r]}{r} \phi '(r)^2=0,
\end{equation}
\begin{equation}\label{mFE22}
    \mathcal{L}(r)+V(r)-\frac{2 \lambda ^2 M}{r^2(2 M-r)}- \frac{\epsilon  \left[2 M\left(\lambda ^2+1\right) -r\right]}{r} \phi '(r)^2=0
\end{equation}
\begin{equation}\label{mFE33}
  \mathcal{L}(r)+V(r)-\frac{\lambda ^2 M (r-M)}{r^2(r-2 M)^2}- \frac{\epsilon  \left[r-2 M\left(\lambda ^2+1\right) \right]}{r} \phi '(r)^2-\frac{4P^2\mathcal{L}_{\mathcal{F}}(r)}{r^4}=0
\end{equation}
\begin{align}\label{scalareq}
 2 \epsilon  \left[\frac{\left(4M^2 \left(\lambda^2+1\right)-3M \left(\lambda^2+2\right) r+2 r^2\right) \phi '(r)}{r (r-2M)}+\left(r-2M \left(\lambda^2+1\right)\right) \phi ''(r)\right]+\frac{r V'(r)}{\phi '(r)}=0
\end{align}

\begin{figure}[H]
    \centering
    \includegraphics[width=0.5\linewidth]{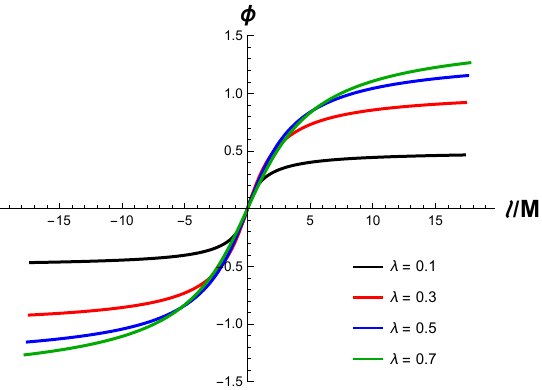}
    \caption{Behavior of the phantom scalar field $\phi(r(\ell))$ for some values of $\lambda$.}
    \label{graficophi}
\end{figure}

Initially, we subtract Eqs. \eqref{mFE11} and \eqref{mFE22}, obtaining the differential equation for $\phi(r)$
\begin{equation}
    r \epsilon  \left[2M \left(\lambda^2+1\right)-r\right] \phi '(r)^2-\frac{M \lambda^2}{r-2M}=0
\end{equation}
whose general solution is
\begin{equation}
    \phi(r)=\phi_0-\sqrt{2\lambda^2}  F\left(\arcsin\left(\sqrt{\frac{2M}{r}}\right)|\lambda^2+1\right).
\end{equation}
The scalar field must be real, and for this to be the case, $\epsilon=-1$, indicating a phantom scalar field. We must fix the integration constant $\phi_0$ such that $\displaystyle \lim_{r \to r_{th}^{+}} \phi(r) = 0$. Thus, the solution for the scalar field is
\begin{equation}
   \phi(r)= \frac{\sqrt{2\lambda^2}  }{\sqrt{\lambda^2+1}} K\left(\frac{1}{\lambda^2+1}\right)-\sqrt{2\lambda^2}  F\left(\arcsin\left(\sqrt{\frac{2M}{r}}\right)|\lambda^2+1\right).
\end{equation}
where $K(m)$ and $F(\alpha|m)$ are the complete and incomplete elliptic integrals of the first kind, respectively. They are defined as:
\begin{equation}
    K(m) = \int_0^{\pi/2}
\frac{d\theta}{\sqrt{1-m\sin^2\theta}} \quad \text{and} \quad F(\alpha|m) = \int_0^\alpha
\frac{d\theta}{\sqrt{1-m\sin^2\theta}}.
\end{equation}
We can see the behavior of $\phi(r(\ell))$ in Fig.~\ref{graficophi}.

Next, in order to determine the potential $V(r)$, we must solve Eq.~\eqref{scalareq}, which takes the form:
\begin{equation}
-M \lambda^2 +r^3 (r-2M) V'(r)=0
\end{equation}
Thus, the general solution for the potential is:
\begin{equation}
V(r)=V_0+\frac{\lambda^2}{4Mr}
+\frac{\lambda^2}{4r^2}
-\frac{\lambda^2}{8M^2}
\ln\left(\frac{r}{r-2M}\right).
\end{equation}
where we fix $V_0=0$. 

\begin{figure}[H]
    \centering
    \includegraphics[width=0.5\linewidth]{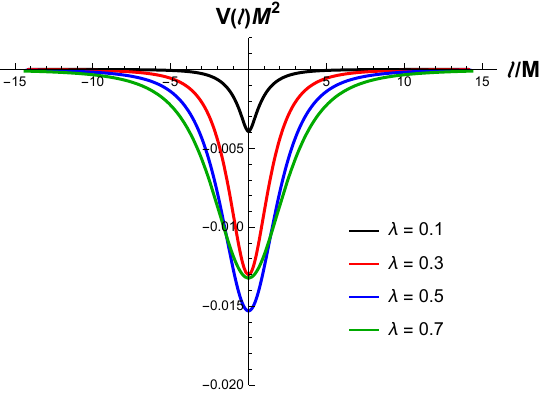}
    \caption{Behavior of the scalar-field potential $V(r(\ell))$ for different values of $\lambda$.}
    \label{potencial}
\end{figure}

We can see its behavior in terms of the coordinate $\ell$ in Fig.~\ref{potencial}, presenting a minimum at the throat $(\ell=0)$ and asymptotically approaching zero.

\begin{figure}[H]
    \centering
\includegraphics[width=0.45\linewidth]{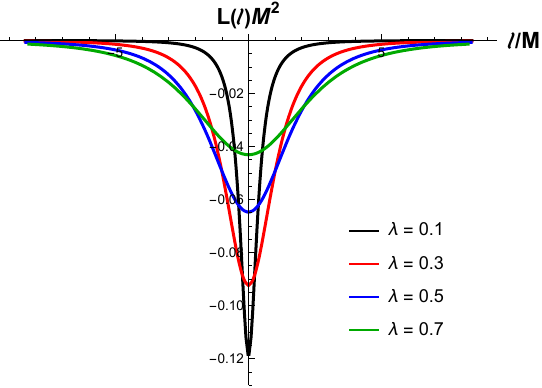}    \includegraphics[width=0.45\linewidth]{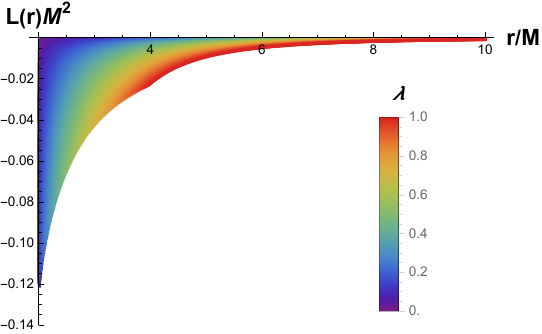}
    \caption{Behavior of the Lagrangian $\mathcal{L}$ in terms of the proper radial coordinate $\ell$ (left) and the radial coordinate $r$ (right), for a continuous range of $\lambda$, $0.1<\lambda<1$, in the latter case.}
    \label{lagrangianamag}
\end{figure}

Next, we determine the functions $\mathcal{L}(r)$ and $\mathcal{L}_{F}$. From Eqs.\eqref{mFE11} and \eqref{mFE33}, we can determine $\mathcal{L}(r)$ and $\mathcal{L}_{\mathcal{F}}(r)$, respectively:
\begin{align}
 \mathcal{L}(r)&= -\frac{\lambda^2}{4M(r-2M)}
    +\frac{\lambda^2}{4r^2}
    +\frac{\lambda^2}{8M^2}
    \ln\left(\frac{r}{r-2M}\right). \\
    \mathcal{L}_{F}(r)&=\frac{\lambda ^2 M r^2 (M-r)}{4 P^2 (r-2 M)^2}
\end{align}
Using the relation $F=2P^2/r^4$, we can invert it to obtain $r(F)$, and thus write the Lagrangian $\mathcal{L}(F)$:
\begin{equation}
 \mathcal{L}(\mathcal{F})=   \frac{\lambda^2}{8M^2}
\left[
\ln\left(
\frac{2^{1/4}\sqrt{P}}
{\left(2^{1/4}\sqrt{P}-2M\mathcal{F}^{1/4}\right)}
\right)
+\frac{2M \mathcal{F}^{1/4}}{2M\mathcal{F}^{1/4}-2^{1/4}\sqrt{P}}
+\frac{M^2\sqrt{2\mathcal{F}}}{P}
\right]
\end{equation}
The functions $\mathcal{L}(r)$ and $\mathcal{L}_{F}(r)$ are consistently related through the equation $L'(r)-\mathcal{L}_{F}(r)F'(r)=0$.
The plot of $\mathcal{L}$ is shown in Fig.~\ref{lagrangianamag}. We see that at the throat the function reaches a minimum, while asymptotically it tends to zero. In terms of the proper radial coordinate $\ell$, the function is symmetric, satisfying $\mathcal{L}(\ell)=\mathcal{L}(-\ell)$, and smoothly connected across the throat at $\ell=0$. This behavior reflects the symmetry of the magnetic configuration with respect to the throat, while the nonlinear electromagnetic contribution varies most significantly in the region surrounding $\ell=0$.

%As expected, in the uncharged limit $P \to 0$, all matter field profiles and potentials identically vanish, consistently recovering the vacuum Schwarzschild spacetime ($T_{\mu\nu} = 0$). Furthermore, since the limit $\lambda \to 0$ also reduces the Damour--Solodukhin wormhole metric to a Schwarzschild black hole, the phenomenological deformation parameter $\lambda$ can be physically reinterpreted as a direct manifestation of the magnetic monopole charge $P$. Given that $\lambda$ is dimensionless while the magnetic charge $P$ carries dimensions of mass ($[P] = M$), we propose the fundamental mapping $P(\lambda) \sim \lambda M$. Under this physical identification, a charge-free configuration ($P = 0$) dynamically forces $\lambda = 0$, eliminating the wormhole throat and restoring the standard Schwarzschild geometry. Finally, setting $M \to 0$ smoothly recovers flat Minkowski spacetime, ensuring the full physical consistency of the constructed solution across all asymptotic limits.

\subsection{Electric Source}
Previously, we obtained a source generated by magnetic monopoles. We now consider a radial electric field, such that the non-vanishing components of the electromagnetic tensor are $F_{01}=-F_{10}=E(r)$, and the invariant is given by
\begin{equation}\label{inveletric}
    \mathcal{F}(r)=-\frac{2 E(r)^2 \left(2 M \left(\lambda^2+1\right)-r\right)}{2 M-r}.
\end{equation}

From Eq.~\eqref{MaxwellNED}, we can write the electric field as
\begin{equation}
  E(r)=  \frac{q}{r^2 \mathcal{L}_\mathcal{F}(r)} \sqrt{\frac{r-2M}{r-2(\lambda^2+1)M}}.
\end{equation}

Since the scalar field must retain its form, the energy-momentum tensor $T_{\mu\nu}^{NED}$ is written as
\begin{equation}
    T_{\mu}^{\nu}=\text{diag}\left(\frac{4 q^2}{r^4 \mathcal{L}_ \mathcal{F}(r)}+\mathcal{L}(r),\frac{4 q^2}{r^4 \mathcal{L}_\mathcal{F}(r)}+\mathcal{L}(r),\mathcal{L}(r),\mathcal{L}(r) \right).
\end{equation}
We proceed analogously to the steps presented previously for the magnetic case. The $33$ component is given by:
\begin{equation}
    \mathcal{L}(r)-\frac{M(r-M) \lambda^2}{r^2 (r-2 M)^2}-\frac{1}{r}\left[2 \left(\lambda ^2+1\right) M-r\right] \phi '(r)^2+V(r)=0
\end{equation}
and, substituting $\phi(r)$ and $V(r)$, we obtain the Lagrangian:
\begin{equation}
\mathcal{L}(r)=\frac{3 \lambda ^2}{4 (r-2 M)^2}-\frac{\lambda ^2 r}{4 M (r-2 M)^2}+\frac{\lambda^2}{8M^2}\ln\left(\frac{r}{r-2M}\right)
\end{equation}
%It is worth emphasizing that the electric configuration does not determine a unique NED Lagrangian $\mathcal{L}(F)$. The reason is that the relation $F=F(r)$ cannot be globally inverted into a single-valued function $r=r(F)$, as a given value of the electromagnetic invariant may be attained at more than one radial position. As a result, expressing $L$ in terms of $F$ through the substitution $r=r(F)$ generally yields multiple branches of the Lagrangian. Hence, the electric NED description of the this geometry requires a multibranch structure and cannot be captured by a single-valued function $\mathcal{L}(F)$ throughout the entire spacetime.

\begin{figure}[H]
    \centering
    \includegraphics[width=0.45\linewidth]{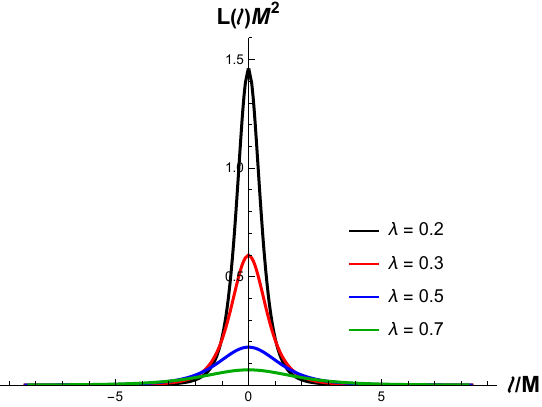}
    \includegraphics[width=0.45\linewidth]{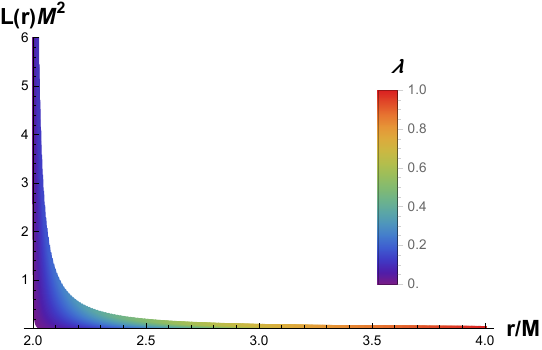}

    \caption{Behavior of the Lagrangian $L$ in terms of the proper radial coordinate $\ell$ (left) and the radial coordinate $r$ (right) for some values of $\lambda$.}
    \label{campoeletric}
\end{figure}

Next, from the $00$ component of the Einstein equations, we obtain $\mathcal{L}_\mathcal{F}(r)$:
\begin{equation}
    \mathcal{L}_\mathcal{F}(r)=\frac{4 q^2 (r-2 M)^2}{\lambda ^2 M r^2 (M-r)}
\end{equation}
and finally, the electric field $E(r)$ and invariant $F(r)$:
\begin{equation}
    E(r)=\frac{\lambda ^2 M (M-r)}{4 q (r-2 M)^2}\sqrt{\frac{2 M-r}{2 \left(\lambda ^2+1\right) M-r}}
\end{equation}
\begin{equation}\label{invariant}
    \mathcal{F}(r)=-\frac{\lambda ^4 M^2 (M-r)^2}{8 q^2 (r-2 M)^4}.
\end{equation}
Near the throat, $E(r\to r_{th}^{+})\propto 1/\sqrt{r-r_{th}}$ with $r_{th}=2M(1+\lambda^2)$, exhibiting a divergent behavior, in contrast to the Lagrangian, which approaches a finite constant near the throat:
\begin{equation}
    \mathcal{L}(r\to r_{th}^{+})=\frac{-2 \lambda ^2+1+2 \lambda ^4 \ln (1+\lambda^{-2}) }{16 \lambda ^2 M^2}.
\end{equation}
With this result, it is possible to invert the relation in Eq.~\eqref{invariant} and obtain an expression for $r(F)$. Thus, we can write the Lagrangian in terms of the invariant $F$:
\begin{align}
    \mathcal{L}(F)&=\frac{\lambda^2 }{8M^2}\log \left(\frac{1}{8} \left(\sqrt{-\frac{2\lambda ^2 M^2 \left(8 \sqrt{2} \sqrt{-F} q+\lambda ^2\right)}{F q^2}}+M \left(\frac{\sqrt{2} \lambda ^2}{\sqrt{-F} q}+16\right)\right)\right) +\frac{\lambda ^2}{4 M^2}+\frac{\lambda ^2 \log (32)}{16 M^2}\\ 
    &-\frac{\lambda^2}{8 M^2}\log \left(\sqrt{-\frac{\lambda ^2 M^2 \left(8 \sqrt{2} \sqrt{-F} q+\lambda ^2\right)}{F q^2}}+\frac{\lambda ^2 M}{\sqrt{-F} q}\right)+\frac{\sqrt{-F} q}{\sqrt{2} M^2}-\frac{\sqrt{-F} q \sqrt{-\frac{\lambda ^2 M^2 \left(8 \sqrt{2} \sqrt{-F} q+\lambda ^2\right)}{F q^2}}}{4 M^3} 
\end{align}
The behavior of $\mathcal{L}$ is shown in Fig.~\ref{campoeletric}.

In the figure above, we observe the regular behavior of the Lagrangian at the throat. In terms of the proper radial coordinate $\ell$, the Lagrangian is symmetric and smoothly connected across the two universes at $\ell=0$.

%As in the magnetic scenario, taking the uncharged limit $q \to 0$ causes the electric field $E(r)$, and the NED Lagrangian $\mathcal{L}(F)$ to vanish identically, thereby recovering the vacuum Schwarzschild spacetime ($T_{\mu\nu} = 0$). In this setup, the electric charge parameter $q$ carries dimensions of mass ($[q] = M$) and is directly mapped to the wormhole deformation parameter via $q \sim \lambda M$. Consequently, the non-linear electric field coupled with the phantom scalar field provides the physical source required to generate and sustain the Damour--Solodukhin throat. In the absence of electric charge ($q \to 0$), the throat radius $r_{\rm th} = 2M\left(1 + q^2/M^2\right)$ smoothly reduces to $r = 2M$, effectively removing the wormhole deformation ($\lambda = 0$) and restoring the standard Schwarzschild geometry. Finally, taking $M \to 0$ yields flat Minkowski spacetime, ensuring the physical consistency of the solution across all asymptotic limits.

\section{Null Geodesics}\label{sec4}
\label{sec:geodesicas}
To study the motion of particles in this spacetime, we investigate the
geodesics. With the temporal coordinate and mass parameter previously
redefined, the geodesic Lagrangian can be written as
\begin{equation}
    \mathcal{L}
    =
    -\left(1-\frac{2M}{r}\right)\dot{\tilde{t}}^2
    +\left(1-\frac{2(1+\lambda^2)M}{r}\right)^{-1}\dot{r}^2
    +r^2\dot{\theta}^2
    +r^2\sin^2\theta\,\dot{\varphi}^2
    =-\delta,
\end{equation}
where the dot denotes differentiation with respect to the affine
parameter, and $\delta$ is a normalization constant, defined as
$\delta=1$ for massive (timelike) particles and $\delta=0$ for massless
(null) particles. Applying the Euler--Lagrange equations, we obtain the
geodesic equations \cite{Damour:2007ap}:
\begin{align}
    &\ddot{\tilde{t}}
    +\frac{2M}
    {r(r-2M)}
    \dot{r}\dot{\tilde{t}}
    =0,
    \\
    &\ddot{r}
    -\frac{(1+\lambda^2)M}
    {r^2\left(1-\frac{2(1+\lambda^2)M}{r}\right)}
    \dot{r}^2
    +\frac{M}{r^2}
    \left(1-\frac{2(1+\lambda^2)M}{r}\right)
    \dot{\tilde{t}}^2-r\left(1-\frac{2(1+\lambda^2)M}{r}\right)
    (\dot{\theta}^2+\sin^2\theta
    \dot{\varphi}^2)
    =0,
    \\
    &\ddot{\theta}
    +\frac{2}{r}\dot{r}\dot{\theta}
    -\sin\theta\cos\theta\,\dot{\varphi}^2
    =0,
    \\
    &\ddot{\varphi}
    +\frac{2}{r}\dot{r}\dot{\varphi}
    +2\cot\theta\,\dot{\theta}\dot{\varphi} \label{geodesicaphi}
    =0.
\end{align}

Due to the spherical symmetry of the spacetime, the motion can be
restricted to the equatorial plane, $\theta=\pi/2$, without loss of
generality. In this case, the conserved quantities associated with the
cyclic coordinates $\tilde{t}$ and $\varphi$ are
\begin{align}
    \widetilde{E} &=
    \left(1-\frac{2M}{r}\right)
    \dot{\tilde{t}}, \label{energia}
    \\
    L &= r^2\dot{\varphi}, \label{phi}
\end{align}
where $\widetilde{E}$ and $L$ denote the conserved energy and angular
momentum of the particle, respectively.

Using these conserved quantities together with the normalization
condition, the radial equation can be written in terms of an effective
potential as
\begin{equation}\label{rdot}
    \dot{r}^{\,2}
    =
    \left(1-\frac{2(1+\lambda^2)M}{r}\right)
    \left[
    \frac{\widetilde{E}^2}
    {1-\frac{2M}{r}}
    -\delta-\frac{L^2}{r^2}
    \right],
\end{equation}
where
\begin{equation}\label{Veff}
    V_{\mathrm{eff}}(r)
    =
    \left(1-\frac{2M}{r}\right)
    \left(\delta+\frac{L^2}{r^2}\right).
\end{equation}

\begin{figure}[H]
    \centering
    \includegraphics[width=0.5\linewidth]{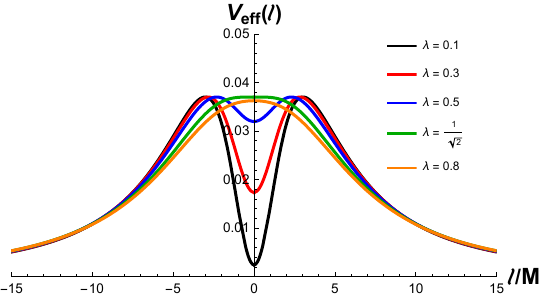} %\hspace{0.1cm}\includegraphics[width=0.44\linewidth]{Veffr.pdf}

    \caption{Behavior of $V_{\rm eff}$ for several values of $\lambda$, in terms of the proper radial coordinate $\ell$. It can be seen that, for $\lambda>1/\sqrt{2}$, the throat ceases to be a minimum of the effective potential and becomes a maximum. At $\lambda=1/\sqrt{2}$, the photon-sphere radius coincides with the throat radius, while for $\lambda>1/\sqrt{2}$ the corresponding photon-sphere radius lies at $r<r_{th}$ and is therefore outside the physical domain of the radial coordinate.}
    \label{veff}
\end{figure}

To analyze the existence of circular orbits, we consider the effective
potential in terms of the proper radial coordinate $\ell$. We seek a
value $\ell_c$ satisfying the condition
$V_{\rm eff}'(\ell_c)=0$. Stable (unstable) circular orbits require
$V_{\rm eff}''(\ell_c)>0$ ($V_{\rm eff}''(\ell_c)<0$), respectively.
From Eq.~\eqref{dl}, and setting $\delta=0$, we can write
\begin{equation}
    V_{\rm eff}'(\ell(r))
    =
    V_{\rm eff}'(r)
    \sqrt{
    1-\frac{2(1+\lambda^2)M}{r}
    } \implies V_{\rm eff}'(\ell(r)) = -\frac{2 L^2 (r-3 M) \sqrt{1-\frac{2 \left(\lambda ^2+1\right) M}{r}}}{r^4}.
\end{equation}
Therefore, we find two circular orbits: one located exactly at the throat,
\begin{equation}
    r_{th}=2\left(1+\lambda^2\right)M,
    \qquad (\ell=0),
\end{equation}
and another at \cite{Ho:2026gmy}
\begin{equation}
    r_{\rm ph}=3M,\label{raiofoton}
\end{equation}
corresponding to a photon sphere. For the photon sphere to lie outside
the throat, its radius must satisfy
$r_{\rm ph}>r_{th}$. Using the expressions above, this condition becomes
\begin{equation}
    3M>
    2\left(1+\lambda^2\right)M,
\end{equation}
which implies
\begin{equation}
    \lambda^2<\frac{1}{2}.
\end{equation}
Therefore, for $\lambda^2<1/2$, the photon sphere is located outside
the throat, whereas for $\lambda^2=1/2$ it coincides with the throat,
$r_{\rm ph}=r_{th}$. For $\lambda^2>1/2$, the corresponding stationary
point lies in the region $r<r_{th}$, which is not part of the physical
domain of the Morris--Thorne coordinates. Evaluating the second derivative with respect to the proper radial
coordinate $\ell$ at the throat, we obtain
\begin{equation}
    V_{\rm eff}''(r=r_{\rm th})
    =
    \frac{\left(1-2\lambda^2\right)L^2}
    {2r_{\rm th}^4\left(\lambda^2+1\right)},
\end{equation}
and therefore the circular orbit at the throat is stable for
$0<\lambda^2<1/2$, whereas for $\lambda^2>1/2$ it is unstable.
At the photon sphere, located at
$r_{\rm ph}=3M$, we instead find
\begin{equation}
   V_{\rm eff}''(r=r_{\rm ph})
   =
   -\frac{2L^2\left(1-2\lambda^2\right)}
   {243M^4}.
\end{equation}
For $\lambda^2<1/2$, for which the photon sphere lies within the
physical domain $r\geq r_{\rm th}$, this quantity is negative,
indicating that the photon-sphere orbit is unstable. At
$\lambda^2=1/2$, the photon sphere coincides with the throat, and the
second derivative vanishes. For $\lambda^2>1/2$, the photon-sphere
radius satisfies $r_{\rm ph}<r_{\rm th}$ and therefore lies outside
the physical domain covered by the radial coordinate $r$. Hence,
whenever the photon sphere lies within the physical domain, the
corresponding circular orbit is unstable. The plot of $V_{\rm eff}$ is
shown in Fig.~\ref{veff}.

Another important point to be studied is the angular geodesics. From them, we can analyze the null trajectories around the object, directly influencing how it would be observed by an external observer when illuminated by an accretion disk. Combining Eqs.~\eqref{phi} and \eqref{rdot}, and using the relation $d\varphi/dr=\dot{\varphi}/\dot{r}$, we obtain the orbital equation:
\begin{equation}\label{trajetoria}
    \frac{d\varphi}{dr}
    =
    \pm
    \frac{b}
    {r^2
    \sqrt{
    \left[1-\frac{r_{th}}{r}\right]
    \left[
    \left(1-\frac{2M}{r}\right)^{-1}
    -\frac{b^2}{r^2}
    \right]
    }},
\end{equation}
where $b\equiv L/\tilde{E}$ is the impact parameter. From this expression, we can analyze the critical turning points associated with circular orbits. From the condition $\dot{r}=0$, we obtain
\begin{equation}
    b(r)=\sqrt{\frac{r^3}{r-2M}},\label{impact_parameter}
\end{equation}
which represents a combination $(r,b)$ for which $\dot{r}=0$. To find the point that minimizes $b$, we impose
\begin{equation}
    b'(r)=0
    \implies
    r=3M,
\end{equation}
which corresponds precisely to the photon sphere. Thus, for $\lambda^2\le 1/2$, the critical impact parameter is given by
\begin{equation}
    b_{\rm crit}=3\sqrt{3}M.\label{bcritico}
\end{equation}
Using this critical parameter, we can analyze the trajectories of photons coming from infinity and passing through the wormhole.

In the regime $\lambda^2<1/2$, for $b>b_{\rm crit}$, a photon coming from infinity reaches a radius $r_0>3M$ and then returns to the same asymptotic region. For $b=b_{\rm crit}$, the photon remains trapped at the photon sphere, orbiting indefinitely. Finally, for $b<b_{\rm crit}$, the photon crosses the photon sphere and consequently reaches the throat, passing through it and entering the other universe. The traversal occurs because, at the throat,
\begin{equation}\label{bgarganta}
    b_{th}
    =
    \frac{2M(\lambda^2+1)^{3/2}}{\lambda}
    >
    b_{\rm crit},
\end{equation}
so that, once the photon crosses the photon sphere, it necessarily reaches the throat and proceeds toward the other universe. 

In this way, Eq.~\eqref{trajetoria} is evaluated in the limits:
\begin{equation}
    \Delta\varphi
    =
    2\int_{r_0}^{\infty}
    \frac{b\,dr}
    {r^2
    \sqrt{
    \left(1-\frac{r_{\rm th}}{r}\right)
    \left[
    \left(1-\frac{2M}{r}\right)^{-1}
    -\frac{b^2}{r^2}
    \right]
    }} \quad \text{for} \quad b>b_{crit},\label{deltaphi1}
\end{equation}
and
\begin{equation}
    \Delta\varphi
    =
    2\int_{r_{\rm th}}^{\infty}
    \frac{b\,dr}
    {r^2
    \sqrt{
    \left(1-\frac{r_{\rm th}}{r}\right)
    \left[
    \left(1-\frac{2M}{r}\right)^{-1}
    -\frac{b^2}{r^2}
    \right]
    }} \quad \text{for}\quad b<b_{crit}.\label{deltaphi2}
\end{equation}

\section{Constraints from the shadow of Sgr A*}\label{sec5}
\label{sec:shadow_sgrA}
 
As mentioned before, the DSWH can closely mimic a Schwarzschild black hole for small $\lambda$~\cite{Damour:2007ap}. Horizon-scale images provide a direct test of this behavior, since the edge of the shadow is determined by the light rings discussed in the previous section. In this section, we describe how the shadow radius of the DSWH is computed and how it is confronted with the EHT observation of Sgr A*~\cite{EventHorizonTelescope:2022wkp,EventHorizonTelescope:2022xqj}. Only the metric~\eqref{metric} is used, so that the analysis applies to any matter content supporting this geometry.

Let the observer be at rest at $r=r_O$, in one of the asymptotic regions. A light ray reaching this observer makes an angle $\alpha$ with the radial direction, which, in the observer's local frame, is given by $\cot\alpha=\sqrt{g_{rr}/g_{\varphi\varphi}}\,(dr/d\varphi)|_{r=r_O}$. Using the orbit equation~\eqref{trajetoria}, this relation becomes $\sin\alpha=b/b(r_O)$, where $b(r)$ is the function in Eq.~\eqref{impact_parameter}. The boundary of the shadow, or critical curve, is formed by the rays with $b=b_{\rm crit}$, which asymptotically approach the unstable light ring; the extent of the dark region in an actual image also depends on the emitting matter, as discussed in Sec.~\ref{sec6}. For a distant observer, the angular radius of the shadow satisfies $\alpha_{\rm sh}\ll1$, and the shadow radius, $r_s\equiv r_O\,\alpha_{\rm sh}$, reads~\cite{Perlick:2021aok}
\begin{equation}
r_s=b_{\rm crit}\sqrt{1-\frac{2M}{r_O}}.
\label{eq:rs}
\end{equation}
The factor $1-r_{th}/r$ in Eq.~\eqref{trajetoria}, which carries the information on the throat, cancels in $\cot\alpha$, so that $\lambda$ enters Eq.~\eqref{eq:rs} only through $b_{\rm crit}$. This is a particular case of a general property of static, spherically symmetric spacetimes, whose shadow depends only on $g_{tt}$ and on the areal radius~\cite{Perlick:2021aok}. Moreover, since $\tilde{t}$ is normalized at infinity, the square root in Eq.~\eqref{eq:rs} differs from unity only by terms of order $M/r_O$, which are of order $10^{-11}$ for Sgr A*, located at $r_O\simeq4\times10^{10}M$. Hence, to excellent accuracy, $r_s=b_{\rm crit}$. The critical impact parameter is the minimum of $b(r)$ in the domain $r\geq r_{th}$. For $\lambda^2\leq1/2$, it is given by Eq.~\eqref{bcritico}. For $\lambda^2>1/2$, the photon sphere lies outside this domain, $b(r)$ increases monotonically for $r\geq r_{th}$, and the minimum is attained at the throat, so that $b_{\rm crit}=b_{th}$, given by Eq.~\eqref{bgarganta}.
 
The EHT bounds on $r_s$ are expressed in units of the mass inferred from the S-star orbits, which, at Newtonian order, probe only $g_{\tilde{t}\tilde{t}}$: for circular timelike orbits of the metric~\eqref{metric}, Kepler's law reads $(d\varphi/d\tilde{t})^2=M/r^3$. We therefore identify this mass with $M$. For comparison, we also consider its identification with the ADM mass, $\wtm=M(1+\lambda^2)$, in which case the relevant ratio is $r_s/\wtm$.
 
For Sgr A*, the EHT Collaboration estimated the deviation $\delta=r_s/(3\sqrt{3}M)-1$ of the shadow radius from the Schwarzschild value using the Keck and VLTI/GRAVITY mass-to-distance priors~\cite{EventHorizonTelescope:2022xqj,Do:2019txf,GRAVITY:2020gka}. Combining the two estimates, $\delta=-0.060\pm0.065$, one obtains~\cite{Vagnozzi:2022moj}
\begin{equation}
4.55\lesssim r_s/M\lesssim5.22\quad(1\sigma),\qquad
4.21\lesssim r_s/M\lesssim5.56\quad(2\sigma),
\label{eq:rs_bounds}
\end{equation}
and the allowed values of $\lambda$ are those for which the theoretical shadow radius~\eqref{eq:rs} lies within these intervals.  This assumes that the calibration used by the EHT to relate the observed emission ring to the shadow also applies to the DS geometry, as is usual in this kind of analysis~\cite{Vagnozzi:2022moj}, so that the limits below should be understood as estimates under this assumption.

\begin{figure}[H]
    \centering
    \includegraphics[width=0.47\linewidth]{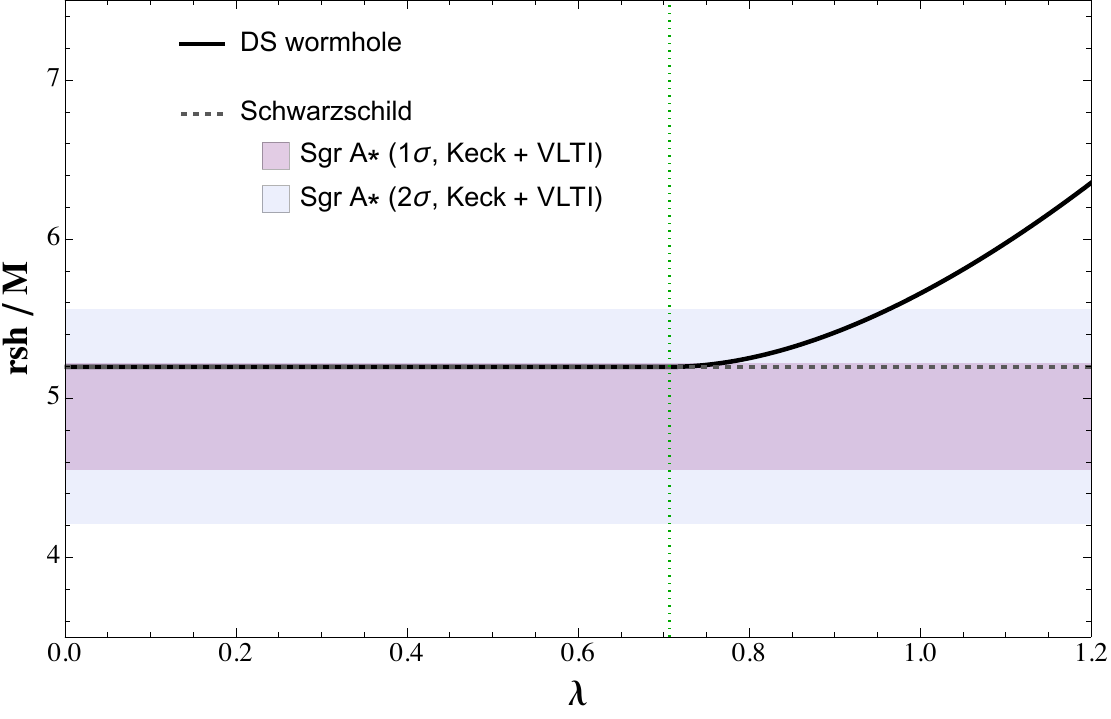}
    \hspace{0.1cm}
    \includegraphics[width=0.47\linewidth]{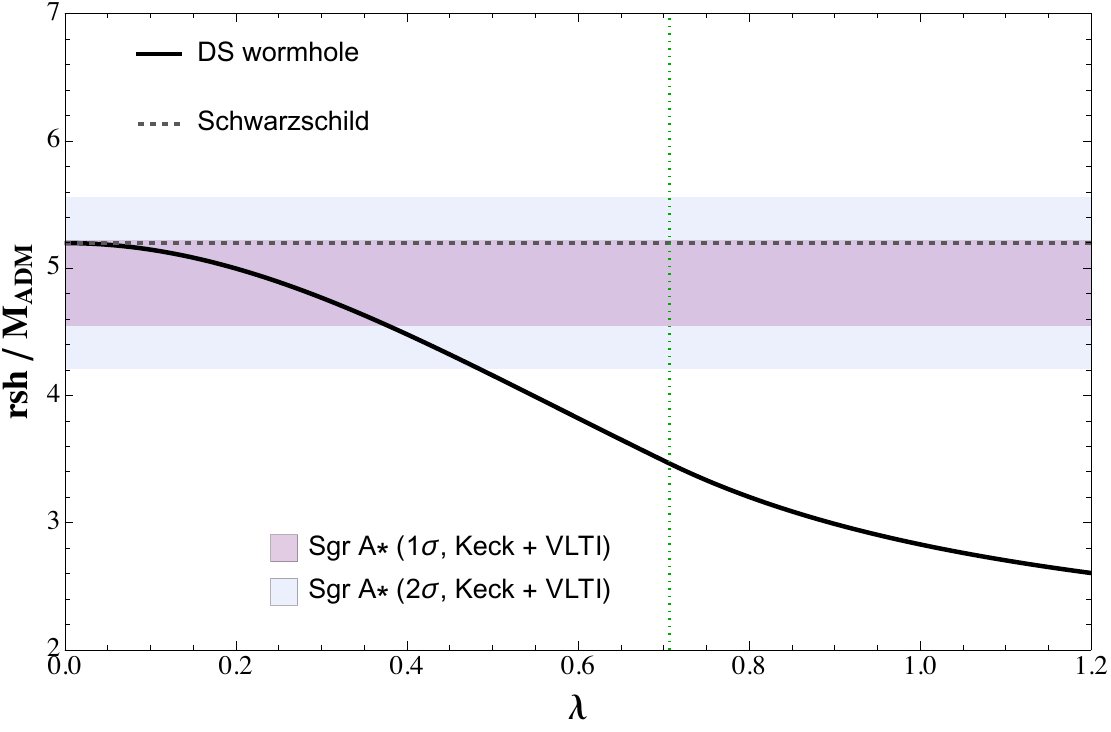}
    \caption{Shadow radius of the DSWH as a function of $\lambda$, normalized by the mass $M$ (left) and by the ADM mass $\wtm=M(1+\lambda^2)$ (right). The shaded regions correspond to the $1\sigma$ and $2\sigma$ intervals of Eq.~\eqref{eq:rs_bounds}, inferred from the EHT observation of Sgr A* with the Keck and VLTI mass-to-distance priors. The dashed line indicates the Schwarzschild value, $3\sqrt{3}$, and the vertical dotted line marks $\lambda=1/\sqrt{2}$, above which the shadow is bounded by the throat instead of the photon sphere at $r=3M$.}
    \label{raiosombra}
\end{figure}

The result of this comparison is shown in Fig.~\ref{raiosombra}. When the shadow radius is expressed in units of $M$ (left panel), the DS curve coincides with the Schwarzschild one for $\lambda^2\leq1/2$, i.e., to the left of the vertical line, as expected from Eq.~\eqref{bcritico}, and it lies inside the $1\sigma$ band. Therefore, the EHT data do not constrain $\lambda$ in this range. For $\lambda^2>1/2$, the shadow is bounded by the throat and grows with $\lambda$, eventually leaving the observational bands through their upper limits. This occurs at $\lambda\simeq0.77$ for the $1\sigma$ interval and at $\lambda\simeq0.96$ for the $2\sigma$ one, which yields $\lambda\lesssim0.77$ ($1\sigma$) and $\lambda\lesssim0.96$ ($2\sigma$), or, in terms of the throat radius, $r_{th}\lesssim3.19M$ and $r_{th}\lesssim3.86M$. Since $r_s/M\geq3\sqrt{3}$ for every $\lambda$, the lower limits in Eq.~\eqref{eq:rs_bounds} play no role.

In the right panel, normalized by the ADM mass, the shadow radius decreases monotonically with $\lambda$, and the relevant limits become the lower ones. They are crossed at $\lambda\simeq0.38$ ($1\sigma$) and $\lambda\simeq0.48$ ($2\sigma$), so that $\lambda\lesssim0.38$ and $\lambda\lesssim0.48$, respectively. In this case, the whole regime $\lambda^2>1/2$ is excluded, since there $r_s/\wtm<2\sqrt{3}\simeq3.46$.

The comparison between the two panels shows that the information extracted from the shadow depends on the mass attributed to Sgr A*. With the normalization by $M$, which is the one probed by the stellar orbits, the EHT observation constrains $\lambda$ only in the regime in which the throat replaces the photon sphere as the light ring that bounds the shadow. Configurations with $\lambda^2\leq1/2$, which include the black hole mimicking regime of small $\lambda$, are compatible with the data and cannot be distinguished from a Schwarzschild black hole by the size of the shadow alone; this would require observables sensitive to $g_{rr}$, such as the structure of the photon rings.

\section{Images of the DSWH with a thin accretion disk}\label{sec6}

The analysis of Sec.~\ref{sec:shadow_sgrA} showed that, for $\lambda^2\leq1/2$, the shadow of the DSWH has the same radius as the Schwarzschild one, because it is fixed by $g_{\tilde t\tilde t}$ alone. The component $g_{rr}$, which carries the throat, nevertheless affects the bending of every light ray through the factor $1-r_{th}/r$ in Eq.~\eqref{trajetoria}, and thus the way in which the matter around the object is lensed. To access this information, we now compute the image of the DSWH surrounded by a thin accretion disk, following Ref.~\cite{Gralla:2019xty}, where the image is decomposed according to the number of times each light ray crosses the disk. Compared with a black hole, two new ingredients appear. First, the throat is not a surface of infinite redshift, since $1-2M/r_{th}=\lambda^2/(1+\lambda^2)$. Second, the rays with $b<b_{\rm crit}$ are not captured: they pass through the throat into the other asymptotic region (Sec.~\ref{sec:geodesicas}), so that a disk extending through the throat also contributes to the image from the other side. The imprints of light rays that cross the throat on the image of a wormhole have been discussed, for instance, in Refs.~\cite{Wielgus:2020uqz,Guerrero:2021pxt,Macedo:2025ipc}.
 
The propagation of light depends only on the metric~\eqref{metric}, so that the ray tracing below applies to any matter content supporting this geometry, while the images also depend on the emission model specified below. We consider four values of the parameter, selected from Fig.~\ref{raiosombra}: $\lambda=0$, which is the Schwarzschild black hole; $\lambda=0.6$, for which $\lambda^2<1/2$ and the shadow radius lies in the $1\sigma$ interval of Eq.~\eqref{eq:rs_bounds}; $\lambda=0.9$, which lies only in the $2\sigma$ interval; and $\lambda=1.2$, which is excluded at $2\sigma$. In the last two cases $\lambda^2>1/2$, and the throat is the light ring that bounds the shadow.
 
The observer is at rest at $r_O=10^3M$ in the region $\ell>0$, and the disk lies on the equatorial plane of both regions, with negligible thickness and opacity. For a face-on disk, each ray reaching the observer moves on a plane orthogonal to the disk, and crosses it whenever the azimuthal angle on this plane, measured from the direction of the observer, reaches $\varphi=(2n+1)\pi/2$, with $n=0,1,2,\ldots$ Starting from the observer, the orbit is given by Eq.~\eqref{trajetoria}. For $b>b_{\rm crit}$ the ray reaches a turning point and goes back to the region $\ell>0$, Eq.~\eqref{deltaphi1}, while for $b<b_{\rm crit}$ it reaches the throat and continues into the region $\ell<0$, Eq.~\eqref{deltaphi2}; for $\lambda=0$, these rays end at the horizon. The crossings with $n=0$, $1$ and $2$ form the direct image, the lensing ring and the photon ring, respectively. We have checked that the crossings with $n=3$, $4$ and $5$ increase the flux by at most $1.1\%$ in the cases considered here, so that higher orders are neglected.

The location of each crossing must be specified with some care when the ray goes through the throat. The areal radius decreases to $r_{th}$ and grows again on the other side, so that the value of $r$ alone does not tell on which side the crossing occurs. We therefore locate the crossings with the proper radial coordinate $\ell$ of Eq.~\eqref{proper}, which runs continuously through the throat and distinguishes the two regions by its sign. The transfer function $\ell_n(b)$ is the value of $\ell$ at the $n$-th crossing of the ray with impact parameter $b$, and $r_n(b)\equiv r(\ell_n(b))$ is the corresponding areal radius. For $\lambda=0$, Eq.~\eqref{proper} gives the proper distance from the horizon, and only $\ell_n\geq0$ occurs.

We assume that the disk emits isotropically in the frame of static observers, with a frequency-integrated intensity $I_{\rm em}$ that depends only on the position. As in Ref.~\cite{Gralla:2019xty}, this phenomenological model does not describe a self-consistent Keplerian disk, but isolates the effect of the geometry on the propagation of light. Since $I_\nu/\nu^3$ is conserved along the ray, and the frequency of a photon emitted at $r$ and received at $r_O$ is shifted by the factor $\sqrt{A(r)/A(r_O)}$, with $A(r)\equiv1-2M/r$ (for emitters on circular geodesics and a face-on observer, this factor would become $\sqrt{(1-3M/r)/A(r_O)}$), each crossing contributes $[A(r_n)/A(r_O)]^2I_{\rm em}$. Neglecting absorption, these contributions add up, and the observed intensity reads
\begin{equation}
I_{\rm obs}(b)=\sum_{n=0}^{2}\left[\frac{A(r_n(b))}{A(r_O)}\right]^2 I_{\rm em}\big(\ell_n(b)\big).
\label{eq:Iobs}
\end{equation}
The function $A(r)$ is the Schwarzschild one, so the redshift at a given areal radius does not depend on $\lambda$. What changes is the place where the observer side of the disk ends: at the throat, the factor $[A/A(r_O)]^2$ takes the finite values $0.070$, $0.20$ and $0.35$ for $\lambda=0.6$, $0.9$ and $1.2$, while it vanishes at the horizon of the Schwarzschild black hole. Light emitted close to the throat, or on the other side, is therefore not suppressed as it would be close to a horizon.
 
For the emission we adopt the Johnson SU distribution~\cite{Johnson:1949,Gralla:2020srx},
\begin{equation}
J_{\rm SU}(r;\mu,\sigma,\gamma)=\frac{\exp\left\{-\frac{1}{2}\left[\gamma+{\rm arcsinh}\left(\frac{r-\mu}{\sigma}\right)\right]^2\right\}}{\sqrt{(r-\mu)^2+\sigma^2}},
\label{eq:JSU}
\end{equation}
in which $\mu$, $\sigma$ and $\gamma$ control the position of the peak, its width and its asymmetry. We take $\gamma=-2$ and $\sigma=M/4$, as in one of the models of Ref.~\cite{Gralla:2020srx} and in several later works~\cite{daSilva:2023jxa,Alencar:2026qeb,Alencar:2026gcl,Silva:2026yao}, and three values of $\mu$, associated with characteristic radii of the geometry. The first is the throat, $r_{th}=2.72M$, $3.62M$ and $4.88M$ for $\lambda=0.6$, $0.9$ and $1.2$, replaced by the horizon, $r_h=2M$, for $\lambda=0$. The second is the light ring that bounds the shadow: the photon sphere, $r=3M$ [Eq.~\eqref{raiofoton}], for $\lambda=0$ and $0.6$, and the throat for $\lambda=0.9$ and $1.2$. The third is the ISCO. The circular timelike orbits follow from the effective potential~\eqref{Veff} with $\delta=1$, which contains only $g_{\tilde t\tilde t}$, so that $r_{\rm ISCO}=6M$, as in the Schwarzschild case (see also Ref.~\cite{Ho:2026gmy}), and the ISCO lies outside the throat for $\lambda^2<2$. Each profile is normalized to unit maximum, which is reached at $r\simeq\mu+0.37M$.

Equation~\eqref{eq:JSU} is written in terms of the areal radius because this is the variable in which the parameters above are defined. In this way, the emission on the observer side is the one usually adopted for black holes, and the differences with respect to the Schwarzschild case are due to the geometry. Replacing $r$ by $\ell$ in Eq.~\eqref{eq:JSU}, while keeping $\mu$, $\sigma$ and $\gamma$, would not amount to a change of coordinates, but to a different disk: since $d\ell/dr=(1-r_{th}/r)^{-1/2}$ diverges at the throat, a width of $M/4$ in $\ell$ corresponds to a much thinner region in $r$ near the throat, and the images on the observer side would also change.

The disk is not truncated at the throat. To include the emission from the matter on the other side, we continue the argument of the same distribution through the throat, using the sign of $\ell$ to identify the side,
\begin{equation}
I_{\rm em}(\ell)=J_{\rm SU}\big(r_{th}+{\rm sgn}(\ell)\,[r(\ell)-r_{th}];\mu,\sigma,\gamma\big),
\label{eq:Iem}
\end{equation}
normalized as above. For $\ell>0$ this is the profile~\eqref{eq:JSU} with $r=r(\ell)$, whereas a point of the other side with areal radius $r$ emits $J_{\rm SU}(2r_{th}-r;\mu,\sigma,\gamma)$, i.e., the tail of the distribution below $r_{th}$. Equation~\eqref{eq:Iem} is continuous at the throat and, since $dr/d\ell=0$ there, its first derivative with respect to $\ell$ is also continuous. It introduces no new parameter, but it is an assumption on the distribution of matter, not a consequence of the symmetry $r(-\ell)=r(\ell)$ of the geometry. This asymmetric continuation describes matter that crosses the throat from the observer side and fades as it moves away from it: for the profile centered at the throat, the emission on the other side falls to $3\%$ and $0.3\%$ of the maximum at $r-r_{th}=0.25M$ and $0.5M$. A reflection-symmetric configuration, $I_{\rm em}(\ell)=J_{\rm SU}(r(\ell);\mu,\sigma,\gamma)$, would instead describe identical disks in both regions. For $\lambda=0$, $I_{\rm em}=J_{\rm SU}(r;\mu,\sigma,\gamma)$, and only the crossings outside the horizon enter Eq.~\eqref{eq:Iobs}.

These choices also determine the variables used in the figures. The transfer functions are shown as $\ell_n(b)$, for the reason given above, with $\ell<0$ for the crossings on the other side. The emitted profiles are shown as functions of the areal radius, the variable in which they are defined; the profile of the other side is plotted against the areal radius of that side, and, for $\lambda=0$, the curves are also drawn for $r<r_h$, a region that never contributes to Eq.~\eqref{eq:Iobs}. The observed intensities are shown as functions of $b$, which, for $r_O=10^3M$, differs from the radius on the observer's screen, $b\sqrt{1-2M/r_O}$ [cf. Eq.~\eqref{eq:rs}], by about $0.1\%$. Each observed profile is normalized such that $\int_0^{12M}I_{\rm obs}\,db=1$, and the two-dimensional images, obtained by rotating $I_{\rm obs}(b)$ around the line of sight, share a single color scale. Since this normalization does not fix the flux, which is proportional to $\int I_{\rm obs}\,b\,db$, the comparisons between panels concern how the light is distributed, and not the luminosity of the sources.
 
\begin{figure}[H]
    \centering
    \includegraphics[width=\linewidth]{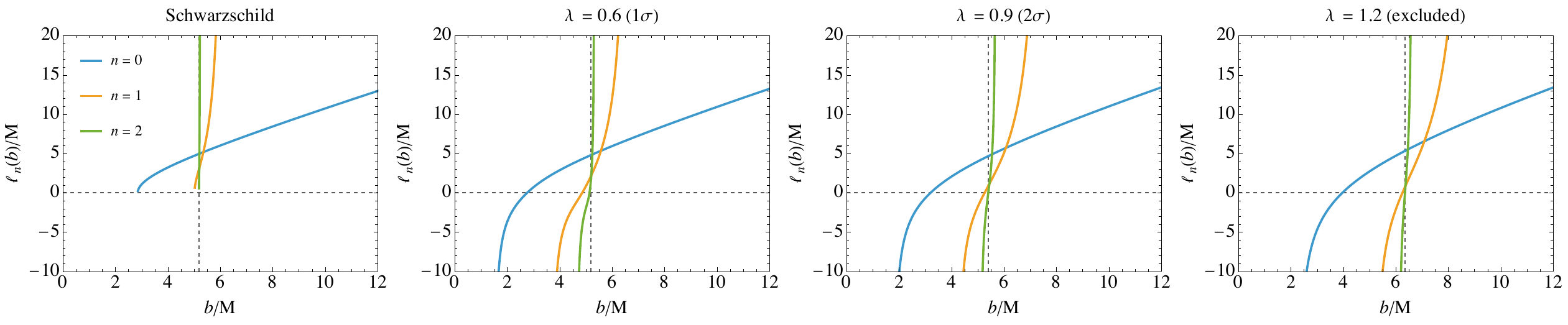}
    \caption{Transfer functions $\ell_n(b)$ for $\lambda=0$ (Schwarzschild), $0.6$, $0.9$ and $1.2$, i.e., the proper radial coordinate of the $n$-th crossing of the disk by the ray with impact parameter $b$: $n=0$ (direct image), $n=1$ (lensing ring) and $n=2$ (photon ring). Negative values of $\ell$ correspond to crossings on the other side of the throat. The horizontal dashed line marks the throat, $\ell=0$, or the horizon for $\lambda=0$, and the vertical one marks $b_{\rm crit}$. The labels in parentheses indicate whether the shadow radius lies in the $1\sigma$ or $2\sigma$ interval of Eq.~\eqref{eq:rs_bounds}, or is excluded.}
    \label{fig:transfer}
\end{figure}

The transfer functions are displayed in Fig.~\ref{fig:transfer}. For the Schwarzschild black hole, they reproduce the structure found in Ref.~\cite{Gralla:2019xty}: the direct image starts at $b\simeq2.85M$, where the ray meets the disk at the horizon, and the lensing and photon rings are restricted to $5.02\lesssim b/M\lesssim6.15$ and $5.19\lesssim b/M\lesssim5.23$. For the DSWH, the curves continue smoothly to $\ell<0$ when $b<b_{\rm crit}$. The direct image reaches the throat at $b\simeq2.76M$, $3.20M$ and $3.96M$ for $\lambda=0.6$, $0.9$ and $1.2$, and, for smaller impact parameters, the ray meets the disk on the other side, with $\ell_0\to-\infty$ as $b$ decreases to $1.49M$, $1.75M$ and $2.18M$, respectively. The lensing and photon rings now occupy much wider intervals. For $\lambda=0.6$, whose critical impact parameter is the Schwarzschild one, they extend over $3.68\lesssim b/M\lesssim6.69$ and $4.64\lesssim b/M\lesssim5.32$. This broadening is the effect of $g_{rr}$ mentioned above: for the same $b$ and $r$, the right-hand side of Eq.~\eqref{trajetoria} exceeds its Schwarzschild value by the factor $[(1-2M/r)/(1-r_{th}/r)]^{1/2}>1$, so that the rays are bent more strongly and cross the disk more often. For $\lambda^2>1/2$, the same structure is shifted to the larger critical parameter $b_{\rm crit}=b_{th}$, and the lensing ring extends over $4.20\lesssim b/M\lesssim7.53$ and $5.14\lesssim b/M\lesssim8.90$ for $\lambda=0.9$ and $1.2$.
 
\begin{figure}[H]
    \centering
    \includegraphics[width=\linewidth]{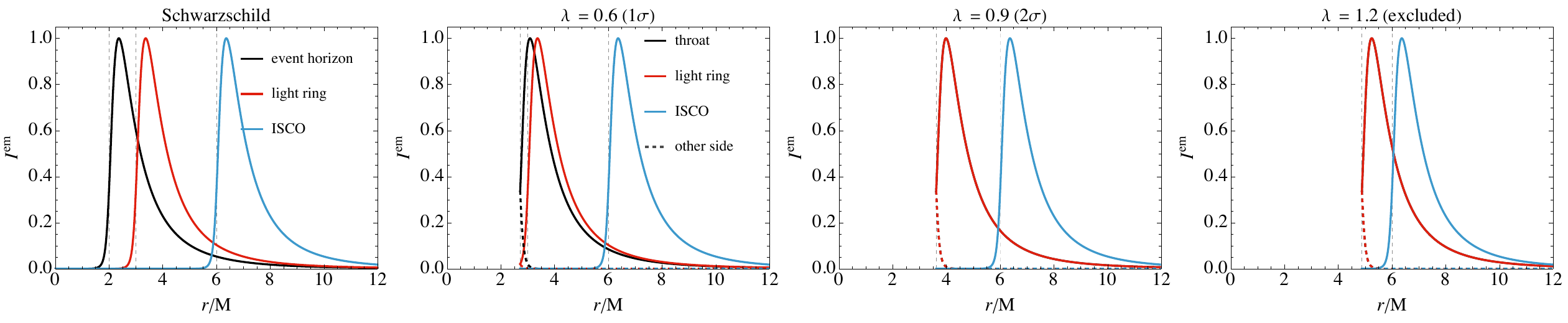}
    \caption{Emitted intensity, Eq.~\eqref{eq:Iem}, normalized to unit maximum, for the profiles centered at the horizon ($\lambda=0$) or at the throat (black), at the light ring (red) and at the ISCO (blue), as functions of the areal radius $r$. Solid curves correspond to the observer side and dashed curves to the other side of the throat, plotted against the areal radius of that side. For $\lambda=0$ the curves are also shown for $r<r_h$, which does not contribute to the observed intensity. The vertical dashed lines mark the horizon or throat, the light ring and the ISCO. For $\lambda=0.9$ and $1.2$ the light ring is at the throat, and the black and red curves coincide. The labels in parentheses are as in Fig.~\ref{fig:transfer}.}
    \label{fig:emitted}
\end{figure}

Figure~\ref{fig:emitted} shows the three emitted profiles. On the observer side they are the usual SU profiles, while on the other side only the profile centered at the throat gives an appreciable emission, restricted to the vicinity of the throat.
 
\begin{figure}[H]
    \centering
    \includegraphics[width=\linewidth]{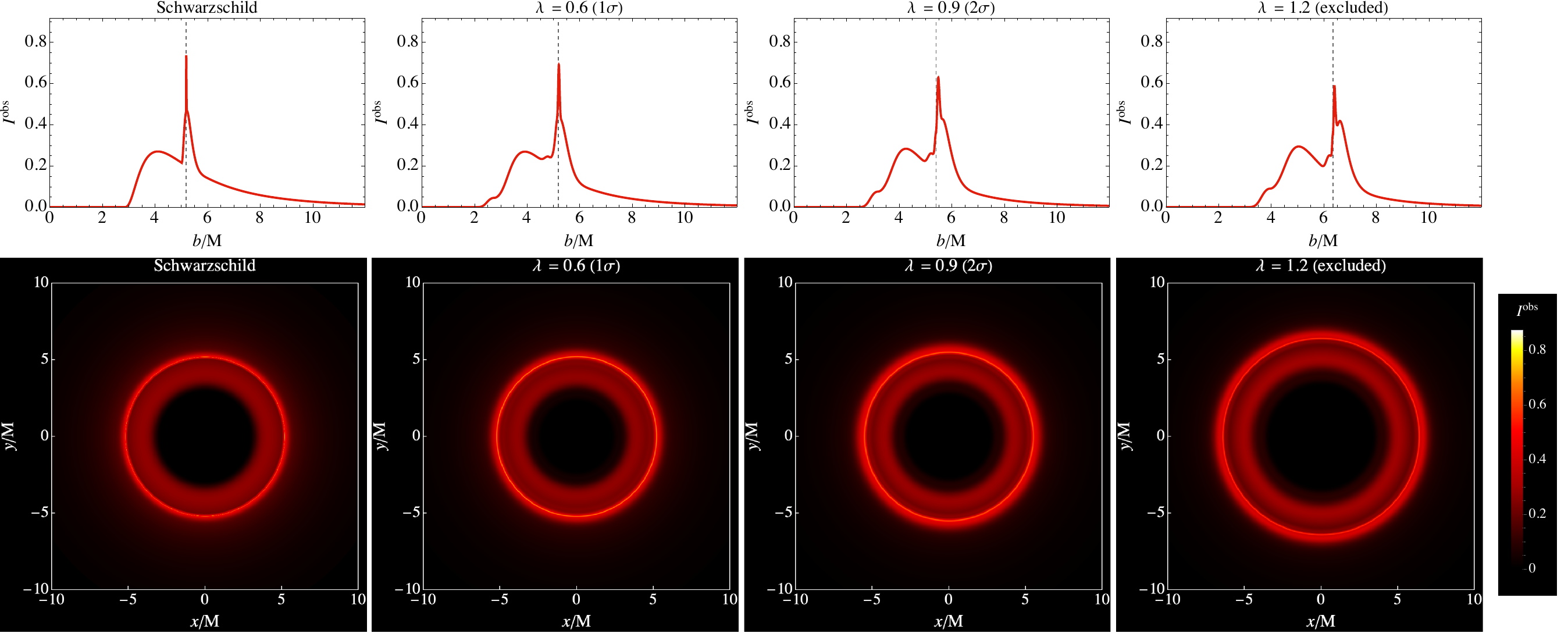}
    \caption{Observed intensity $I_{\rm obs}(b)$ (top) and image (bottom) for the emission profile centered at the horizon ($\lambda=0$) or at the throat ($\lambda=0.6$, $0.9$ and $1.2$). Each profile is normalized such that $\int_0^{12M}I_{\rm obs}\,db=1$, and the vertical dashed line marks $b_{\rm crit}$. The images share the color scale shown on the right, and $x$ and $y$ are Cartesian coordinates on the observer's screen, with $\sqrt{x^2+y^2}=b\sqrt{1-2M/r_O}$. The labels in parentheses are as in Fig.~\ref{fig:transfer}.}
    \label{fig:obs_throat}
\end{figure}

\begin{figure}[H]
    \centering
    \includegraphics[width=\linewidth]{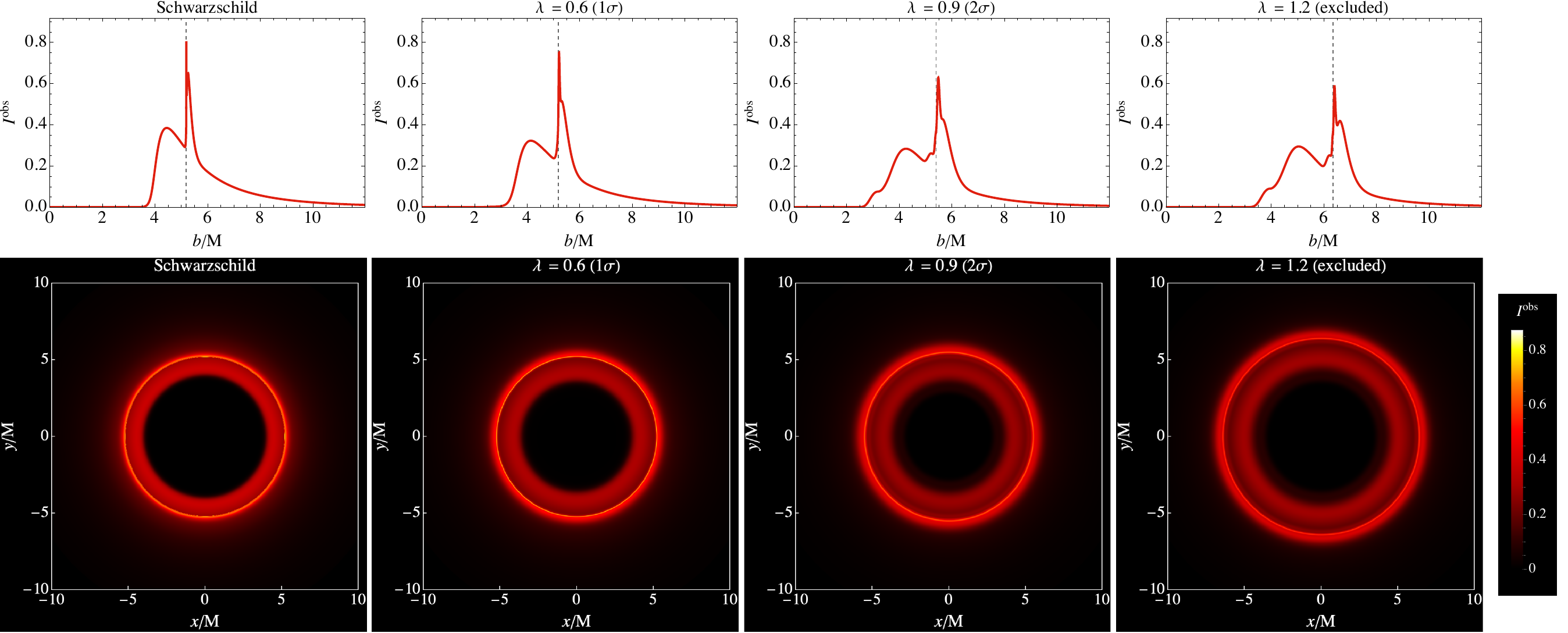}
    \caption{Same as Fig.~\ref{fig:obs_throat}, for the emission profile centered at the light ring: the photon sphere, $r=3M$, for $\lambda=0$ and $0.6$, and the throat for $\lambda=0.9$ and $1.2$, in which case the panels coincide with those of Fig.~\ref{fig:obs_throat}.}
    \label{fig:obs_lr}
\end{figure}

\begin{figure}[H]
    \centering
    \includegraphics[width=\linewidth]{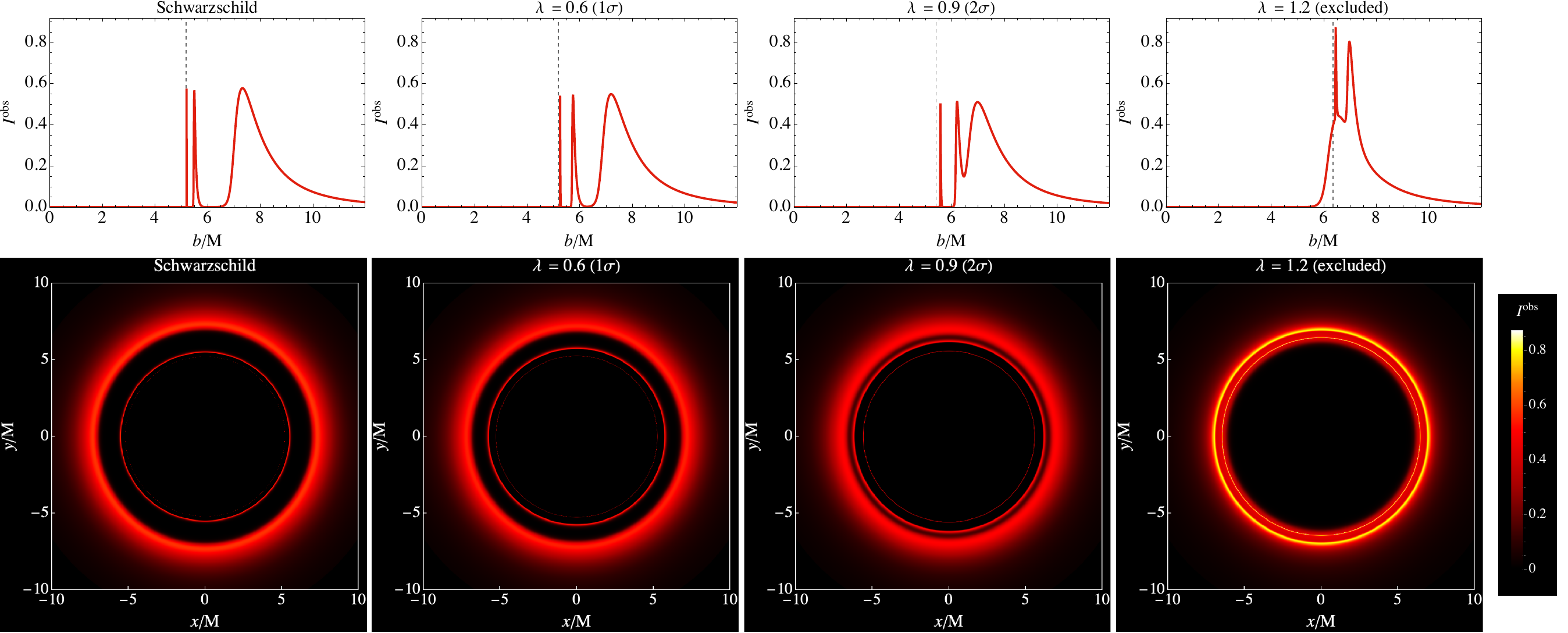}
    \caption{Same as Fig.~\ref{fig:obs_throat}, for the emission profile centered at the ISCO, $r_{\rm ISCO}=6M$.}
    \label{fig:obs_isco}
\end{figure}

The observed intensities and images are shown in Figs.~\ref{fig:obs_throat}--\ref{fig:obs_isco}. For the profiles centered at the horizon or throat and at the light ring, the brightest feature is a narrow ring just outside $b_{\rm crit}$, formed by the lensing and photon rings on top of the direct emission. Its position follows the critical curve, i.e., the shadow radius of Fig.~\ref{raiosombra}: it is the Schwarzschild one for $\lambda=0.6$ and moves to $b\simeq5.41M$ and $6.35M$ for $\lambda=0.9$ and $1.2$. The main imprint of the throat is the weight of these rings, which grows with the wider intervals of Fig.~\ref{fig:transfer}. The ISCO profile isolates this effect, since its emission on the observer side has the same form for all values of $\lambda$ and is negligible near the throat: the lensing and photon rings carry $3.6\%$ of the flux $\int_0^{12M}I_{\rm obs}\,b\,db$ for Schwarzschild, and $5.8\%$, $9.3\%$ and $15.0\%$ for $\lambda=0.6$, $0.9$ and $1.2$. For the two inner profiles the effect is stronger: already at $\lambda=0.6$, where the shadow radius is the Schwarzschild one, this fraction roughly doubles, from $7.9\%$ to $15.4\%$ for the light-ring profile and from $8.8\%$ to $18.5\%$ for the profile centered at the horizon or throat.
 
The throat also affects the center of the image and, for large $\lambda$, the separation between the rings and the direct emission. Unlike the critical curve, the dark region at the center depends on the emission~\cite{Gralla:2019xty}. In the Schwarzschild case, the rays with $b<2.85M$ fall into the horizon without crossing the disk, and the intensity vanishes at the image of the horizon because of the redshift. For the wormhole, these rays cross the throat and can meet the disk on the other side, while the inner edge of the direct emission on the observer side is the image of the throat, seen with a finite redshift. For the profile centered at the throat, the intensity stays below $1\%$ of its maximum only for $b<2.34M$, $2.69M$ and $3.37M$ for $\lambda=0.6$, $0.9$ and $1.2$, compared with $b<2.97M$ for Schwarzschild, but the other side provides only $1.7\%$, $2.9\%$ and $3.6\%$ of the flux, and a negligible amount for the light-ring profile with $\lambda=0.6$ and for the ISCO profile. For the latter, the rings appear as thin circles inside the dark region, separated from the direct emission, which starts at $b\simeq6.6M$ for Schwarzschild. This separation decreases as $\lambda$ grows, and for $\lambda=1.2$ the critical impact parameter, $6.35M$, exceeds that of the direct image of the ISCO, $b\simeq6.07M$, so that the rings overlap with the direct emission and produce the highest normalized intensity of the set.

For $\lambda=0.6$, allowed at $1\sigma$, the images are therefore very similar to those of a Schwarzschild black hole, and the differences reside in the weight of the rings and in the faint emission at the center. For $\lambda=0.9$ and $1.2$, the most visible change is the size of the image, which, for $\lambda=1.2$, is already excluded by the shadow of Sgr A*. Some of these features depend only on the geometry, and others also on the emission model. The critical curve is located at $b_{\rm crit}$, which depends only on $g_{\tilde t\tilde t}$ for $\lambda^2\leq1/2$ and also on the position of the throat for $\lambda^2>1/2$, while the intervals occupied by the lensing and photon rings depend on $g_{rr}$ through Eq.~\eqref{trajetoria}; both statements hold for any matter content that supports the metric~\eqref{metric}. The precise position and height of the intensity peak, the weight of the rings and the amount of light from the other side also depend on the emission. In particular, the small contribution of the other side is a property of the asymmetric continuation~\eqref{eq:Iem}: identical disks in both regions would raise it to about $20\%$ of the flux for the profile centered at the throat. The enhancement of the rings, on the other hand, comes from the factor $1-r_{th}/r$ in Eq.~\eqref{trajetoria}, so that the mechanism operates for every $\lambda\neq0$, although its magnitude depends on the emission. It is the main imprint of the throat on the thin-disk images for $\lambda=0.6$, where the shadow size is the Schwarzschild one, and it could be probed by observations able to isolate the photon ring~\cite{Johnson:2019ljv}.

\section{Accretion on the DSWH}\label{sec7}

In this section, we study the accretion of matter onto the DSWH.
Accretion is the process by which matter from the surroundings of a
compact object is captured and transported by the gravitational field.
Although accretion onto black holes and stars has been extensively
studied in the literature (See \cite{Chaverra:2015bya,Barrero-Gonzalez:2026xiu,Bahamonde:2015uwa}), the corresponding process in wormhole
geometries remains comparatively less explored \cite{Debnath2014,Yusupova2021,Crispim:2024dgd,Rueda:2023val}.

\subsection{General relations}
Here, we consider the accretion of matter onto the geometry
\eqref{metric}. We assume that the accreting matter is
described by a perfect, isotropic fluid, and is characterized by a particle density $n$, an energy density $\epsilon$, a pressure $p$, entropy per particle $s$, and temperature $\Theta$. Furthermore, we define a function describing the enthalpy per particle $h(p,s)$ that satisfies the first law of thermodynamics. Thus, the fluid is described by the equations \cite{Chaverra:2015bya}
\begin{eqnarray}
      T_{\alpha\beta}
    &=&
    (\varepsilon+p)u_\alpha u_\beta
    +p g_{\alpha\beta},\\
    d\epsilon &=& h dn + n\Theta ds,
\end{eqnarray}
where
$u^\alpha$ is the four-velocity of the fluid. In spherical coordinates,
we consider a stationary and radial flow,
\begin{equation}
    u^\alpha
    =
    \frac{dx^\alpha}{d\tau}
    =
    (u^t,u^r,0,0),
\end{equation}
where $\tau$ is the proper time along the fluid worldlines.

In the present treatment, we neglect the backreaction of the accreting
matter on the wormhole geometry. In other words, the geometry is assumed
to remain fixed throughout the accretion process. Although this is a
simplifying assumption, it allows us to investigate the properties of
the fluid flow on the prescribed wormhole background.

The four-velocity satisfies the normalization condition
$u^\alpha u_\alpha=-1$. For the metric \eqref{morristhorne}, this condition
gives
\begin{equation}
    u^t
    =
    e^{-\Phi(r)}
    \sqrt{
        1+
        \frac{(u^r)^2}
        {1-\frac{b(r)}{r}}
    }.
\end{equation}
For a future-directed flow, we take $u^t>0$, while the sign of the
radial component distinguishes ingoing and outgoing flows,
\begin{equation}
    u^r<0
    \quad\text{(ingoing)},
    \qquad
    u^r>0
    \quad\text{(outgoing)}.
\end{equation}
Unlike the case of black holes, both ingoing and outgoing flows can in
principle be considered in a wormhole spacetime, since there is no event
horizon preventing matter from crossing the throat and emerging into the
other asymptotic region. In what follows, however, we restrict our
attention to the ingoing flow, for which $u^r<0$. The analysis of the
outgoing branch is analogous.

The conservation equation
\begin{equation}
    \nabla_\mu T^{\mu\nu}=0,
\end{equation}
provides the energy-flux conservation law. In particular, using the
stationarity of the spacetime, one obtains
\begin{equation}\label{eqconT}
    \frac{
        e^{2\Phi(r)}
        (\varepsilon+p)
        u^r
        r^3
    }
    {r-b(r)}
    \sqrt{
        (u^r)^2
        +1-\frac{b(r)}{r}
    }
    =C_1,
\end{equation}
where $C_1$ is an integration constant.

The projection of the conservation equation along the fluid
four-velocity,
\begin{equation}
    u_\nu\nabla_\mu T^{\mu\nu}=0,
\end{equation}
gives the relativistic energy equation
\begin{equation}
    u^\alpha\nabla_\alpha\varepsilon
    +
    (\varepsilon+p)
    \nabla_\alpha u^\alpha
    =0.
\end{equation}
For the stationary radial flow considered here, this equation becomes
\begin{equation}
    \frac{\varepsilon'}{\varepsilon+p}
    +
    \frac{(u^r)'}{u^r}
    +
    \Phi'
    -
    \frac{b-rb'}{2r(r-b)}
    +
    \frac{2}{r}
    =0.
\end{equation}

We now introduce the particle-number current
\begin{equation}
    J^\alpha=n u^\alpha,
\end{equation}
where $n$ denotes the particle-number density in the fluid rest frame \cite{Chaverra:2015bya}.
Assuming that the flow is adiabatic and isentropic ($ds=0$), particle-number
conservation,
\begin{equation}
    \nabla_\alpha J^\alpha=0,
\end{equation}
together with the first law of thermodynamics implies
\begin{equation}
    d\varepsilon
    =
    \frac{\varepsilon+p}{n}\,dn,
\end{equation}
or equivalently,
\begin{equation}
    \frac{dn}{n}
    =
    \frac{d\varepsilon}{\varepsilon+p}.
\end{equation}

The conservation of the particle-number current yields
\begin{equation}\label{eqflux}
    \frac{
        n u^r r^2 e^{\Phi}
    }
    {\sqrt{1-\frac{b(r)}{r}}}
    =C_2,
\end{equation}
where $C_2$ is a constant associated with the conserved radial particle
flux. Equivalently, using the thermodynamic relation above, Eq.~\eqref{eqflux}
can be written as
\begin{equation}\label{eqcont}
    \frac{
        u^r r^2
    }
    {\sqrt{1-\frac{b(r)}{r}}}
    \exp\left[
        \Phi
        +
        \int
        \frac{d\varepsilon}{\varepsilon+p(\varepsilon)}
    \right]
    =\widetilde C_2,
\end{equation}
where $\widetilde C_2$ differs from $C_2$ only by a constant
normalization factor.

Dividing Eq.~\eqref{eqconT} by Eq.~\eqref{eqcont}, we obtain
\begin{equation}
    (\varepsilon+p)
    \sqrt{
        1+
        \frac{(u^r)^2}
        {1-\frac{b(r)}{r}}
    }
    \exp\left[
        \Phi
        -
        \int
        \frac{d\varepsilon}{\varepsilon+p(\varepsilon)}
    \right]
    =C_3,
\end{equation}
where $C_3$ is a constant.

It is convenient to introduce the specific enthalpy \cite{Chaverra:2015bya},
\begin{equation}
    h\equiv\frac{\varepsilon+p}{n}.
\end{equation}
Using
\begin{equation}
    \frac{dn}{n}\label{termo}
    =
    \frac{d\varepsilon}{\varepsilon+p},
\end{equation}
the previous relation can be written in the more transparent form
\begin{equation}\label{bernouli}
    h\,e^{\Phi}
    \sqrt{
        1+
        \frac{(u^r)^2}
        {1-\frac{b(r)}{r}}
    }
    =C_3',
\end{equation}
where $C_3'$ is another integration constant.

We now define the squared sound speed of the fluid as
\begin{equation}
    v_s^2
    \equiv
    \frac{dp}{d\varepsilon}.
\end{equation}
For an isentropic flow, the first law gives
\begin{equation}
    dh=\frac{dp}{n},
\end{equation}
and therefore
\begin{equation}
    d\ln h
    =
    \frac{dp}{\varepsilon+p}
    =
    v_s^2
    \frac{d\varepsilon}{\varepsilon+p}
    =
    v_s^2\frac{dn}{n}.
\end{equation}

Taking the differential of Eq.~\eqref{eqflux}, we obtain
\begin{equation}
    \frac{dn}{n}
    +
    \frac{du^r}{u^r}
    +
    \left(
        \Phi'
        -
        \frac{B'}{2B}
        +
        \frac{2}{r}
    \right)dr
    =0,
\end{equation}
where, for convenience, we have defined
\begin{equation}
    B(r)\equiv
    1-\frac{b(r)}{r}.
\end{equation}
On the other hand, differentiating the Bernoulli-type relation gives
\begin{equation}
    d\ln h
    +\Phi' dr
    +
    \frac{
        u^rdu^r+\frac{1}{2}B'dr
    }
    {(u^r)^2+B}
    -
    \frac{B'}{2B}dr
    =0.
\end{equation}
Combining the last two equations and eliminating $dn/n$, we obtain
\begin{equation}\label{eqaccretion}
\left[
v_s^2
-
\frac{(u^r)^2}
{
    (u^r)^2+1-\frac{b(r)}{r}
}
\right]
\frac{du^r}{u^r}
+
\left[
(v_s^2-1)
\left(
\Phi'
-
\frac{b-rb'}{2r(r-b)}
\right)
+
\frac{2v_s^2}{r}
-
\frac{b-rb'}
{
    2r^2
    \left[
        (u^r)^2+1-\frac{b(r)}{r}
    \right]
}
\right]dr
=0.
\end{equation}

The critical point of the accretion flow is obtained by requiring the
simultaneous vanishing of the coefficients of $du^r$ and $dr$ in
Eq.~\eqref{eqaccretion}. Denoting quantities evaluated at the critical
radius $r=r_c$ by a subscript $c$, we obtain
\begin{equation}\label{eqcritical1}
    {v_s}_c^2
    =
    \frac{(u_c^r)^2}
    {(u_c^r)^2+B_c},
\end{equation}
and
\begin{equation}\label{eqcritical2}
    ({v_s}_c^2-1)
    \left(
        \Phi'_c
        -
        \frac{B'_c}{2B_c}
    \right)
    +
    \frac{2{v_s}_c^2}{r_c}
    -
    \frac{B'_c}
    {2\left[(u_c^r)^2+B_c\right]}
    =0.
\end{equation}

From Eq.~\eqref{eqcritical1}, we find
\begin{equation}
    (u_c^r)^2
    =
    \frac{{v_s}_c^2}{1-{v_s}_c^2}B_c,
\end{equation}
and consequently
\begin{equation}
    (u_c^r)^2+B_c
    =
    \frac{B_c}{1-{v_s}_c^2}.
\end{equation}
Substituting these relations into Eq.~\eqref{eqcritical2}, the terms
proportional to $B'_c$ cancel, yielding
\begin{equation}
    -(1-{v_s}_c^2)\Phi'_c
    +
    \frac{2{v_s}_c^2}{r_c}
    =0.
\end{equation}
Therefore, the squared sound speed at the critical point is
\begin{equation}\label{eqacritical}
    {v_s}_c^2
    =
    \frac{r_c\Phi'_c}
    {2+r_c\Phi'_c}.
\end{equation}

Using Eq.~\eqref{eqcritical1}, the radial velocity at the critical point
is then given by
\begin{equation}\label{equcriticalvelocity}
    (u_c^r)^2
    =
    \frac{r_cB_c\Phi'_c}{2}
    =
    \frac{r_c-b(r_c)}{2}\Phi'_c.
\end{equation}
For the ingoing branch, we select the negative root,
\begin{equation}
    u_c^r
    =
    -\sqrt{
        \frac{r_c-b(r_c)}{2}\Phi'(r_c)
    }.
\end{equation}
\subsection{Accretion Flow onto DSWH}

We now specialize the above results to the DSWH
(Eq.~\eqref{metric}) geometry. Before proceeding, it is useful to
clarify the thermodynamic description adopted here. Accretion onto the
DSWH has already been investigated in the
literature (see Ref.~\cite{Yusupova:2025niy}). In that treatment, the
quantity denoted by \(\rho\), introduced as the energy density of the
perfect fluid, is also used as the density entering the conserved
matter flux. While this identification is appropriate for
pressureless matter, for which the energy density is proportional to
the particle-number density, it is not generally justified for a
relativistic fluid with nonzero pressure, where these quantities play
distinct thermodynamic roles. A consistent relativistic-fluid
description therefore requires the particle-number density and the
energy density to be treated as distinct variables, with their
relation determined by the thermodynamic properties of the fluid
\cite{Chaverra:2015bya,Disconzi:2023rtt,Rocha:2023ilf}. We adopt this
formulation in the present work. This distinction is particularly
relevant for the barotropic fluids considered below, including
quintessence, phantom energy, dust, and stiff matter.
We assume a linear barotropic equation of state,
\begin{equation}
    p=w\varepsilon,
\end{equation}
where \(w\) is a constant. In this case, the squared sound speed introduced above is simply
\begin{equation}
    v_s^2
    =
    \frac{dp}{d\varepsilon}
    =
    w.
\end{equation}
Thus, the parameter \(w\) directly measures the squared sound speed of the accreting fluid within the present isentropic description.

The thermodynamic relation \eqref{termo} can be integrated as
\begin{equation}
    n(\varepsilon)
    =
    n_0\,\varepsilon^{1/(1+w)},
\end{equation}
where \(n_0\) is a positive integration constant. The specific enthalpy therefore takes the form
\begin{equation}
    h(\varepsilon)
    =
    \frac{\varepsilon+p}{n}
    =
    \frac{1+w}{n_0}
    \varepsilon^{w/(1+w)}.
\end{equation}

For the ingoing branch, \(u^r<0\), and since the other factors are positive in the region \(r>r_{\rm th}\), one has \(C_2<0\). Defining \(\mathcal{C}_2\equiv -C_2>0\), the relation \eqref{eqflux} can be written as
\begin{equation}
    n
    =
    \frac{
        \mathcal{C}_2
    }
    {(-u^r)r^2}
    \sqrt{
        \frac{B(r)}{A(r)}
    },
    \qquad
    B(r)
    \equiv
    1-\frac{2(1+\lambda^2)M}{r},
    \,
    A(r)
    \equiv
    1-\frac{2M}{r}
    =
    e^{2\Phi(r)}.
\end{equation}
Consequently, the energy density is obtained explicitly as
\begin{equation}\label{eq:epsilonDS}
    \varepsilon(r,u^r)
    =
    \left[
        \frac{\mathcal{C}_2}
        {n_0(-u^r)r^2}
        \sqrt{
            \frac{
                1-\frac{2(1+\lambda^2)M}{r}
            }{
                1-\frac{2M}{r}
            }
        }
    \right]^{1+w},
\end{equation}
while the pressure is
\begin{equation}\label{eq:pDS}
    p(r,u^r)
    =
    w
    \left[
        \frac{\mathcal{C}_2}
        {n_0(-u^r)r^2}
        \sqrt{
            \frac{
                1-\frac{2(1+\lambda^2)M}{r}
            }{
                1-\frac{2M}{r}
            }
        }
    \right]^{1+w}.
\end{equation}
Thus, once the radial component of the four-velocity is determined,
both thermodynamic variables follow directly from
Eqs.~\eqref{eq:epsilonDS} and \eqref{eq:pDS}.

The Bernoulli-type relation can also be specialized to the DS
geometry. Substituting the expressions above for the specific
enthalpy and the redshift function into \eqref{bernouli}
and absorbing the constant normalizations into a single positive
constant \(\mathcal{K}\), one obtains the following algebraic relation
for the radial component of the four-velocity:
\begin{equation}\label{eq:urDS}
    (u^r)^2
    +
    1-\frac{2(1+\lambda^2)M}{r}
    =
    \mathcal{K}^2
    r^{4w}
    (-u^r)^{2w}
    \left[
        \frac{
            1-\frac{2(1+\lambda^2)M}{r}
        }{
            1-\frac{2M}{r}
        }
    \right]^{1-w},
\end{equation}
where
\begin{equation}
    \mathcal{K}
    =
    \frac{
        C_3' n_0^{1+w}
    }
    {(1+w)\mathcal{C}_2^{\,w}}.
\end{equation}
Equation~\eqref{eq:urDS} determines the stationary radial velocity
profile implicitly for an arbitrary constant \(w\). In general, the
equation cannot be solved for \(u^r(r)\) in elementary closed form for
arbitrary \(w\), although explicit expressions are obtained for specific
values of the equation-of-state parameter.

The critical-point condition derived above can also be written
explicitly for the DS geometry. Using the expression for \(\Phi'(r)\),
Eq.~\eqref{eqacritical} gives
\begin{equation}\label{eq:csDS}
{v_s}_c^2
=
\frac{M}{2r_c-3M}.
\end{equation}
Since \(v_s^2=w\) for the linear equation of state considered here, the
critical radius is determined by
\begin{equation}\label{eq:rcDS}
r_c
=
\frac{
M(1+3w)
}
{
2w
}.
\end{equation}
The corresponding radial component of the four-velocity follows from
Eq.~\eqref{equcriticalvelocity},
\begin{equation}\label{eq:ucDS}
(u_c^r)^2
=
\frac{
M\left[r_c-2(1+\lambda^2)M\right]
}
{
2r_c(r_c-2M)
}.
\end{equation}
For the ingoing branch, we therefore take
\begin{equation}\label{eq:ucDSingoing}
u_c^r
=
-
\sqrt{
\frac{
M\left[r_c-2(1+\lambda^2)M\right]
}
{
2r_c(r_c-2M)
}
}.
\end{equation}

In the limit $\lambda \to 0$, where the DSWH spacetime continuously reduces to the standard Schwarzschild black hole metric, our critical flow relations and hydrodynamic equations perfectly recover the classical relativistic accretion theory originally established by Michel \cite{Michel1972} and systematically detailed by Rezzolla and Zanotti \cite{Rezzolla2013}. Specifically, setting $\lambda = 0$ yields $b(r) = 2M$ and $e^{2\Phi(r)} = 1 - 2M/r$, under which the governing differential relation \eqref{eqaccretion} collapses to the standard Michel accretion equation on a Schwarzschild background. Consequently, the critical sound speed \eqref{eq:csDS}, and the critical radial four-velocity \eqref{eq:ucDS} simplify to 
\begin{align}
    v_{s,c}^2 &= \frac{M}{2r_c - 3M}, \\
    (u_c^r)^2 &= \frac{M}{2r_c}
\end{align}
respectively, in exact agreement with the established literature for non-rotating black hole accretion. The critical radius \eqref{eq:rcDS} remains unchanged, since Eq. \eqref{eq:rcDS} does not depend on $\lambda$ in this case.

It is worth emphasizing that the critical-point condition has a
direct sonic interpretation. Indeed, Eq.~\eqref{eqcritical1} identifies
the point at which the radial flow reaches the local sound speed.
Therefore, the critical radius identified above corresponds to the
sonic point of the stationary accretion flow.
For the linear barotropic equation of state \(p=w\varepsilon\), the present
isentropic perfect-fluid description gives \(v_s^2=w\). The existence of
a conventional sonic point in the physical exterior region requires
\(r_c>r_{\rm th}\).
For \(w>0\), this condition yields
\begin{equation}\label{eq:soniccondition}
0<w<\frac{1}{1+4\lambda^2}.
\end{equation}
Thus, although the formal expression for \(r_c\) is independent of
\(\lambda\), the existence of an exterior sonic point is restricted by
the throat radius, and the allowed range of \(w\) becomes narrower as
\(\lambda\) increases.

For \(w=0\), the critical radius is pushed to infinity,
\begin{equation}
r_c\rightarrow\infty,
\end{equation}
and therefore there is no finite sonic point. For \(w<0\), the formal
critical radius obtained from Eq.~\eqref{eq:rcDS} either becomes
negative or lies below the throat, depending on the value of \(w\).
Consequently, negative-\(w\) fluids do not admit a conventional exterior
sonic point within the present perfect-fluid, linear-EOS treatment.

For a linear barotropic equation of state, our isentropic perfect-fluid assumption inherently yields $v_s^2=w$. While more general dark-energy models allow the background equation-of-state parameter and the physical sound speed to differ, they are identical by construction in our framework. Consequently, for $w<0$, the adiabatic sound-speed squared is negative. Rather than supporting a propagating sound mode, this signals a hydrodynamic instability \cite{Kim2005,BeanFlanaganTrodden2008} and precludes the existence of a conventional exterior sonic point.

\begin{figure}[H]
\centering
\includegraphics[width=0.47\linewidth]{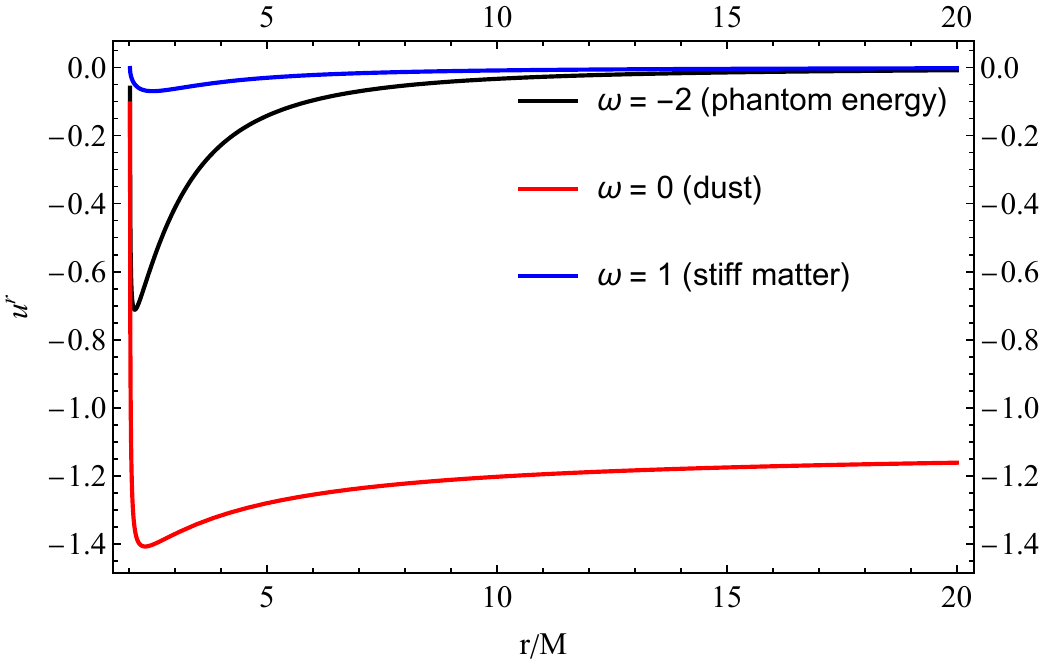} \hspace{0.1cm}\includegraphics[width=0.47\linewidth]{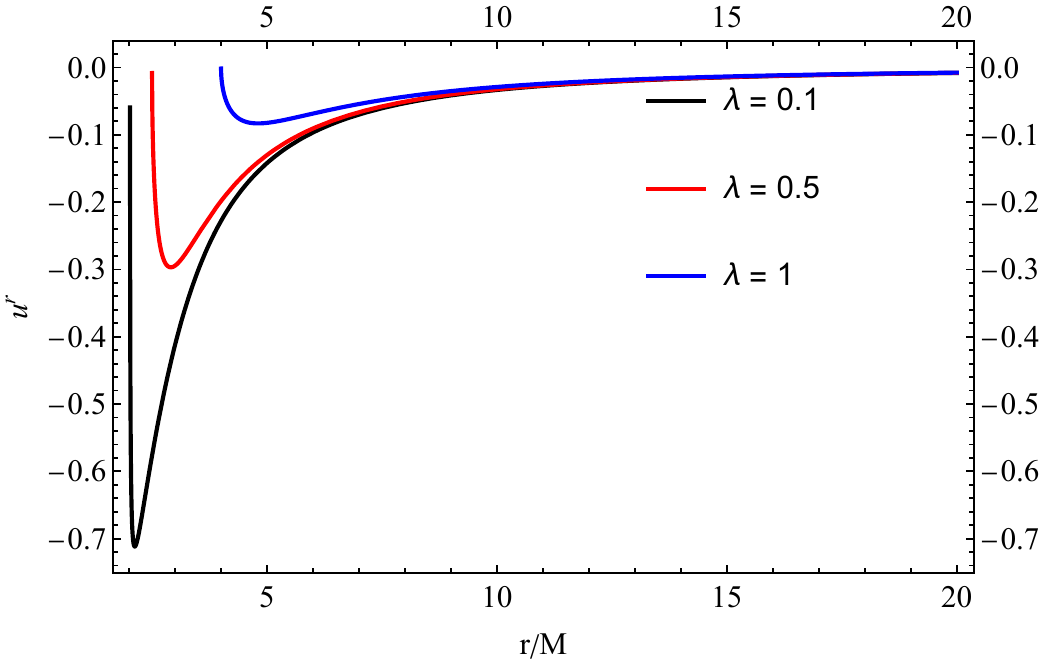}
\caption{Radial component of the fluid four-velocity, $u^r$, for the
stationary accretion flow onto the DSWH. In the left
panel, $\lambda=0.1$ is fixed and the cases $w=-2$ (phantom energy),
$w=0$ (dust), and $w=1$ (stiff matter) are shown, with the
integration constants $\mathcal{K}=10$, $1.5$, and $1.02$, respectively.
In the right panel, the phantom case $w=-2$ is considered with
$\mathcal{K}=10$, while the wormhole parameter is varied as
$\lambda=0.1$, $0.5$, and $1$.}
\label{ur}
\end{figure}

Nevertheless, fluids with $w<0$ are routinely employed in the literature as effective descriptions of dark energy and phantom accretion---including accretion onto Morris--Thorne and Ellis--Bronnikov wormholes \cite{Debnath2014,Yusupova2021}. Since the radial velocity profile is determined directly from Eq.~\eqref{eq:urDS}, obtaining a stationary solution does not require a finite sonic point. Therefore, we retain the $w=-2$ case as an effective phantom-fluid description within the stationary flow equations, despite its inherent instability.
Fig.~\ref{ur} (left) illustrates the resulting radial profiles of the
four-velocity for \(\lambda=0.1\). For the chosen integration constants,
the phantom and stiff-matter configurations approach \(u^r\rightarrow0\)
at large radius, whereas the dust solution approaches a constant
asymptotic value. The phantom case exhibits a stronger variation of the
radial four-velocity, reaching a larger magnitude in the region close
to the throat before decreasing toward the asymptotic region, whereas
the dust solution presents an intermediate behavior and the stiff-matter
solution remains closer to zero over most of the exterior region. Since
different values of the integration constant \(\mathcal{K}\) are adopted
for the three equations of state, the relative amplitudes of the curves
should be regarded as illustrative rather than as a direct quantitative
comparison between the different fluids. For the phantom case, Fig.~\ref{ur}
(right) also shows the effect of varying the wormhole parameter \(\lambda\),
with larger values of \(\lambda\) reducing the magnitude of \(u^r\) over
the exterior region for the chosen boundary conditions.

\begin{figure}[H]
    \centering
    \includegraphics[width=0.5\linewidth]{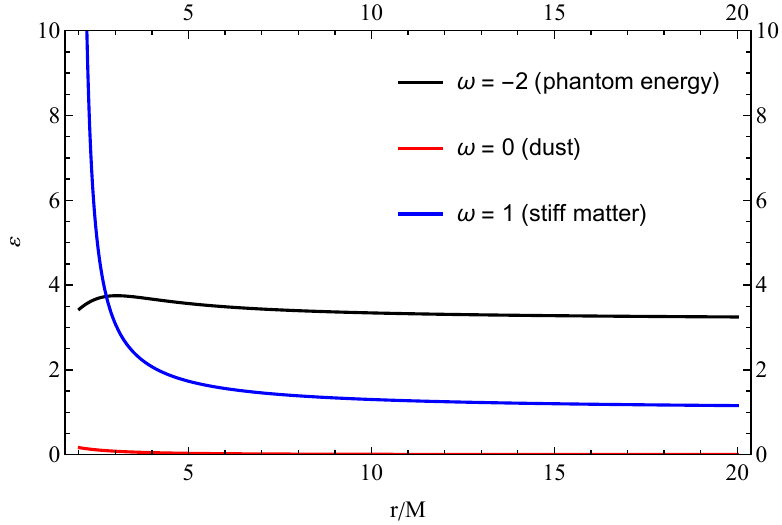}
    \caption{Energy density as a function of $r$. $\lambda=0.1$ is fixed and the cases $w=-2$ (phantom energy),
    $w=0$ (dust), and $w=1$ (stiff matter) are shown, with the
    integration constants $\mathcal{K}=10$, $1.5$, and $1.02$, respectively, and $\mathcal{C}_2=n_0=1$.}
    \label{fig:e}
\end{figure}

In Fig.~\ref{fig:e}, we present the radial profiles of the energy density $\varepsilon$ for the different fluids considered ($w = -2, 0, 1$). To physically interpret this behavior, it is crucial to note that the DSWH metric acts as a continuous deformation of the Schwarzschild geometry. The parameter $\lambda$ controls the magnitude of this spatial deformation such that the $\lambda \to 0$ limit exactly recovers the Schwarzschild black hole solution, with the throat position coinciding with the classical event horizon ($r_{\rm th} \to 2M$).

To explicitly demonstrate that the energy density profile $\varepsilon(r)$ is strictly independent of the DSWH deformation parameter $\lambda$, we eliminate the radial four-velocity component $u^r(r)$ between the particle flux conservation equation and the generalized Bernoulli relation.

From the particle-current conservation equation, Eq.~\eqref{eqflux}, together with the thermodynamic relation $n(\varepsilon) = n_0 \varepsilon^{1/(1+w)}$, the radial four-velocity component for the ingoing branch ($u^r < 0$) can be expressed in terms of $\varepsilon$ as
\begin{equation}\label{eq:ur_implicit_proof}
    -u^r = \frac{\mathcal{C}_2}{n_0 r^2} \varepsilon^{-\frac{1}{1+w}} \sqrt{\frac{B(r)}{A(r)}}.
\end{equation}

Next, we substitute this velocity field into the Bernoulli-type relation, Eq.~\eqref{bernouli}. Recalling that $e^{\Phi(r)} = \sqrt{A(r)}$ and using the specific enthalpy $h(\varepsilon) = \frac{1+w}{n_0}\varepsilon^{\frac{w}{1+w}}$, Eq.~\eqref{bernouli} takes the explicit form
\begin{equation}\label{eq:bernouli_expanded_proof}
    \frac{1+w}{n_0} \varepsilon^{\frac{w}{1+w}} \sqrt{A(r)} \, \frac{\sqrt{(u^r)^2 + B(r)}}{\sqrt{B(r)}} = C_3'.
\end{equation}
By evaluating the expression $(u^r)^2 + B(r)$ using Eq.~\eqref{eq:ur_implicit_proof}, we factor out $B(r)$:
\begin{align}
    (u^r)^2 + B(r) &= \left( \frac{\mathcal{C}_2}{n_0 r^2} \varepsilon^{-\frac{1}{1+w}} \sqrt{\frac{B(r)}{A(r)}} \right)^2 + B(r) \nonumber \\
    &= B(r) \left[ 1 + \frac{\mathcal{C}_2^2}{n_0^2 r^4 A(r)} \varepsilon^{-\frac{2}{1+w}} \right].
\end{align}
Substituting this decomposition back into Eq.~\eqref{eq:bernouli_expanded_proof}, the spatial metric function $\sqrt{B(r)}$ in the numerator and denominator cancels out identically:
\begin{equation}
    \frac{1+w}{n_0} \varepsilon^{\frac{w}{1+w}} \sqrt{A(r)} \sqrt{1 + \frac{\mathcal{C}_2^2}{n_0^2 r^4 A(r)} \varepsilon^{-\frac{2}{1+w}}} = C_3'.
\end{equation}

Finally, squaring both sides and distributing the factor $A(r)$ yields the master algebraic equation governing the radial distribution of the energy density:
\begin{equation}\label{eq:master_epsilon_proof}
    \left(\frac{1+w}{n_0}\right)^2 \left(1 - \frac{2M}{r}\right) \varepsilon(r)^{\frac{2w}{1+w}} + \frac{(1+w)^2 \mathcal{C}_2^2}{n_0^4 r^4} \varepsilon(r)^{\frac{2(w-1)}{1+w}} = (C_3')^2.
\end{equation}

Equation~\eqref{eq:master_epsilon_proof} determines $\varepsilon(r)$ implicitly as a function of the radial coordinate $r$, the fluid parameter $w$, the central mass $M$, and integration constants. Crucially, the function $B(r)$, and therefore the wormhole deformation parameter $\lambda$, has completely vanished from Eq.~\eqref{eq:master_epsilon_proof}. Because $F(r) = 1 - 2M/r$ coincides exactly with the Schwarzschild time-metric coefficient, it follows rigorously that for any radial coordinate $r > r_{\rm th}$, the energy density $\varepsilon(r)$ (and consequently $p(r)$ and $n(r)$) around a DSWH is strictly identical to that surrounding a Schwarzschild black hole of equal mass.

Visually and physically, this means that the energy density of the accreted matter onto a DSWH at any radius $r > r_{\rm th}$ is identical to the density that would accumulate around a Schwarzschild black hole. The sole effect manifested by $\lambda$ on the thermodynamic scalars is to truncate the exterior spacetime domain earlier, shifting the position of the absorption surface (the throat) to $r_{\rm th} = 2M(1+\lambda^2)$.

The physical interpretation underlying this scalar invariance lies in the kinematics of the flow. For the energy and particle currents to remain globally conserved, the radial velocity $u^r$, implicitly determined by Eq.~\eqref{eq:urDS}, acts as a compensatory mechanism. The velocity field $u^r(r)$ entirely absorbs the effects of the geometric alteration of the spatial volumes encoded in $B(r)$. In short, the fluid alters its rate of infall, accelerating or decelerating in the exact proportion required to cancel the spatial deformation generated by $\lambda$, thus keeping the energy density $\varepsilon$ and the particle number density $n$ completely immune to the intrinsic nature of the wormhole.

As the flow approaches the throat ($r \to r_{\rm th}$), the geometry imposes an effective bottleneck. Since the conservation of the macroscopic flux demands a balance between density and velocity, the thermodynamics of each fluid dictate drastically different responses. For fluids such as stiff matter ($w=1$), the flow inertia causes the magnitude of $u^r$ to decay significantly near the throat, forcing an abrupt peak in matter accumulation. In contrast, for dust ($w=0$) and especially for phantom energy ($w=-2$), the velocity field allows for a more fluid transit, resulting in a much smoother and controlled density profile in the immediate vicinity of the throat before the matter emerges into the opposite asymptotic region.

\section{Conclusions}\label{sec8}
In this work, we have investigated the physical realization and observational properties of DSWH within GR, combining the reconstruction of their supporting matter fields with an analysis of photon propagation, optical images, and relativistic accretion. Our results provide a unified description of the matter content and phenomenology of this geometry, clarifying the extent to which it can reproduce the properties of a Schwarzschild black hole.

We first constructed an explicit classical matter source capable of supporting the DSWH, consisting of a phantom scalar field coupled to NED. Both magnetic and electric configurations were considered, providing a field-theoretical realization of the wormhole within Einstein gravity. The phantom scalar field supplies the violation of the null energy condition required to maintain the traversable throat, while the nonlinear electromagnetic sector contributes to the remaining components of the energy--momentum tensor. This construction complements previous investigations based on quantum-field contributions to the stress--energy tensor and provides a classical framework for studying the physical origin of the geometry.

We then examined the null geodesic structure and the observational constraints imposed by Event Horizon Telescope measurements of Sgr A*. A central result is that, for $\lambda^2\leq 1/2$, the critical impact parameter and the corresponding shadow radius coincide exactly with those of a Schwarzschild black hole when expressed in terms of the mass parameter $M$. Consequently, shadow-size measurements alone cannot distinguish the two geometries throughout this regime. For larger deformations, the wormhole throat determines the critical impact parameter, leading to departures from the Schwarzschild prediction. We also showed that the inferred constraints on $\lambda$ depend on the mass normalization adopted in the comparison with observations, emphasizing the importance of distinguishing the mass parameter entering the temporal metric component from the ADM mass.

The analysis of thin-disk images further illustrates that the equality of shadow radii does not imply complete optical equivalence. Although the temporal component of the metric retains its Schwarzschild form, the radial component modifies photon trajectories and gravitational lensing. Using a Johnson SU emissivity profile, we examined how these effects influence the apparent disk structure and the contributions from different photon trajectories. The resulting images approach the Schwarzschild appearance in the small-deformation regime, while larger values of $\lambda$ produce more pronounced differences associated with the wormhole geometry. Thus, the detailed structure of lensed emission provides information beyond that contained in the shadow radius alone.

Finally, we revisited stationary relativistic accretion using a thermodynamically consistent treatment of barotropic fluids, accounting for the contribution of pressure through the specific enthalpy. Remarkably, we found that the energy density and particle number density are exactly independent of $\lambda$ throughout the common exterior region, provided the same boundary conditions are imposed. The radial four-velocity, however, retains an explicit dependence on the deformation parameter, compensating for the modification of the radial geometry. Furthermore, the existence of an exterior sonic point requires $0<w<(1+4\lambda^2)^{-1}$, showing that the location of the throat restricts the range of equations of state admitting such transitions.

Taken together, these results reveal that the ability of DSWHs to mimic Schwarzschild black holes depends strongly on the physical observable under consideration. While the shadow radius and the thermodynamic profiles of stationary accretion can exhibit exact degeneracies over a finite range of the deformation parameter, photon trajectories, detailed disk images, and fluid kinematics remain sensitive to the radial geometry. In particular, these classical degeneracies do not require the exponentially small values of $\lambda$ associated with mimicking black hole behavior over extremely long dynamical timescales. Future investigations incorporating more realistic accretion flows, radiative transfer, and perturbative stability could help determine whether the remaining differences provide observationally distinguishable signatures of the wormhole throat.

\acknowledgments{G. A. acknowledges financial support from the Conselho Nacional de Desenvolvimento Científico e Tecnológico (CNPq) through a PQ-B fellowship (No. 308182/2025-1) and from the Fundação Cearense de Apoio ao Desenvolvimento Científico e Tecnológico (FUNCAP). V. H. U. B. acknowledges financial support from FUNCAP under Contract No. BMD-0008-10054.01.06/26. This study was financed in part by the Coordenação de Aperfeiçoamento de Pessoal de Nível Superior – Brasil (CAPES) – Finance Code 001. M.S. is supported by CNPq/PDE 200218/2025-5.}
\bibliographystyle{apsrev4-1}
\bibliography{referencias}

\begin{thebibliography}{78}%
\makeatletter
\providecommand \@ifxundefined [1]{%
 \@ifx{#1\undefined}
}%
\providecommand \@ifnum [1]{%
 \ifnum #1\expandafter \@firstoftwo
 \else \expandafter \@secondoftwo
 \fi
}%
\providecommand \@ifx [1]{%
 \ifx #1\expandafter \@firstoftwo
 \else \expandafter \@secondoftwo
 \fi
}%
\providecommand \natexlab [1]{#1}%
\providecommand \enquote  [1]{``#1''}%
\providecommand \bibnamefont  [1]{#1}%
\providecommand \bibfnamefont [1]{#1}%
\providecommand \citenamefont [1]{#1}%
\providecommand \href@noop [0]{\@secondoftwo}%
\providecommand \href [0]{\begingroup \@sanitize@url \@href}%
\providecommand \@href[1]{\@@startlink{#1}\@@href}%
\providecommand \@@href[1]{\endgroup#1\@@endlink}%
\providecommand \@sanitize@url [0]{\catcode `\\12\catcode `\$12\catcode `\&12\catcode `\#12\catcode `\^12\catcode `\_12\catcode `\%12\relax}%
\providecommand \@@startlink[1]{}%
\providecommand \@@endlink[0]{}%
\providecommand \url  [0]{\begingroup\@sanitize@url \@url }%
\providecommand \@url [1]{\endgroup\@href {#1}{\urlprefix }}%
\providecommand \urlprefix  [0]{URL }%
\providecommand \Eprint [0]{\href }%
\providecommand \doibase [0]{http://dx.doi.org/}%
\providecommand \selectlanguage [0]{\@gobble}%
\providecommand \bibinfo  [0]{\@secondoftwo}%
\providecommand \bibfield  [0]{\@secondoftwo}%
\providecommand \translation [1]{[#1]}%
\providecommand \BibitemOpen [0]{}%
\providecommand \bibitemStop [0]{}%
\providecommand \bibitemNoStop [0]{.\EOS\space}%
\providecommand \EOS [0]{\spacefactor3000\relax}%
\providecommand \BibitemShut  [1]{\csname bibitem#1\endcsname}%
\let\auto@bib@innerbib\@empty
%</preamble>
\bibitem [{\citenamefont {Morris}\ and\ \citenamefont {Thorne}(1988)}]{Morris:1988cz}%
  \BibitemOpen
  \bibfield  {author} {\bibinfo {author} {\bibfnamefont {M.~S.}\ \bibnamefont {Morris}}\ and\ \bibinfo {author} {\bibfnamefont {K.~S.}\ \bibnamefont {Thorne}},\ }\href {\doibase 10.1119/1.15620} {\bibfield  {journal} {\bibinfo  {journal} {Am. J. Phys.}\ }\textbf {\bibinfo {volume} {56}},\ \bibinfo {pages} {395} (\bibinfo {year} {1988})}\BibitemShut {NoStop}%
\bibitem [{\citenamefont {Einstein}\ and\ \citenamefont {Rosen}(1935)}]{Einstein:1935tc}%
  \BibitemOpen
  \bibfield  {author} {\bibinfo {author} {\bibfnamefont {A.}~\bibnamefont {Einstein}}\ and\ \bibinfo {author} {\bibfnamefont {N.}~\bibnamefont {Rosen}},\ }\href {\doibase 10.1103/PhysRev.48.73} {\bibfield  {journal} {\bibinfo  {journal} {Phys. Rev.}\ }\textbf {\bibinfo {volume} {48}},\ \bibinfo {pages} {73} (\bibinfo {year} {1935})}\BibitemShut {NoStop}%
\bibitem [{\citenamefont {Ellis}(1973)}]{Ellis:1973yv}%
  \BibitemOpen
  \bibfield  {author} {\bibinfo {author} {\bibfnamefont {H.~G.}\ \bibnamefont {Ellis}},\ }\href {\doibase 10.1063/1.1666161} {\bibfield  {journal} {\bibinfo  {journal} {J. Math. Phys.}\ }\textbf {\bibinfo {volume} {14}},\ \bibinfo {pages} {104} (\bibinfo {year} {1973})}\BibitemShut {NoStop}%
\bibitem [{\citenamefont {Bronnikov}(1973)}]{Bronnikov:1973fh}%
  \BibitemOpen
  \bibfield  {author} {\bibinfo {author} {\bibfnamefont {K.~A.}\ \bibnamefont {Bronnikov}},\ }\href@noop {} {\bibfield  {journal} {\bibinfo  {journal} {Acta Phys. Polon. B}\ }\textbf {\bibinfo {volume} {4}},\ \bibinfo {pages} {251} (\bibinfo {year} {1973})}\BibitemShut {NoStop}%
\bibitem [{\citenamefont {Crispim}\ \emph {et~al.}(2024)\citenamefont {Crispim}, \citenamefont {Alencar},\ and\ \citenamefont {Muniz}}]{Crispim:2024dgd}%
  \BibitemOpen
  \bibfield  {author} {\bibinfo {author} {\bibfnamefont {T.~M.}\ \bibnamefont {Crispim}}, \bibinfo {author} {\bibfnamefont {G.}~\bibnamefont {Alencar}}, \ and\ \bibinfo {author} {\bibfnamefont {C.~R.}\ \bibnamefont {Muniz}},\ }\href@noop {} {\  (\bibinfo {year} {2024})},\ \Eprint {http://arxiv.org/abs/2410.11147} {arXiv:2410.11147 [gr-qc]} \BibitemShut {NoStop}%
\bibitem [{\citenamefont {de~S.~Silva}\ \emph {et~al.}(2026)\citenamefont {de~S.~Silva}, \citenamefont {Pereira}, \citenamefont {Bragato}, \citenamefont {Rodrigues}, \citenamefont {Fabris},\ and\ \citenamefont {Belich}}]{deSSilva:2025ncg}%
  \BibitemOpen
  \bibfield  {author} {\bibinfo {author} {\bibfnamefont {M.~V.}\ \bibnamefont {de~S.~Silva}}, \bibinfo {author} {\bibfnamefont {C.~F.~S.}\ \bibnamefont {Pereira}}, \bibinfo {author} {\bibfnamefont {B.}~\bibnamefont {Bragato}}, \bibinfo {author} {\bibfnamefont {M.~E.}\ \bibnamefont {Rodrigues}}, \bibinfo {author} {\bibfnamefont {J.~C.}\ \bibnamefont {Fabris}}, \ and\ \bibinfo {author} {\bibfnamefont {H.}~\bibnamefont {Belich}},\ }\href {\doibase 10.1088/1361-6382/ae85a8} {\bibfield  {journal} {\bibinfo  {journal} {Class. Quant. Grav.}\ }\textbf {\bibinfo {volume} {43}},\ \bibinfo {pages} {145004} (\bibinfo {year} {2026})},\ \Eprint {http://arxiv.org/abs/2512.11018} {arXiv:2512.11018 [gr-qc]} \BibitemShut {NoStop}%
\bibitem [{\citenamefont {Crispim}\ \emph {et~al.}(2025)\citenamefont {Crispim}, \citenamefont {de~S.~Silva}, \citenamefont {Alencar}, \citenamefont {Muniz},\ and\ \citenamefont {S{\'a}ez-Chill{\'o}n~G{\'o}mez}}]{Crispim:2024lzf}%
  \BibitemOpen
  \bibfield  {author} {\bibinfo {author} {\bibfnamefont {T.~M.}\ \bibnamefont {Crispim}}, \bibinfo {author} {\bibfnamefont {M.~V.}\ \bibnamefont {de~S.~Silva}}, \bibinfo {author} {\bibfnamefont {G.}~\bibnamefont {Alencar}}, \bibinfo {author} {\bibfnamefont {C.~R.}\ \bibnamefont {Muniz}}, \ and\ \bibinfo {author} {\bibfnamefont {D.}~\bibnamefont {S{\'a}ez-Chill{\'o}n~G{\'o}mez}},\ }\href {\doibase 10.1088/1361-6382/adc654} {\bibfield  {journal} {\bibinfo  {journal} {Class. Quant. Grav.}\ }\textbf {\bibinfo {volume} {42}},\ \bibinfo {pages} {085005} (\bibinfo {year} {2025})},\ \Eprint {http://arxiv.org/abs/2412.05236} {arXiv:2412.05236 [gr-qc]} \BibitemShut {NoStop}%
\bibitem [{\citenamefont {Su}\ \emph {et~al.}(2025)\citenamefont {Su}, \citenamefont {Hao}, \citenamefont {Fang},\ and\ \citenamefont {Wang}}]{Su:2024gxp}%
  \BibitemOpen
  \bibfield  {author} {\bibinfo {author} {\bibfnamefont {X.}~\bibnamefont {Su}}, \bibinfo {author} {\bibfnamefont {C.-H.}\ \bibnamefont {Hao}}, \bibinfo {author} {\bibfnamefont {T.-F.}\ \bibnamefont {Fang}}, \ and\ \bibinfo {author} {\bibfnamefont {Y.-Q.}\ \bibnamefont {Wang}},\ }\href {\doibase 10.1140/epjc/s10052-025-14729-2} {\bibfield  {journal} {\bibinfo  {journal} {Eur. Phys. J. C}\ }\textbf {\bibinfo {volume} {85}},\ \bibinfo {pages} {1040} (\bibinfo {year} {2025})},\ \Eprint {http://arxiv.org/abs/2407.09591} {arXiv:2407.09591 [gr-qc]} \BibitemShut {NoStop}%
\bibitem [{\citenamefont {Matyjasek}(2020)}]{Matyjasek:2020cmi}%
  \BibitemOpen
  \bibfield  {author} {\bibinfo {author} {\bibfnamefont {J.}~\bibnamefont {Matyjasek}},\ }\href {\doibase 10.1103/PhysRevD.102.024082} {\bibfield  {journal} {\bibinfo  {journal} {Phys. Rev. D}\ }\textbf {\bibinfo {volume} {102}},\ \bibinfo {pages} {024082} (\bibinfo {year} {2020})},\ \Eprint {http://arxiv.org/abs/2005.07077} {arXiv:2005.07077 [gr-qc]} \BibitemShut {NoStop}%
\bibitem [{\citenamefont {Simpson}\ and\ \citenamefont {Visser}(2019)}]{Simpson:2018tsi}%
  \BibitemOpen
  \bibfield  {author} {\bibinfo {author} {\bibfnamefont {A.}~\bibnamefont {Simpson}}\ and\ \bibinfo {author} {\bibfnamefont {M.}~\bibnamefont {Visser}},\ }\href {\doibase 10.1088/1475-7516/2019/02/042} {\bibfield  {journal} {\bibinfo  {journal} {JCAP}\ }\textbf {\bibinfo {volume} {02}},\ \bibinfo {pages} {042} (\bibinfo {year} {2019})},\ \Eprint {http://arxiv.org/abs/1812.07114} {arXiv:1812.07114 [gr-qc]} \BibitemShut {NoStop}%
\bibitem [{\citenamefont {Bronnikov}\ \emph {et~al.}(2012)\citenamefont {Bronnikov}, \citenamefont {Konoplya},\ and\ \citenamefont {Zhidenko}}]{Bronnikov:2012ch}%
  \BibitemOpen
  \bibfield  {author} {\bibinfo {author} {\bibfnamefont {K.~A.}\ \bibnamefont {Bronnikov}}, \bibinfo {author} {\bibfnamefont {R.~A.}\ \bibnamefont {Konoplya}}, \ and\ \bibinfo {author} {\bibfnamefont {A.}~\bibnamefont {Zhidenko}},\ }\href {\doibase 10.1103/PhysRevD.86.024028} {\bibfield  {journal} {\bibinfo  {journal} {Phys. Rev. D}\ }\textbf {\bibinfo {volume} {86}},\ \bibinfo {pages} {024028} (\bibinfo {year} {2012})},\ \Eprint {http://arxiv.org/abs/1205.2224} {arXiv:1205.2224 [gr-qc]} \BibitemShut {NoStop}%
\bibitem [{\citenamefont {Bronnikov}\ and\ \citenamefont {Galiakhmetov}(2015)}]{Bronnikov:2015pha}%
  \BibitemOpen
  \bibfield  {author} {\bibinfo {author} {\bibfnamefont {K.~A.}\ \bibnamefont {Bronnikov}}\ and\ \bibinfo {author} {\bibfnamefont {A.~M.}\ \bibnamefont {Galiakhmetov}},\ }\href {\doibase 10.1134/S0202289315040027} {\bibfield  {journal} {\bibinfo  {journal} {Grav. Cosmol.}\ }\textbf {\bibinfo {volume} {21}},\ \bibinfo {pages} {283} (\bibinfo {year} {2015})},\ \Eprint {http://arxiv.org/abs/1508.01114} {arXiv:1508.01114 [gr-qc]} \BibitemShut {NoStop}%
\bibitem [{\citenamefont {Bolokhov}\ \emph {et~al.}(2012)\citenamefont {Bolokhov}, \citenamefont {Bronnikov},\ and\ \citenamefont {Skvortsova}}]{Bolokhov:2012kn}%
  \BibitemOpen
  \bibfield  {author} {\bibinfo {author} {\bibfnamefont {S.~V.}\ \bibnamefont {Bolokhov}}, \bibinfo {author} {\bibfnamefont {K.~A.}\ \bibnamefont {Bronnikov}}, \ and\ \bibinfo {author} {\bibfnamefont {M.~V.}\ \bibnamefont {Skvortsova}},\ }\href {\doibase 10.1088/0264-9381/29/24/245006} {\bibfield  {journal} {\bibinfo  {journal} {Class. Quant. Grav.}\ }\textbf {\bibinfo {volume} {29}},\ \bibinfo {pages} {245006} (\bibinfo {year} {2012})},\ \Eprint {http://arxiv.org/abs/1208.4619} {arXiv:1208.4619 [gr-qc]} \BibitemShut {NoStop}%
\bibitem [{\citenamefont {Bronnikov}\ and\ \citenamefont {Galiakhmetov}(2016)}]{Bronnikov:2016xvj}%
  \BibitemOpen
  \bibfield  {author} {\bibinfo {author} {\bibfnamefont {K.~A.}\ \bibnamefont {Bronnikov}}\ and\ \bibinfo {author} {\bibfnamefont {A.~M.}\ \bibnamefont {Galiakhmetov}},\ }\href {\doibase 10.1103/PhysRevD.94.124006} {\bibfield  {journal} {\bibinfo  {journal} {Phys. Rev. D}\ }\textbf {\bibinfo {volume} {94}},\ \bibinfo {pages} {124006} (\bibinfo {year} {2016})},\ \Eprint {http://arxiv.org/abs/1607.07791} {arXiv:1607.07791 [gr-qc]} \BibitemShut {NoStop}%
\bibitem [{\citenamefont {Lobo}(2005)}]{Lobo:2005us}%
  \BibitemOpen
  \bibfield  {author} {\bibinfo {author} {\bibfnamefont {F.~S.~N.}\ \bibnamefont {Lobo}},\ }\href {\doibase 10.1103/PhysRevD.71.084011} {\bibfield  {journal} {\bibinfo  {journal} {Phys. Rev. D}\ }\textbf {\bibinfo {volume} {71}},\ \bibinfo {pages} {084011} (\bibinfo {year} {2005})},\ \Eprint {http://arxiv.org/abs/gr-qc/0502099} {arXiv:gr-qc/0502099} \BibitemShut {NoStop}%
\bibitem [{\citenamefont {Lobo}\ and\ \citenamefont {Oliveira}(2009)}]{Lobo:2009ip}%
  \BibitemOpen
  \bibfield  {author} {\bibinfo {author} {\bibfnamefont {F.~S.~N.}\ \bibnamefont {Lobo}}\ and\ \bibinfo {author} {\bibfnamefont {M.~A.}\ \bibnamefont {Oliveira}},\ }\href {\doibase 10.1103/PhysRevD.80.104012} {\bibfield  {journal} {\bibinfo  {journal} {Phys. Rev. D}\ }\textbf {\bibinfo {volume} {80}},\ \bibinfo {pages} {104012} (\bibinfo {year} {2009})},\ \Eprint {http://arxiv.org/abs/0909.5539} {arXiv:0909.5539 [gr-qc]} \BibitemShut {NoStop}%
\bibitem [{\citenamefont {Mehdizadeh}\ \emph {et~al.}(2015)\citenamefont {Mehdizadeh}, \citenamefont {Kord~Zangeneh},\ and\ \citenamefont {Lobo}}]{Mehdizadeh:2015jra}%
  \BibitemOpen
  \bibfield  {author} {\bibinfo {author} {\bibfnamefont {M.~R.}\ \bibnamefont {Mehdizadeh}}, \bibinfo {author} {\bibfnamefont {M.}~\bibnamefont {Kord~Zangeneh}}, \ and\ \bibinfo {author} {\bibfnamefont {F.~S.~N.}\ \bibnamefont {Lobo}},\ }\href {\doibase 10.1103/PhysRevD.91.084004} {\bibfield  {journal} {\bibinfo  {journal} {Phys. Rev. D}\ }\textbf {\bibinfo {volume} {91}},\ \bibinfo {pages} {084004} (\bibinfo {year} {2015})},\ \Eprint {http://arxiv.org/abs/1501.04773} {arXiv:1501.04773 [gr-qc]} \BibitemShut {NoStop}%
\bibitem [{\citenamefont {Garcia}\ and\ \citenamefont {Lobo}(2010)}]{Garcia:2010xb}%
  \BibitemOpen
  \bibfield  {author} {\bibinfo {author} {\bibfnamefont {N.~M.}\ \bibnamefont {Garcia}}\ and\ \bibinfo {author} {\bibfnamefont {F.~S.~N.}\ \bibnamefont {Lobo}},\ }\href {\doibase 10.1103/PhysRevD.82.104018} {\bibfield  {journal} {\bibinfo  {journal} {Phys. Rev. D}\ }\textbf {\bibinfo {volume} {82}},\ \bibinfo {pages} {104018} (\bibinfo {year} {2010})},\ \Eprint {http://arxiv.org/abs/1007.3040} {arXiv:1007.3040 [gr-qc]} \BibitemShut {NoStop}%
\bibitem [{\citenamefont {Lobo}(2017)}]{Lobo:2017cay}%
  \BibitemOpen
  \bibinfo {editor} {\bibfnamefont {F.~S.~N.}\ \bibnamefont {Lobo}},\ ed.,\ \href {\doibase 10.1007/978-3-319-55182-1} {\emph {\bibinfo {title} {{Wormholes, Warp Drives and Energy Conditions}}}},\ \bibinfo {series} {Fundamental Theories of Physics}, Vol.\ \bibinfo {volume} {189}\ (\bibinfo  {publisher} {Springer},\ \bibinfo {year} {2017})\ \Eprint {http://arxiv.org/abs/2103.05610} {arXiv:2103.05610 [gr-qc]} \BibitemShut {NoStop}%
\bibitem [{\citenamefont {de~S.~Silva}\ \emph {et~al.}(2025)\citenamefont {de~S.~Silva}, \citenamefont {Alencar}, \citenamefont {Costa~Filho}, \citenamefont {Neves},\ and\ \citenamefont {Muniz}}]{deSSilva:2024gdc}%
  \BibitemOpen
  \bibfield  {author} {\bibinfo {author} {\bibfnamefont {M.~V.}\ \bibnamefont {de~S.~Silva}}, \bibinfo {author} {\bibfnamefont {G.}~\bibnamefont {Alencar}}, \bibinfo {author} {\bibfnamefont {R.~N.}\ \bibnamefont {Costa~Filho}}, \bibinfo {author} {\bibfnamefont {R.~M.~P.}\ \bibnamefont {Neves}}, \ and\ \bibinfo {author} {\bibfnamefont {C.~R.}\ \bibnamefont {Muniz}},\ }\href {\doibase 10.1140/epjp/s13360-025-06214-2} {\bibfield  {journal} {\bibinfo  {journal} {Eur. Phys. J. Plus}\ }\textbf {\bibinfo {volume} {140}},\ \bibinfo {pages} {289} (\bibinfo {year} {2025})},\ \Eprint {http://arxiv.org/abs/2411.12063} {arXiv:2411.12063 [gr-qc]} \BibitemShut {NoStop}%
\bibitem [{\citenamefont {Bronnikov}(2018)}]{Bronnikov:2017sgg}%
  \BibitemOpen
  \bibfield  {author} {\bibinfo {author} {\bibfnamefont {K.~A.}\ \bibnamefont {Bronnikov}},\ }\href {\doibase 10.1142/S0218271818410055} {\bibfield  {journal} {\bibinfo  {journal} {Int. J. Mod. Phys. D}\ }\textbf {\bibinfo {volume} {27}},\ \bibinfo {pages} {1841005} (\bibinfo {year} {2018})},\ \Eprint {http://arxiv.org/abs/1711.00087} {arXiv:1711.00087 [gr-qc]} \BibitemShut {NoStop}%
\bibitem [{\citenamefont {Bronnikov}\ and\ \citenamefont {Walia}(2022)}]{Bronnikov:2021uta}%
  \BibitemOpen
  \bibfield  {author} {\bibinfo {author} {\bibfnamefont {K.~A.}\ \bibnamefont {Bronnikov}}\ and\ \bibinfo {author} {\bibfnamefont {R.~K.}\ \bibnamefont {Walia}},\ }\href {\doibase 10.1103/PhysRevD.105.044039} {\bibfield  {journal} {\bibinfo  {journal} {Phys. Rev. D}\ }\textbf {\bibinfo {volume} {105}},\ \bibinfo {pages} {044039} (\bibinfo {year} {2022})},\ \Eprint {http://arxiv.org/abs/2112.13198} {arXiv:2112.13198 [gr-qc]} \BibitemShut {NoStop}%
\bibitem [{\citenamefont {Alencar}\ \emph {et~al.}(2026{\natexlab{a}})\citenamefont {Alencar}, \citenamefont {Crispim}, \citenamefont {D{\'a}rlla},\ and\ \citenamefont {Landim}}]{Alencar:2026kqx}%
  \BibitemOpen
  \bibfield  {author} {\bibinfo {author} {\bibfnamefont {G.}~\bibnamefont {Alencar}}, \bibinfo {author} {\bibfnamefont {T.}~\bibnamefont {Crispim}}, \bibinfo {author} {\bibfnamefont {R.}~\bibnamefont {D{\'a}rlla}}, \ and\ \bibinfo {author} {\bibfnamefont {R.}~\bibnamefont {Landim}},\ }\href@noop {} {\  (\bibinfo {year} {2026}{\natexlab{a}})},\ \Eprint {http://arxiv.org/abs/2610.14357} {arXiv:2610.14357 [gr-qc]} \BibitemShut {NoStop}%
\bibitem [{\citenamefont {Alencar}\ and\ \citenamefont {Crispim}(2026)}]{Alencar:2026hjf}%
  \BibitemOpen
  \bibfield  {author} {\bibinfo {author} {\bibfnamefont {G.}~\bibnamefont {Alencar}}\ and\ \bibinfo {author} {\bibfnamefont {T.~M.}\ \bibnamefont {Crispim}},\ }\href@noop {} {\  (\bibinfo {year} {2026})},\ \Eprint {http://arxiv.org/abs/2603.30003} {arXiv:2603.30003 [gr-qc]} \BibitemShut {NoStop}%
\bibitem [{\citenamefont {Damour}\ and\ \citenamefont {Solodukhin}(2007)}]{Damour:2007ap}%
  \BibitemOpen
  \bibfield  {author} {\bibinfo {author} {\bibfnamefont {T.}~\bibnamefont {Damour}}\ and\ \bibinfo {author} {\bibfnamefont {S.~N.}\ \bibnamefont {Solodukhin}},\ }\href {\doibase 10.1103/PhysRevD.76.024016} {\bibfield  {journal} {\bibinfo  {journal} {Phys. Rev. D}\ }\textbf {\bibinfo {volume} {76}},\ \bibinfo {pages} {024016} (\bibinfo {year} {2007})},\ \Eprint {http://arxiv.org/abs/0704.2667} {arXiv:0704.2667 [gr-qc]} \BibitemShut {NoStop}%
\bibitem [{\citenamefont {Solodukhin}(2026)}]{Solodukhin:2026mux}%
  \BibitemOpen
  \bibfield  {author} {\bibinfo {author} {\bibfnamefont {S.~N.}\ \bibnamefont {Solodukhin}},\ }\href@noop {} {\  (\bibinfo {year} {2026})},\ \Eprint {http://arxiv.org/abs/2610.11664} {arXiv:2610.11664 [hep-th]} \BibitemShut {NoStop}%
\bibitem [{\citenamefont {Tsukamoto}(2023)}]{Tsukamoto:2023rzd}%
  \BibitemOpen
  \bibfield  {author} {\bibinfo {author} {\bibfnamefont {N.}~\bibnamefont {Tsukamoto}},\ }\href {\doibase 10.1088/1361-6382/acfb6e} {\bibfield  {journal} {\bibinfo  {journal} {Class. Quant. Grav.}\ }\textbf {\bibinfo {volume} {40}},\ \bibinfo {pages} {228001} (\bibinfo {year} {2023})},\ \Eprint {http://arxiv.org/abs/2307.11303} {arXiv:2307.11303 [gr-qc]} \BibitemShut {NoStop}%
\bibitem [{\citenamefont {Tsukamoto}(2020)}]{Tsukamoto:2020uay}%
  \BibitemOpen
  \bibfield  {author} {\bibinfo {author} {\bibfnamefont {N.}~\bibnamefont {Tsukamoto}},\ }\href {\doibase 10.1103/PhysRevD.101.104021} {\bibfield  {journal} {\bibinfo  {journal} {Phys. Rev. D}\ }\textbf {\bibinfo {volume} {101}},\ \bibinfo {pages} {104021} (\bibinfo {year} {2020})},\ \bibinfo {note} {[Erratum: Phys.Rev.D 106, 049901 (2020), Erratum: Phys.Rev.D 106, 049901 (2022)]},\ \Eprint {http://arxiv.org/abs/2004.00822} {arXiv:2004.00822 [gr-qc]} \BibitemShut {NoStop}%
\bibitem [{\citenamefont {Tsukamoto}(2026)}]{Tsukamoto:2026beg}%
  \BibitemOpen
  \bibfield  {author} {\bibinfo {author} {\bibfnamefont {N.}~\bibnamefont {Tsukamoto}},\ }\href {\doibase 10.1103/lg8p-2j8t} {\bibfield  {journal} {\bibinfo  {journal} {Phys. Rev. D}\ }\textbf {\bibinfo {volume} {114}},\ \bibinfo {pages} {084047} (\bibinfo {year} {2026})},\ \Eprint {http://arxiv.org/abs/2608.23269} {arXiv:2608.23269 [gr-qc]} \BibitemShut {NoStop}%
\bibitem [{\citenamefont {Ho}\ \emph {et~al.}(2026)\citenamefont {Ho}, \citenamefont {Ko}, \citenamefont {Hsieh},\ and\ \citenamefont {Lee}}]{Ho:2026gmy}%
  \BibitemOpen
  \bibfield  {author} {\bibinfo {author} {\bibfnamefont {S.-C.}\ \bibnamefont {Ho}}, \bibinfo {author} {\bibfnamefont {Y.-C.}\ \bibnamefont {Ko}}, \bibinfo {author} {\bibfnamefont {T.}~\bibnamefont {Hsieh}}, \ and\ \bibinfo {author} {\bibfnamefont {D.-S.}\ \bibnamefont {Lee}},\ }\href {\doibase 10.1088/1475-7516/2026/09/069} {\bibfield  {journal} {\bibinfo  {journal} {JCAP}\ }\textbf {\bibinfo {volume} {09}},\ \bibinfo {pages} {069} (\bibinfo {year} {2026})},\ \Eprint {http://arxiv.org/abs/2605.22187} {arXiv:2605.22187 [gr-qc]} \BibitemShut {NoStop}%
\bibitem [{\citenamefont {Bueno}\ \emph {et~al.}(2018)\citenamefont {Bueno}, \citenamefont {Cano}, \citenamefont {Goelen}, \citenamefont {Hertog},\ and\ \citenamefont {Vercnocke}}]{Bueno:2017hyj}%
  \BibitemOpen
  \bibfield  {author} {\bibinfo {author} {\bibfnamefont {P.}~\bibnamefont {Bueno}}, \bibinfo {author} {\bibfnamefont {P.~A.}\ \bibnamefont {Cano}}, \bibinfo {author} {\bibfnamefont {F.}~\bibnamefont {Goelen}}, \bibinfo {author} {\bibfnamefont {T.}~\bibnamefont {Hertog}}, \ and\ \bibinfo {author} {\bibfnamefont {B.}~\bibnamefont {Vercnocke}},\ }\href {\doibase 10.1103/PhysRevD.97.024040} {\bibfield  {journal} {\bibinfo  {journal} {Phys. Rev. D}\ }\textbf {\bibinfo {volume} {97}},\ \bibinfo {pages} {024040} (\bibinfo {year} {2018})},\ \Eprint {http://arxiv.org/abs/1711.00391} {arXiv:1711.00391 [gr-qc]} \BibitemShut {NoStop}%
\bibitem [{\citenamefont {Yusupova}\ \emph {et~al.}(2025)\citenamefont {Yusupova}, \citenamefont {Karimov},\ and\ \citenamefont {Bhattacharya}}]{Yusupova:2025niy}%
  \BibitemOpen
  \bibfield  {author} {\bibinfo {author} {\bibfnamefont {R.~M.}\ \bibnamefont {Yusupova}}, \bibinfo {author} {\bibfnamefont {R.~K.}\ \bibnamefont {Karimov}}, \ and\ \bibinfo {author} {\bibfnamefont {A.}~\bibnamefont {Bhattacharya}},\ }\href {\doibase 10.1134/S0202289324700476} {\bibfield  {journal} {\bibinfo  {journal} {Grav. Cosmol.}\ }\textbf {\bibinfo {volume} {31}},\ \bibinfo {pages} {37} (\bibinfo {year} {2025})},\ \Eprint {http://arxiv.org/abs/2607.00388} {arXiv:2607.00388 [gr-qc]} \BibitemShut {NoStop}%
\bibitem [{\citenamefont {Qian}\ \emph {et~al.}(2025)\citenamefont {Qian}, \citenamefont {Pan}, \citenamefont {Li}, \citenamefont {Guo}, \citenamefont {Lin},\ and\ \citenamefont {Yue}}]{Qian:2025lau}%
  \BibitemOpen
  \bibfield  {author} {\bibinfo {author} {\bibfnamefont {W.-L.}\ \bibnamefont {Qian}}, \bibinfo {author} {\bibfnamefont {Q.}~\bibnamefont {Pan}}, \bibinfo {author} {\bibfnamefont {G.-R.}\ \bibnamefont {Li}}, \bibinfo {author} {\bibfnamefont {H.}~\bibnamefont {Guo}}, \bibinfo {author} {\bibfnamefont {K.}~\bibnamefont {Lin}}, \ and\ \bibinfo {author} {\bibfnamefont {R.-H.}\ \bibnamefont {Yue}},\ }\href {\doibase 10.1002/asna.20240117} {\bibfield  {journal} {\bibinfo  {journal} {Astron. Nachr.}\ }\textbf {\bibinfo {volume} {346}},\ \bibinfo {pages} {e20240117} (\bibinfo {year} {2025})}\BibitemShut {NoStop}%
\bibitem [{\citenamefont {Qian}\ \emph {et~al.}(2024)\citenamefont {Qian}, \citenamefont {Pan}, \citenamefont {Wang},\ and\ \citenamefont {Yue}}]{Qian:2024zvq}%
  \BibitemOpen
  \bibfield  {author} {\bibinfo {author} {\bibfnamefont {W.-L.}\ \bibnamefont {Qian}}, \bibinfo {author} {\bibfnamefont {Q.}~\bibnamefont {Pan}}, \bibinfo {author} {\bibfnamefont {B.}~\bibnamefont {Wang}}, \ and\ \bibinfo {author} {\bibfnamefont {R.-H.}\ \bibnamefont {Yue}},\ }\href {\doibase 10.1016/j.physletb.2024.138874} {\bibfield  {journal} {\bibinfo  {journal} {Phys. Lett. B}\ }\textbf {\bibinfo {volume} {856}},\ \bibinfo {pages} {138874} (\bibinfo {year} {2024})},\ \Eprint {http://arxiv.org/abs/2402.05485} {arXiv:2402.05485 [gr-qc]} \BibitemShut {NoStop}%
\bibitem [{\citenamefont {Bambi}\ and\ \citenamefont {Freese}(2009)}]{Bambi:2008jg}%
  \BibitemOpen
  \bibfield  {author} {\bibinfo {author} {\bibfnamefont {C.}~\bibnamefont {Bambi}}\ and\ \bibinfo {author} {\bibfnamefont {K.}~\bibnamefont {Freese}},\ }\href {\doibase 10.1103/PhysRevD.79.043002} {\bibfield  {journal} {\bibinfo  {journal} {Phys. Rev. D}\ }\textbf {\bibinfo {volume} {79}},\ \bibinfo {pages} {043002} (\bibinfo {year} {2009})},\ \Eprint {http://arxiv.org/abs/0812.1328} {arXiv:0812.1328 [astro-ph]} \BibitemShut {NoStop}%
\bibitem [{\citenamefont {Ortiz}\ \emph {et~al.}(2015)\citenamefont {Ortiz}, \citenamefont {Sarbach},\ and\ \citenamefont {Zannias}}]{Ortiz_2015}%
  \BibitemOpen
  \bibfield  {author} {\bibinfo {author} {\bibfnamefont {N.}~\bibnamefont {Ortiz}}, \bibinfo {author} {\bibfnamefont {O.}~\bibnamefont {Sarbach}}, \ and\ \bibinfo {author} {\bibfnamefont {T.}~\bibnamefont {Zannias}},\ }\href {\doibase 10.1103/PhysRevD.92.044035} {\bibfield  {journal} {\bibinfo  {journal} {Phys. Rev. D}\ }\textbf {\bibinfo {volume} {92}},\ \bibinfo {pages} {044035} (\bibinfo {year} {2015})},\ \Eprint {http://arxiv.org/abs/1505.07017} {arXiv:1505.07017 [gr-qc]} \BibitemShut {NoStop}%
\bibitem [{\citenamefont {Shaikh}\ \emph {et~al.}(2019)\citenamefont {Shaikh}, \citenamefont {Kocherlakota}, \citenamefont {Narayan},\ and\ \citenamefont {Joshi}}]{Shaikh_2018}%
  \BibitemOpen
  \bibfield  {author} {\bibinfo {author} {\bibfnamefont {R.}~\bibnamefont {Shaikh}}, \bibinfo {author} {\bibfnamefont {P.}~\bibnamefont {Kocherlakota}}, \bibinfo {author} {\bibfnamefont {R.}~\bibnamefont {Narayan}}, \ and\ \bibinfo {author} {\bibfnamefont {P.~S.}\ \bibnamefont {Joshi}},\ }\href {\doibase 10.1093/mnras/sty2624} {\bibfield  {journal} {\bibinfo  {journal} {Mon. Not. Roy. Astron. Soc.}\ }\textbf {\bibinfo {volume} {482}},\ \bibinfo {pages} {52} (\bibinfo {year} {2019})},\ \Eprint {http://arxiv.org/abs/1802.08060} {arXiv:1802.08060 [astro-ph.HE]} \BibitemShut {NoStop}%
\bibitem [{\citenamefont {Bambi}(2013)}]{Bambi:2013nla}%
  \BibitemOpen
  \bibfield  {author} {\bibinfo {author} {\bibfnamefont {C.}~\bibnamefont {Bambi}},\ }\href {\doibase 10.1103/PhysRevD.87.107501} {\bibfield  {journal} {\bibinfo  {journal} {Phys. Rev. D}\ }\textbf {\bibinfo {volume} {87}},\ \bibinfo {pages} {107501} (\bibinfo {year} {2013})},\ \Eprint {http://arxiv.org/abs/1304.5691} {arXiv:1304.5691 [gr-qc]} \BibitemShut {NoStop}%
\bibitem [{\citenamefont {Azreg-A{\"\i}nou}(2015)}]{Azreg-Ainou:2014dwa}%
  \BibitemOpen
  \bibfield  {author} {\bibinfo {author} {\bibfnamefont {M.}~\bibnamefont {Azreg-A{\"\i}nou}},\ }\href {\doibase 10.1088/1475-7516/2015/07/037} {\bibfield  {journal} {\bibinfo  {journal} {JCAP}\ }\textbf {\bibinfo {volume} {07}},\ \bibinfo {pages} {037} (\bibinfo {year} {2015})},\ \Eprint {http://arxiv.org/abs/1412.8282} {arXiv:1412.8282 [gr-qc]} \BibitemShut {NoStop}%
\bibitem [{\citenamefont {Ishkaeva}\ and\ \citenamefont {Sushkov}(2023)}]{Ishkaeva:2023xny}%
  \BibitemOpen
  \bibfield  {author} {\bibinfo {author} {\bibfnamefont {V.~A.}\ \bibnamefont {Ishkaeva}}\ and\ \bibinfo {author} {\bibfnamefont {S.~V.}\ \bibnamefont {Sushkov}},\ }\href {\doibase 10.1103/PhysRevD.108.084054} {\bibfield  {journal} {\bibinfo  {journal} {Phys. Rev. D}\ }\textbf {\bibinfo {volume} {108}},\ \bibinfo {pages} {084054} (\bibinfo {year} {2023})},\ \Eprint {http://arxiv.org/abs/2308.02268} {arXiv:2308.02268 [gr-qc]} \BibitemShut {NoStop}%
\bibitem [{\citenamefont {Alloqulov}\ \emph {et~al.}(2024)\citenamefont {Alloqulov}, \citenamefont {Atamurotov}, \citenamefont {Abdujabbarov}, \citenamefont {Ahmedov},\ and\ \citenamefont {Khamidov}}]{Alloqulov:2024olb}%
  \BibitemOpen
  \bibfield  {author} {\bibinfo {author} {\bibfnamefont {M.}~\bibnamefont {Alloqulov}}, \bibinfo {author} {\bibfnamefont {F.}~\bibnamefont {Atamurotov}}, \bibinfo {author} {\bibfnamefont {A.}~\bibnamefont {Abdujabbarov}}, \bibinfo {author} {\bibfnamefont {B.}~\bibnamefont {Ahmedov}}, \ and\ \bibinfo {author} {\bibfnamefont {V.}~\bibnamefont {Khamidov}},\ }\href {\doibase 10.1088/1674-1137/ad1677} {\bibfield  {journal} {\bibinfo  {journal} {Chin. Phys. C}\ }\textbf {\bibinfo {volume} {48}},\ \bibinfo {pages} {025104} (\bibinfo {year} {2024})}\BibitemShut {NoStop}%
\bibitem [{\citenamefont {Guerrero}\ \emph {et~al.}(2021{\natexlab{a}})\citenamefont {Guerrero}, \citenamefont {Olmo}, \citenamefont {Rubiera-Garcia},\ and\ \citenamefont {G{\'o}mez}}]{Guerrero:2021ues}%
  \BibitemOpen
  \bibfield  {author} {\bibinfo {author} {\bibfnamefont {M.}~\bibnamefont {Guerrero}}, \bibinfo {author} {\bibfnamefont {G.~J.}\ \bibnamefont {Olmo}}, \bibinfo {author} {\bibfnamefont {D.}~\bibnamefont {Rubiera-Garcia}}, \ and\ \bibinfo {author} {\bibfnamefont {D.~S.-C.}\ \bibnamefont {G{\'o}mez}},\ }\href {\doibase 10.1088/1475-7516/2021/08/036} {\bibfield  {journal} {\bibinfo  {journal} {JCAP}\ }\textbf {\bibinfo {volume} {08}},\ \bibinfo {pages} {036} (\bibinfo {year} {2021}{\natexlab{a}})},\ \Eprint {http://arxiv.org/abs/2105.15073} {arXiv:2105.15073 [gr-qc]} \BibitemShut {NoStop}%
\bibitem [{\citenamefont {Shaikh}(2018)}]{Shaikh:2018kfv}%
  \BibitemOpen
  \bibfield  {author} {\bibinfo {author} {\bibfnamefont {R.}~\bibnamefont {Shaikh}},\ }\href {\doibase 10.1103/PhysRevD.98.024044} {\bibfield  {journal} {\bibinfo  {journal} {Phys. Rev. D}\ }\textbf {\bibinfo {volume} {98}},\ \bibinfo {pages} {024044} (\bibinfo {year} {2018})},\ \Eprint {http://arxiv.org/abs/1803.11422} {arXiv:1803.11422 [gr-qc]} \BibitemShut {NoStop}%
\bibitem [{\citenamefont {Akiyama}\ \emph {et~al.}(2019)\citenamefont {Akiyama} \emph {et~al.}}]{EventHorizonTelescope:2019dse}%
  \BibitemOpen
  \bibfield  {author} {\bibinfo {author} {\bibfnamefont {K.}~\bibnamefont {Akiyama}} \emph {et~al.} (\bibinfo {collaboration} {Event Horizon Telescope}),\ }\href {\doibase 10.3847/2041-8213/ab0ec7} {\bibfield  {journal} {\bibinfo  {journal} {Astrophys. J. Lett.}\ }\textbf {\bibinfo {volume} {875}},\ \bibinfo {pages} {L1} (\bibinfo {year} {2019})},\ \Eprint {http://arxiv.org/abs/1906.11238} {arXiv:1906.11238 [astro-ph.GA]} \BibitemShut {NoStop}%
\bibitem [{\citenamefont {Akiyama}\ \emph {et~al.}(2022{\natexlab{a}})\citenamefont {Akiyama} \emph {et~al.}}]{EventHorizonTelescope:2022wkp}%
  \BibitemOpen
  \bibfield  {author} {\bibinfo {author} {\bibfnamefont {K.}~\bibnamefont {Akiyama}} \emph {et~al.} (\bibinfo {collaboration} {Event Horizon Telescope}),\ }\href {\doibase 10.3847/2041-8213/ac6674} {\bibfield  {journal} {\bibinfo  {journal} {Astrophys. J. Lett.}\ }\textbf {\bibinfo {volume} {930}},\ \bibinfo {pages} {L12} (\bibinfo {year} {2022}{\natexlab{a}})},\ \Eprint {http://arxiv.org/abs/2311.08680} {arXiv:2311.08680 [astro-ph.HE]} \BibitemShut {NoStop}%
\bibitem [{\citenamefont {De~Falco}\ \emph {et~al.}(2020)\citenamefont {De~Falco}, \citenamefont {Battista}, \citenamefont {Capozziello},\ and\ \citenamefont {De~Laurentis}}]{DeFalco:2020afv}%
  \BibitemOpen
  \bibfield  {author} {\bibinfo {author} {\bibfnamefont {V.}~\bibnamefont {De~Falco}}, \bibinfo {author} {\bibfnamefont {E.}~\bibnamefont {Battista}}, \bibinfo {author} {\bibfnamefont {S.}~\bibnamefont {Capozziello}}, \ and\ \bibinfo {author} {\bibfnamefont {M.}~\bibnamefont {De~Laurentis}},\ }\href {\doibase 10.1103/PhysRevD.101.104037} {\bibfield  {journal} {\bibinfo  {journal} {Phys. Rev. D}\ }\textbf {\bibinfo {volume} {101}},\ \bibinfo {pages} {104037} (\bibinfo {year} {2020})},\ \Eprint {http://arxiv.org/abs/2004.14849} {arXiv:2004.14849 [gr-qc]} \BibitemShut {NoStop}%
\bibitem [{\citenamefont {De~Falco}\ and\ \citenamefont {Capozziello}(2023)}]{DeFalco:2023twb}%
  \BibitemOpen
  \bibfield  {author} {\bibinfo {author} {\bibfnamefont {V.}~\bibnamefont {De~Falco}}\ and\ \bibinfo {author} {\bibfnamefont {S.}~\bibnamefont {Capozziello}},\ }\href {\doibase 10.1103/PhysRevD.108.104030} {\bibfield  {journal} {\bibinfo  {journal} {Phys. Rev. D}\ }\textbf {\bibinfo {volume} {108}},\ \bibinfo {pages} {104030} (\bibinfo {year} {2023})},\ \Eprint {http://arxiv.org/abs/2308.05440} {arXiv:2308.05440 [gr-qc]} \BibitemShut {NoStop}%
\bibitem [{\citenamefont {Jana}\ \emph {et~al.}(2024)\citenamefont {Jana}, \citenamefont {Sharma},\ and\ \citenamefont {Ghosh}}]{Jana:2024pdh}%
  \BibitemOpen
  \bibfield  {author} {\bibinfo {author} {\bibfnamefont {S.}~\bibnamefont {Jana}}, \bibinfo {author} {\bibfnamefont {V.}~\bibnamefont {Sharma}}, \ and\ \bibinfo {author} {\bibfnamefont {S.}~\bibnamefont {Ghosh}},\ }\href@noop {} {\  (\bibinfo {year} {2024})},\ \Eprint {http://arxiv.org/abs/2411.10804} {arXiv:2411.10804 [gr-qc]} \BibitemShut {NoStop}%
\bibitem [{\citenamefont {Amir}\ \emph {et~al.}(2019)\citenamefont {Amir}, \citenamefont {Jusufi}, \citenamefont {Banerjee},\ and\ \citenamefont {Hansraj}}]{Amir:2018pcu}%
  \BibitemOpen
  \bibfield  {author} {\bibinfo {author} {\bibfnamefont {M.}~\bibnamefont {Amir}}, \bibinfo {author} {\bibfnamefont {K.}~\bibnamefont {Jusufi}}, \bibinfo {author} {\bibfnamefont {A.}~\bibnamefont {Banerjee}}, \ and\ \bibinfo {author} {\bibfnamefont {S.}~\bibnamefont {Hansraj}},\ }\href {\doibase 10.1088/1361-6382/ab42be} {\bibfield  {journal} {\bibinfo  {journal} {Class. Quant. Grav.}\ }\textbf {\bibinfo {volume} {36}},\ \bibinfo {pages} {215007} (\bibinfo {year} {2019})},\ \Eprint {http://arxiv.org/abs/1806.07782} {arXiv:1806.07782 [gr-qc]} \BibitemShut {NoStop}%
\bibitem [{\citenamefont {Bronnikov}\ \emph {et~al.}(2023)\citenamefont {Bronnikov}, \citenamefont {Rodrigues},\ and\ \citenamefont {de~S.~Silva}}]{Bronnikov:2023aya}%
  \BibitemOpen
  \bibfield  {author} {\bibinfo {author} {\bibfnamefont {K.~A.}\ \bibnamefont {Bronnikov}}, \bibinfo {author} {\bibfnamefont {M.~E.}\ \bibnamefont {Rodrigues}}, \ and\ \bibinfo {author} {\bibfnamefont {M.~V.}\ \bibnamefont {de~S.~Silva}},\ }\href {\doibase 10.1103/PhysRevD.108.024065} {\bibfield  {journal} {\bibinfo  {journal} {Phys. Rev. D}\ }\textbf {\bibinfo {volume} {108}},\ \bibinfo {pages} {024065} (\bibinfo {year} {2023})},\ \Eprint {http://arxiv.org/abs/2305.19296} {arXiv:2305.19296 [gr-qc]} \BibitemShut {NoStop}%
\bibitem [{\citenamefont {Akiyama}\ \emph {et~al.}(2022{\natexlab{b}})\citenamefont {Akiyama} \emph {et~al.}}]{EventHorizonTelescope:2022xqj}%
  \BibitemOpen
  \bibfield  {author} {\bibinfo {author} {\bibfnamefont {K.}~\bibnamefont {Akiyama}} \emph {et~al.} (\bibinfo {collaboration} {Event Horizon Telescope}),\ }\href {\doibase 10.3847/2041-8213/ac6756} {\bibfield  {journal} {\bibinfo  {journal} {Astrophys. J. Lett.}\ }\textbf {\bibinfo {volume} {930}},\ \bibinfo {pages} {L17} (\bibinfo {year} {2022}{\natexlab{b}})},\ \Eprint {http://arxiv.org/abs/2311.09484} {arXiv:2311.09484 [astro-ph.HE]} \BibitemShut {NoStop}%
\bibitem [{\citenamefont {Perlick}\ and\ \citenamefont {Tsupko}(2022)}]{Perlick:2021aok}%
  \BibitemOpen
  \bibfield  {author} {\bibinfo {author} {\bibfnamefont {V.}~\bibnamefont {Perlick}}\ and\ \bibinfo {author} {\bibfnamefont {O.~Y.}\ \bibnamefont {Tsupko}},\ }\href {\doibase 10.1016/j.physrep.2021.10.004} {\bibfield  {journal} {\bibinfo  {journal} {Phys. Rept.}\ }\textbf {\bibinfo {volume} {947}},\ \bibinfo {pages} {1} (\bibinfo {year} {2022})},\ \Eprint {http://arxiv.org/abs/2105.07101} {arXiv:2105.07101 [gr-qc]} \BibitemShut {NoStop}%
\bibitem [{\citenamefont {Do}\ \emph {et~al.}(2019)\citenamefont {Do} \emph {et~al.}}]{Do:2019txf}%
  \BibitemOpen
  \bibfield  {author} {\bibinfo {author} {\bibfnamefont {T.}~\bibnamefont {Do}} \emph {et~al.},\ }\href {\doibase 10.1126/science.aav8137} {\bibfield  {journal} {\bibinfo  {journal} {Science}\ }\textbf {\bibinfo {volume} {365}},\ \bibinfo {pages} {664} (\bibinfo {year} {2019})},\ \Eprint {http://arxiv.org/abs/1907.10731} {arXiv:1907.10731 [astro-ph.GA]} \BibitemShut {NoStop}%
\bibitem [{\citenamefont {Abuter}\ \emph {et~al.}(2020)\citenamefont {Abuter} \emph {et~al.}}]{GRAVITY:2020gka}%
  \BibitemOpen
  \bibfield  {author} {\bibinfo {author} {\bibfnamefont {R.}~\bibnamefont {Abuter}} \emph {et~al.} (\bibinfo {collaboration} {GRAVITY}),\ }\href {\doibase 10.1051/0004-6361/202037813} {\bibfield  {journal} {\bibinfo  {journal} {Astron. Astrophys.}\ }\textbf {\bibinfo {volume} {636}},\ \bibinfo {pages} {L5} (\bibinfo {year} {2020})},\ \Eprint {http://arxiv.org/abs/2004.07187} {arXiv:2004.07187 [astro-ph.GA]} \BibitemShut {NoStop}%
\bibitem [{\citenamefont {Vagnozzi}\ \emph {et~al.}(2023)\citenamefont {Vagnozzi} \emph {et~al.}}]{Vagnozzi:2022moj}%
  \BibitemOpen
  \bibfield  {author} {\bibinfo {author} {\bibfnamefont {S.}~\bibnamefont {Vagnozzi}} \emph {et~al.},\ }\href {\doibase 10.1088/1361-6382/acd97b} {\bibfield  {journal} {\bibinfo  {journal} {Class. Quant. Grav.}\ }\textbf {\bibinfo {volume} {40}},\ \bibinfo {pages} {165007} (\bibinfo {year} {2023})},\ \Eprint {http://arxiv.org/abs/2205.07787} {arXiv:2205.07787 [gr-qc]} \BibitemShut {NoStop}%
\bibitem [{\citenamefont {Gralla}\ \emph {et~al.}(2019)\citenamefont {Gralla}, \citenamefont {Holz},\ and\ \citenamefont {Wald}}]{Gralla:2019xty}%
  \BibitemOpen
  \bibfield  {author} {\bibinfo {author} {\bibfnamefont {S.~E.}\ \bibnamefont {Gralla}}, \bibinfo {author} {\bibfnamefont {D.~E.}\ \bibnamefont {Holz}}, \ and\ \bibinfo {author} {\bibfnamefont {R.~M.}\ \bibnamefont {Wald}},\ }\href {\doibase 10.1103/PhysRevD.100.024018} {\bibfield  {journal} {\bibinfo  {journal} {Phys. Rev. D}\ }\textbf {\bibinfo {volume} {100}},\ \bibinfo {pages} {024018} (\bibinfo {year} {2019})},\ \Eprint {http://arxiv.org/abs/1906.00873} {arXiv:1906.00873 [astro-ph.HE]} \BibitemShut {NoStop}%
\bibitem [{\citenamefont {Wielgus}\ \emph {et~al.}(2020)\citenamefont {Wielgus}, \citenamefont {Horak}, \citenamefont {Vincent},\ and\ \citenamefont {Abramowicz}}]{Wielgus:2020uqz}%
  \BibitemOpen
  \bibfield  {author} {\bibinfo {author} {\bibfnamefont {M.}~\bibnamefont {Wielgus}}, \bibinfo {author} {\bibfnamefont {J.}~\bibnamefont {Horak}}, \bibinfo {author} {\bibfnamefont {F.}~\bibnamefont {Vincent}}, \ and\ \bibinfo {author} {\bibfnamefont {M.}~\bibnamefont {Abramowicz}},\ }\href {\doibase 10.1103/PhysRevD.102.084044} {\bibfield  {journal} {\bibinfo  {journal} {Phys. Rev. D}\ }\textbf {\bibinfo {volume} {102}},\ \bibinfo {pages} {084044} (\bibinfo {year} {2020})},\ \Eprint {http://arxiv.org/abs/2008.10130} {arXiv:2008.10130 [gr-qc]} \BibitemShut {NoStop}%
\bibitem [{\citenamefont {Guerrero}\ \emph {et~al.}(2021{\natexlab{b}})\citenamefont {Guerrero}, \citenamefont {Olmo},\ and\ \citenamefont {Rubiera-Garcia}}]{Guerrero:2021pxt}%
  \BibitemOpen
  \bibfield  {author} {\bibinfo {author} {\bibfnamefont {M.}~\bibnamefont {Guerrero}}, \bibinfo {author} {\bibfnamefont {G.~J.}\ \bibnamefont {Olmo}}, \ and\ \bibinfo {author} {\bibfnamefont {D.}~\bibnamefont {Rubiera-Garcia}},\ }\href {\doibase 10.1088/1475-7516/2021/04/066} {\bibfield  {journal} {\bibinfo  {journal} {JCAP}\ }\textbf {\bibinfo {volume} {04}},\ \bibinfo {pages} {066} (\bibinfo {year} {2021}{\natexlab{b}})},\ \Eprint {http://arxiv.org/abs/2102.00840} {arXiv:2102.00840 [gr-qc]} \BibitemShut {NoStop}%
\bibitem [{\citenamefont {Macedo}\ \emph {et~al.}(2026)\citenamefont {Macedo}, \citenamefont {Rosa}, \citenamefont {Rubiera-Garcia},\ and\ \citenamefont {Rueda}}]{Macedo:2025ipc}%
  \BibitemOpen
  \bibfield  {author} {\bibinfo {author} {\bibfnamefont {C.~F.~B.}\ \bibnamefont {Macedo}}, \bibinfo {author} {\bibfnamefont {J.~L.}\ \bibnamefont {Rosa}}, \bibinfo {author} {\bibfnamefont {D.}~\bibnamefont {Rubiera-Garcia}}, \ and\ \bibinfo {author} {\bibfnamefont {A.}~\bibnamefont {Rueda}},\ }\href {\doibase 10.1103/wzxv-9wk1} {\bibfield  {journal} {\bibinfo  {journal} {Phys. Rev. D}\ }\textbf {\bibinfo {volume} {113}},\ \bibinfo {pages} {064004} (\bibinfo {year} {2026})},\ \Eprint {http://arxiv.org/abs/2510.19677} {arXiv:2510.19677 [gr-qc]} \BibitemShut {NoStop}%
\bibitem [{\citenamefont {Johnson}(1949)}]{Johnson:1949}%
  \BibitemOpen
  \bibfield  {author} {\bibinfo {author} {\bibfnamefont {N.~L.}\ \bibnamefont {Johnson}},\ }\href {\doibase 10.1093/biomet/36.1-2.149} {\bibfield  {journal} {\bibinfo  {journal} {Biometrika}\ }\textbf {\bibinfo {volume} {36}},\ \bibinfo {pages} {149} (\bibinfo {year} {1949})}\BibitemShut {NoStop}%
\bibitem [{\citenamefont {Gralla}\ \emph {et~al.}(2020)\citenamefont {Gralla}, \citenamefont {Lupsasca},\ and\ \citenamefont {Marrone}}]{Gralla:2020srx}%
  \BibitemOpen
  \bibfield  {author} {\bibinfo {author} {\bibfnamefont {S.~E.}\ \bibnamefont {Gralla}}, \bibinfo {author} {\bibfnamefont {A.}~\bibnamefont {Lupsasca}}, \ and\ \bibinfo {author} {\bibfnamefont {D.~P.}\ \bibnamefont {Marrone}},\ }\href {\doibase 10.1103/PhysRevD.102.124004} {\bibfield  {journal} {\bibinfo  {journal} {Phys. Rev. D}\ }\textbf {\bibinfo {volume} {102}},\ \bibinfo {pages} {124004} (\bibinfo {year} {2020})},\ \Eprint {http://arxiv.org/abs/2008.03879} {arXiv:2008.03879 [gr-qc]} \BibitemShut {NoStop}%
\bibitem [{\citenamefont {da~Silva}\ \emph {et~al.}(2023)\citenamefont {da~Silva}, \citenamefont {Lobo}, \citenamefont {Olmo},\ and\ \citenamefont {Rubiera-Garcia}}]{daSilva:2023jxa}%
  \BibitemOpen
  \bibfield  {author} {\bibinfo {author} {\bibfnamefont {L.~F.~D.}\ \bibnamefont {da~Silva}}, \bibinfo {author} {\bibfnamefont {F.~S.~N.}\ \bibnamefont {Lobo}}, \bibinfo {author} {\bibfnamefont {G.~J.}\ \bibnamefont {Olmo}}, \ and\ \bibinfo {author} {\bibfnamefont {D.}~\bibnamefont {Rubiera-Garcia}},\ }\href {\doibase 10.1103/PhysRevD.108.084055} {\bibfield  {journal} {\bibinfo  {journal} {Phys. Rev. D}\ }\textbf {\bibinfo {volume} {108}},\ \bibinfo {pages} {084055} (\bibinfo {year} {2023})},\ \Eprint {http://arxiv.org/abs/2307.06778} {arXiv:2307.06778 [gr-qc]} \BibitemShut {NoStop}%
\bibitem [{\citenamefont {Alencar}\ \emph {et~al.}(2026{\natexlab{b}})\citenamefont {Alencar}, \citenamefont {Crispim}, \citenamefont {S{\'a}ez-Chill{\'o}n~G{\'o}mez},\ and\ \citenamefont {de~S.~Silva}}]{Alencar:2026qeb}%
  \BibitemOpen
  \bibfield  {author} {\bibinfo {author} {\bibfnamefont {G.}~\bibnamefont {Alencar}}, \bibinfo {author} {\bibfnamefont {T.~M.}\ \bibnamefont {Crispim}}, \bibinfo {author} {\bibfnamefont {D.}~\bibnamefont {S{\'a}ez-Chill{\'o}n~G{\'o}mez}}, \ and\ \bibinfo {author} {\bibfnamefont {M.~V.}\ \bibnamefont {de~S.~Silva}},\ }\href {\doibase 10.1016/j.dark.2026.102427} {\bibfield  {journal} {\bibinfo  {journal} {Phys. Dark Univ.}\ }\textbf {\bibinfo {volume} {53}},\ \bibinfo {pages} {102427} (\bibinfo {year} {2026}{\natexlab{b}})},\ \Eprint {http://arxiv.org/abs/2603.05543} {arXiv:2603.05543 [gr-qc]} \BibitemShut {NoStop}%
\bibitem [{\citenamefont {Alencar}\ \emph {et~al.}(2026{\natexlab{c}})\citenamefont {Alencar}, \citenamefont {Bronnikov}, \citenamefont {Crispim}, \citenamefont {S{\'a}ez-Chill{\'o}n~G{\'o}mez},\ and\ \citenamefont {de~S.~Silva}}]{Alencar:2026gcl}%
  \BibitemOpen
  \bibfield  {author} {\bibinfo {author} {\bibfnamefont {G.}~\bibnamefont {Alencar}}, \bibinfo {author} {\bibfnamefont {K.~A.}\ \bibnamefont {Bronnikov}}, \bibinfo {author} {\bibfnamefont {T.~M.}\ \bibnamefont {Crispim}}, \bibinfo {author} {\bibfnamefont {D.}~\bibnamefont {S{\'a}ez-Chill{\'o}n~G{\'o}mez}}, \ and\ \bibinfo {author} {\bibfnamefont {M.~V.}\ \bibnamefont {de~S.~Silva}},\ }\href@noop {} {\  (\bibinfo {year} {2026}{\natexlab{c}})},\ \Eprint {http://arxiv.org/abs/2609.14199} {arXiv:2609.14199 [gr-qc]} \BibitemShut {NoStop}%
\bibitem [{\citenamefont {Silva}\ \emph {et~al.}(2026)\citenamefont {Silva}, \citenamefont {Junior}, \citenamefont {Pereira},\ and\ \citenamefont {S{\'a}ez-Chill{\'o}n~G{\'o}mez}}]{Silva:2026yao}%
  \BibitemOpen
  \bibfield  {author} {\bibinfo {author} {\bibfnamefont {M.~V. d.~S.}\ \bibnamefont {Silva}}, \bibinfo {author} {\bibfnamefont {E.~L.~B.}\ \bibnamefont {Junior}}, \bibinfo {author} {\bibfnamefont {C.~F.~S.}\ \bibnamefont {Pereira}}, \ and\ \bibinfo {author} {\bibfnamefont {D.}~\bibnamefont {S{\'a}ez-Chill{\'o}n~G{\'o}mez}},\ }\href@noop {} {\  (\bibinfo {year} {2026})},\ \Eprint {http://arxiv.org/abs/2610.04173} {arXiv:2610.04173 [gr-qc]} \BibitemShut {NoStop}%
\bibitem [{\citenamefont {Johnson}\ \emph {et~al.}(2020)\citenamefont {Johnson} \emph {et~al.}}]{Johnson:2019ljv}%
  \BibitemOpen
  \bibfield  {author} {\bibinfo {author} {\bibfnamefont {M.~D.}\ \bibnamefont {Johnson}} \emph {et~al.},\ }\href {\doibase 10.1126/sciadv.aaz1310} {\bibfield  {journal} {\bibinfo  {journal} {Sci. Adv.}\ }\textbf {\bibinfo {volume} {6}},\ \bibinfo {pages} {eaaz1310} (\bibinfo {year} {2020})},\ \Eprint {http://arxiv.org/abs/1907.04329} {arXiv:1907.04329 [astro-ph.IM]} \BibitemShut {NoStop}%
\bibitem [{\citenamefont {Chaverra}\ and\ \citenamefont {Sarbach}(2015)}]{Chaverra:2015bya}%
  \BibitemOpen
  \bibfield  {author} {\bibinfo {author} {\bibfnamefont {E.}~\bibnamefont {Chaverra}}\ and\ \bibinfo {author} {\bibfnamefont {O.}~\bibnamefont {Sarbach}},\ }\href {\doibase 10.1088/0264-9381/32/15/155006} {\bibfield  {journal} {\bibinfo  {journal} {Class. Quant. Grav.}\ }\textbf {\bibinfo {volume} {32}},\ \bibinfo {pages} {155006} (\bibinfo {year} {2015})},\ \Eprint {http://arxiv.org/abs/1501.01641} {arXiv:1501.01641 [gr-qc]} \BibitemShut {NoStop}%
\bibitem [{\citenamefont {Barrero-Gonz{\'a}lez}\ and\ \citenamefont {Cruz-Osorio}(2026)}]{Barrero-Gonzalez:2026xiu}%
  \BibitemOpen
  \bibfield  {author} {\bibinfo {author} {\bibfnamefont {D.}~\bibnamefont {Barrero-Gonz{\'a}lez}}\ and\ \bibinfo {author} {\bibfnamefont {A.}~\bibnamefont {Cruz-Osorio}},\ }\href@noop {} {\  (\bibinfo {year} {2026})},\ \Eprint {http://arxiv.org/abs/2609.38508} {arXiv:2609.38508 [astro-ph.HE]} \BibitemShut {NoStop}%
\bibitem [{\citenamefont {Bahamonde}\ and\ \citenamefont {Jamil}(2015)}]{Bahamonde:2015uwa}%
  \BibitemOpen
  \bibfield  {author} {\bibinfo {author} {\bibfnamefont {S.}~\bibnamefont {Bahamonde}}\ and\ \bibinfo {author} {\bibfnamefont {M.}~\bibnamefont {Jamil}},\ }\href {\doibase 10.1140/epjc/s10052-015-3734-9} {\bibfield  {journal} {\bibinfo  {journal} {Eur. Phys. J. C}\ }\textbf {\bibinfo {volume} {75}},\ \bibinfo {pages} {508} (\bibinfo {year} {2015})},\ \Eprint {http://arxiv.org/abs/1508.07944} {arXiv:1508.07944 [gr-qc]} \BibitemShut {NoStop}%
\bibitem [{\citenamefont {Debnath}(2014)}]{Debnath2014}%
  \BibitemOpen
  \bibfield  {author} {\bibinfo {author} {\bibfnamefont {U.}~\bibnamefont {Debnath}},\ }\href {\doibase 10.1140/epjc/s10052-014-2869-4} {\bibfield  {journal} {\bibinfo  {journal} {European Physical Journal C}\ }\textbf {\bibinfo {volume} {74}},\ \bibinfo {pages} {2869} (\bibinfo {year} {2014})}\BibitemShut {NoStop}%
\bibitem [{\citenamefont {Yusupova}\ \emph {et~al.}(2021)\citenamefont {Yusupova}, \citenamefont {Karimov}, \citenamefont {Izmailov},\ and\ \citenamefont {Nandi}}]{Yusupova2021}%
  \BibitemOpen
  \bibfield  {author} {\bibinfo {author} {\bibfnamefont {R.~M.}\ \bibnamefont {Yusupova}}, \bibinfo {author} {\bibfnamefont {R.~K.}\ \bibnamefont {Karimov}}, \bibinfo {author} {\bibfnamefont {R.~N.}\ \bibnamefont {Izmailov}}, \ and\ \bibinfo {author} {\bibfnamefont {K.~K.}\ \bibnamefont {Nandi}},\ }\href {\doibase 10.3390/universe7060177} {\bibfield  {journal} {\bibinfo  {journal} {Universe}\ }\textbf {\bibinfo {volume} {7}},\ \bibinfo {pages} {177} (\bibinfo {year} {2021})}\BibitemShut {NoStop}%
\bibitem [{\citenamefont {Rueda}\ and\ \citenamefont {Contreras}(2023)}]{Rueda:2023val}%
  \BibitemOpen
  \bibfield  {author} {\bibinfo {author} {\bibfnamefont {A.}~\bibnamefont {Rueda}}\ and\ \bibinfo {author} {\bibfnamefont {E.}~\bibnamefont {Contreras}},\ }\href {\doibase 10.1016/j.aop.2023.169540} {\bibfield  {journal} {\bibinfo  {journal} {Annals Phys.}\ }\textbf {\bibinfo {volume} {459}},\ \bibinfo {pages} {169540} (\bibinfo {year} {2023})},\ \Eprint {http://arxiv.org/abs/2311.08344} {arXiv:2311.08344 [gr-qc]} \BibitemShut {NoStop}%
\bibitem [{\citenamefont {Disconzi}(2024)}]{Disconzi:2023rtt}%
  \BibitemOpen
  \bibfield  {author} {\bibinfo {author} {\bibfnamefont {M.~M.}\ \bibnamefont {Disconzi}},\ }\href {\doibase 10.1007/s41114-024-00052-x} {\bibfield  {journal} {\bibinfo  {journal} {Living Rev. Rel.}\ }\textbf {\bibinfo {volume} {27}},\ \bibinfo {pages} {6} (\bibinfo {year} {2024})},\ \Eprint {http://arxiv.org/abs/2308.09844} {arXiv:2308.09844 [math.AP]} \BibitemShut {NoStop}%
\bibitem [{\citenamefont {Rocha}\ \emph {et~al.}(2024)\citenamefont {Rocha}, \citenamefont {Wagner}, \citenamefont {Denicol}, \citenamefont {Noronha},\ and\ \citenamefont {Rischke}}]{Rocha:2023ilf}%
  \BibitemOpen
  \bibfield  {author} {\bibinfo {author} {\bibfnamefont {G.~S.}\ \bibnamefont {Rocha}}, \bibinfo {author} {\bibfnamefont {D.}~\bibnamefont {Wagner}}, \bibinfo {author} {\bibfnamefont {G.~S.}\ \bibnamefont {Denicol}}, \bibinfo {author} {\bibfnamefont {J.}~\bibnamefont {Noronha}}, \ and\ \bibinfo {author} {\bibfnamefont {D.~H.}\ \bibnamefont {Rischke}},\ }\href {\doibase 10.3390/e26030189} {\bibfield  {journal} {\bibinfo  {journal} {Entropy}\ }\textbf {\bibinfo {volume} {26}},\ \bibinfo {pages} {189} (\bibinfo {year} {2024})},\ \Eprint {http://arxiv.org/abs/2311.15063} {arXiv:2311.15063 [nucl-th]} \BibitemShut {NoStop}%
\bibitem [{\citenamefont {Michel}(1972)}]{Michel1972}%
  \BibitemOpen
  \bibfield  {author} {\bibinfo {author} {\bibfnamefont {F.~C.}\ \bibnamefont {Michel}},\ }\href {\doibase 10.1007/BF00649949} {\bibfield  {journal} {\bibinfo  {journal} {Astrophysics and Space Science}\ }\textbf {\bibinfo {volume} {15}},\ \bibinfo {pages} {153} (\bibinfo {year} {1972})}\BibitemShut {NoStop}%
\bibitem [{\citenamefont {Rezzolla}\ and\ \citenamefont {Zanotti}(2013)}]{Rezzolla2013}%
  \BibitemOpen
  \bibfield  {author} {\bibinfo {author} {\bibfnamefont {L.}~\bibnamefont {Rezzolla}}\ and\ \bibinfo {author} {\bibfnamefont {O.}~\bibnamefont {Zanotti}},\ }\href@noop {} {\emph {\bibinfo {title} {Relativistic Hydrodynamics}}}\ (\bibinfo  {publisher} {Oxford University Press},\ \bibinfo {address} {Oxford},\ \bibinfo {year} {2013})\BibitemShut {NoStop}%
\bibitem [{\citenamefont {Kim}(2005)}]{Kim2005}%
  \BibitemOpen
  \bibfield  {author} {\bibinfo {author} {\bibfnamefont {H.}~\bibnamefont {Kim}},\ }\href {\doibase 10.1111/j.1365-2966.2005.09593.x} {\bibfield  {journal} {\bibinfo  {journal} {Monthly Notices of the Royal Astronomical Society}\ }\textbf {\bibinfo {volume} {364}},\ \bibinfo {pages} {813} (\bibinfo {year} {2005})}\BibitemShut {NoStop}%
\bibitem [{\citenamefont {Bean}\ \emph {et~al.}(2008)\citenamefont {Bean}, \citenamefont {Flanagan},\ and\ \citenamefont {Trodden}}]{BeanFlanaganTrodden2008}%
  \BibitemOpen
  \bibfield  {author} {\bibinfo {author} {\bibfnamefont {R.}~\bibnamefont {Bean}}, \bibinfo {author} {\bibfnamefont {E.~E.}\ \bibnamefont {Flanagan}}, \ and\ \bibinfo {author} {\bibfnamefont {M.}~\bibnamefont {Trodden}},\ }\href {\doibase 10.1103/PhysRevD.78.023009} {\bibfield  {journal} {\bibinfo  {journal} {Physical Review D}\ }\textbf {\bibinfo {volume} {78}},\ \bibinfo {pages} {023009} (\bibinfo {year} {2008})}\BibitemShut {NoStop}%
\end{thebibliography}%
\end{document}